\documentclass[letterpaper,11pt]{article}
\pdfoutput=1

\usepackage{jheppub}
\usepackage{multirow}
\usepackage[T1]{fontenc}
\usepackage[utf8]{inputenc}
\usepackage{subfig}
\usepackage[countmax]{subfloat}

\usepackage{amssymb}
\usepackage{amsmath}
\usepackage{cancel}
\usepackage{xcolor}
\usepackage{graphicx}
\usepackage{verbatim}
\usepackage{slashed}
\usepackage{epstopdf} %converting to PDF

\usepackage[normalem]{ulem}

\usepackage{pdfsync}
\def\cB{\mathcal{B}}
\def\cC{\mathcal{C}}

\def\cI{\mathcal{I}}
\def\cJ{\mathcal{J}}

\def\cL{\mathcal{L}}

\def\cN{\mathcal{N}}
\def\cO{\mathcal{O}}

\def\cP{\mathcal{P}}
\def\cJ{\mathcal{J}}

\def\tr{{\rm tr}}

\def\nn{{\nonumber}}
\def\mcdot{\!\cdot\!}

\newcommand{\Eq}[1]{Equation~\eqref{#1}}

\newcommand{\bea}{\begin{eqnarray}}
\newcommand{\eea}{\end{eqnarray}}
\def\beq{\begin{equation}}
\def\eeq{\end{equation}}

\def\sss{\scriptscriptstyle}

\def\gae{{\gamma_{\sss E}}}

\def\ugr{\gamma^{\sss R}}

\def\im{\mathrm{i}}

\newcommand{\gMuB}{\gamma_B}                     % Beam mu anom dim
\newcommand{\gMuS}{\gamma_S}                     % Soft mu anom dim
\newcommand{\gNu}{\gamma_\nu}                           % Nu anom dim: Only one, we choose gNu = gNu(soft)
\newcommand{\dlogmu}[1]{\mu \frac{\del}{\del\mu} #1}
\newcommand{\dlognu}[1]{\nu \frac{\del}{\del\nu} #1}

\newcommand{\del}{\mathrm{d}}

\newcommand{\bt}{\vec{b}_\perp}

\newcommand{\as}{\alpha_s}
\newcommand{\GammaC}{\Gamma_\text{cusp}}

\newcommand{\Ecm}{E_\mathrm{cm}}

\newcommand{\TEEC}{{\text{TEEC}}}

\DeclareRobustCommand{\Sec}[1]{Sec.~\ref{#1}}

\DeclareRobustCommand{\App}[1]{App.~\ref{#1}}

\DeclareRobustCommand{\Fig}[1]{Fig.~\ref{#1}}

\DeclareRobustCommand{\Eq}[1]{Eq.~(\ref{#1})}

\allowdisplaybreaks[3]

\newcommand{\eq}[1]{Eq.~\eqref{eq:#1}}
\newcommand{\eqs}[2]{Eqs.~\eqref{eq:#1} and \eqref{eq:#2}}
\renewcommand{\sec}[1]{Sec.~\ref{sec:#1}}
\newcommand{\secs}[2]{Secs.~\ref{sec:#1} and \ref{sec:#2}}

\newcommand{\Mae}[3]{\bigl\langle#1\bigr\rvert#2\bigr\rvert#3\bigr\rangle}

\newcommand{\df}{d}

\newcommand\bn{{\bar n}}

\newcommand{\e}{\epsilon}
\newcommand{\eps}{\epsilon}

\newcommand{\w}{\omega}

\newcommand{\brk}{\nonumber\\&}
\newcommand{\nbrk}{\nonumber\\}

\newcommand{\gsoft}{\gamma^{s}}

\def\ugr{\gamma^{r}}

\newcommand{\EEC}{{\text{EEC}}}

\def\Ord{\mathcal{O}}

\def\var{z_\phi}

\preprint{\vbox{\hbox{MIT--CTP 5662}}}

\title{The Transverse Energy-Energy Correlator at Next-to-Next-to-Next-to-Leading Logarithm}

\author[a]{Anjie Gao,}
\affiliation[a]{Center for Theoretical Physics, Massachusetts Institute of Technology, Cambridge, MA 02139, USA}

\author[b]{Hai Tao Li,}
\affiliation[b]{School of Physics, Shandong University, Jinan, Shandong 250100, China}

\author[c]{Ian Moult,}
\affiliation[c]{Department of Physics, Yale University, New Haven, CT 06511, USA\vspace{0.5ex}}

\author[d,e]{Hua Xing Zhu}
\affiliation[d]{School of Physics, Peking University, Beijing, 100871, China}
\affiliation[e]{Center for High Energy Physics, Peking University, Beijing 100871, China}

\emailAdd{anjiegao@mit.edu}
\emailAdd{haitao.li@sdu.edu.cn}
\emailAdd{ian.moult@yale.edu}
\emailAdd{zhuhx@pku.edu.cn}

\abstract{We present an operator based factorization formula for the transverse energy-energy correlator in the back-to-back (dijet) region, and uncover its remarkable perturbative simplicity and relation to transverse momentum dynamics. 
This simplicity enables us to achieve next-to-next-to-next-to leading logarithmic (N$^3$LL) accuracy for a hadron collider dijet event shape for the first time.
Our factorization formula applies to color singlet,  $W/Z/\gamma~+$ jet, and dijet production, providing a natural generalization of transverse momentum observables to one- and two-jet final states. 
This provides a laboratory for precision studies of QCD and transverse momentum dynamics at hadron colliders, as well as an opportunity for understanding factorization and its violation in a perturbatively well controlled setting.}

\begin{document}

%\todaytime
\maketitle
\flushbottom
%\newpage

%%%%%%%%%%%%%%%%%%%%%%%%%%%%%%%%%%%%%%%%%%%%%%%%%%%%%%%%%%%%%%%%%%%%%%%%%%%%%%%%
\section{Introduction}
\label{sec:introduction}
%%%%%%%%%%%%%%%%%%%%%%%%%%%%%%%%%%%%%%%%%%%%%%%%%%%%%%%%%%%%%%%%%%%%%%%%%%%%%%%%

Hadron colliders provide a rich environment for studying Quantum Chromodynamics (QCD). Of particular interest are observables that are amenable to first principles perturbative calculations, of which inclusive event shapes are some of the most well known examples. A large number of event shape variables have been precisely measured at the LHC (see e.g. \cite{Khachatryan:2011dx,Aad:2012np,Aad:2012fza,Chatrchyan:2013tna,Khachatryan:2014ika,ATLAS:2015yaa,Aaboud:2017fml,CMS:2018svp,ATLAS:2023tgo}), and have been used for applications ranging from extracting the value of the strong coupling constant \cite{ATLAS:2023tgo},  to deriving constraints on potential new colored particles \cite{Kaplan:2008pt,Llorente:2018wup}.

Due to remarkable theoretical progress, dijet event shapes in $e^+e^-$ colliders are now under excellent theoretical control, with state-of-the-art predictions incorporating   next-to-next-to-next-to-leading logarithmic (N$^3$LL) \cite{Becher:2008cf,Abbate:2010xh,Chien:2010kc,Hoang:2014wka} or N$^4$LL resummation \cite{Duhr:2022yyp} of singular contributions, next-to-next-to-leading order (NNLO) fixed order calculations \cite{GehrmannDeRidder:2007hr,Gehrmann-DeRidder:2007nzq,Weinzierl:2008iv,Weinzierl:2009ms}, and non-perturbative power corrections \cite{Abbate:2010xh}.  However, event shapes at hadron colliders are much less well understood. This is due not only to their increased perturbative complexity, but also due to a lack of understanding of the applicability of factorization. In particular, it is known that Glauber effects invalidate standard factorization formulas \cite{Collins:2007nk,Collins:2007jp,Bomhof:2007su,Rogers:2010dm,Buffing:2013dxa,Gaunt:2014ska,Zeng:2015iba,Catani:2011st,Schwartz:2017nmr,Forshaw:2008cq,Forshaw:2006fk,Martinez:2018ffw,Angeles-Martinez:2016dph,Forshaw:2012bi,Angeles-Martinez:2015rna,Schwartz:2018obd,Rothstein:2016bsq,Forshaw:2021fxs}, both through spectator interactions, and through the invalidation of collinear factorization for spacelike splittings with multiple colored collinear Wilson lines \cite{Dixon:2019lnw}. Alternatively, this makes precision measurements and high order calculations of hadron collider event shapes of significant interest for understanding a variety of aspects related to QCD factorization.

For hadron collider event shapes, there has been spectacular recent progress on the fixed order side, with the NNLO calculation of hadron collider event shapes \cite{Alvarez:2023fhi}. However, there has been much less progress in the study of resummation and factorization for hadron collider event shapes. Next-to-leading logarithmic  (NLL) resummation has been achieved for a large number of observables \cite{Banfi:2004nk,Banfi:2010xy}. Next-to-next-to-leading logarithmic  (NNLL) resummation has only been achieved for zero-jet \cite{Stewart:2010pd,Becher:2015lmy,Becher:2015gsa,Becher:2015lmy,Becher:2015gsa}, and  one-jet event shapes \cite{Jouttenus:2013hs}. N$^3$LL resummation was recently achieved for one-jettiness \cite{Alioli:2023rxx}. However,  the most interesting dynamics occurs when there are multiple incoming and outgoing colored particles, which first occurs for dijet event shapes.  Due to the simultaneous complexity of the color flow and the observables typically considered, relatively little progress has been made in extending resummed calculations for dijet event shapes to higher perturbative orders. However, these observables are interesting both practically, as well as theoretically, for studying the generic structure of factorization theorems.

In this paper we uncover the perturbative simplicity of the transverse energy-energy correlator (TEEC) dijet event shape observable and exploit this to achieve an unprecedented N$^3$LL accuracy for a hadron collider dijet event shape, extending the NNLL results presented in \cite{Gao:2019ojf}, and describing some additional aspects of the calculation. While the TEEC has been measured at the LHC \cite{ATLAS:2015yaa,Aaboud:2017fml,ATLAS:2023tgo} (we note that this measurement is on jets instead of hadrons)  and used to extract $\alpha_s$, it has received relatively little theoretical attention (see, however, an NLO calculation of the TEEC for jets \cite{Ali:2012rn} and the recent remarkable NNLO calculation  \cite{Alvarez:2023fhi}.). For an interesting recent application of the TEEC to the study of saturation at the future EIC, see \cite{Li:2021txc,Li:2020bub,Kang:2023oqj,Cao:2023qat}.  On the other hand, the energy-energy correlator (EEC) $e^+e^-$ event shape has recently received significant theoretical attention, including analytic fixed order calculation at NLO \cite{Dixon:2018qgp,Luo:2019nig,Gao:2020vyx} in QCD, and to NNLO in $\cN=4$  \cite{Belitsky:2013xxa,Belitsky:2013bja,Belitsky:2013ofa,Henn:2019gkr}, a factorization and derivation of the singular structure in the back-to-back limit \cite{Moult:2018jzp,Ebert:2020sfi}, an understanding of the all orders structure in the collinear limit \cite{Dixon:2019uzg,Kologlu:2019mfz,Korchemsky:2019nzm,Chen:2020vvp,Chen:2023zzh}, numerical calculations and extractions of $\alpha_s$ \cite{DelDuca:2016ily,Tulipant:2017ybb}, and a fixed order calculation of the three-point correlator in the collinear limit \cite{Chen:2019bpb} and at general angles \cite{Yan:2022cye,Yang:2022tgm}. They have also been applied in a wide range of phenomenological applications from high energy to nuclear physics \cite{Komiske:2022enw,Holguin:2022epo,Liu:2022wop,Liu:2023aqb,Cao:2023oef,Devereaux:2023vjz,Andres:2022ovj,Andres:2023xwr,Craft:2022kdo,Lee:2022ige}. We will illustrate how many of these nice features carry over to the back-to-back limit in the  hadron collider case.

In this paper we derive an operator based factorization formula for the TEEC in the back-to-back (dijet) region, which was first presented without derivation in \cite{Gao:2019ojf}. This factorization is derived using soft collinear effective theory (SCET) \cite{Bauer:2000ew, Bauer:2000yr, Bauer:2001ct, Bauer:2001yt}, and takes a remarkably simple form, being essentially a projection of transverse momentum factorization onto a scattering plane. In particular, the factorization formula involves the standard transverse momentum dependent (TMD) beam functions, as well as the TMD fragmentation functions, making it interesting for studying TMD dynamics.  We describe the structure of the renormalization group evolution, give all ingredients required to achieve N$^3$LL resummation, and present numerical results. 

Our calculation incorporates a large number of the most precisely known perturbative ingredients in QCD, namely the:
\begin{itemize}
\item Three loop quadrupole soft anomalous dimension \cite{Almelid:2015jia,Almelid:2017qju},
\item Three loop rapidity anomalous dimension \cite{Li:2016ctv},
\item Four loop cusp anomalous dimension \cite{Moch:2018wjh,Moch:2017uml,Davies:2016jie,Henn:2019swt},
\item NNLO TMD PDFs \cite{Gehrmann:2012ze,Gehrmann:2014yya,Echevarria:2016scs,Luebbert:2016itl,Luo:2019hmp,Luo:2019bmw},
\item NNLO TMD Fragmentation functions \cite{Echevarria:2016scs,Luo:2019hmp,Luo:2019bmw},
\item NNLO TEEC Soft function \cite{Gao:2019ojf},
\item NNLO $2\to 2$ scattering amplitudes \cite{Anastasiou:2000kg, Anastasiou:2000ue,Glover:2001af, Bern:2002tk, Bern:2003ck, Glover:2003cm,Glover:2004si, DeFreitas:2004kmi},
\end{itemize}
which illustrates the remarkable power of factorization, as well as the complexity of event shapes in a hadron collider environment. 

Since the publication of our original result for the TEEC \cite{Gao:2019ojf}, there has been significant progress in perturbative calculations at hadron colliders, in particular the calculation of $2\to 3$ jet cross section \cite{Czakon:2021mjy} and hadron collider event shapes \cite{Alvarez:2023fhi} at NNLO (building on significant progress in the understanding of the properties of $2\to 3$ amplitudes at NNLO, see e.g. \cite{Abreu:2023rco,Chicherin:2021dyp,Chicherin:2020oor,Badger:2019djh,Chicherin:2018old,Chicherin:2018mue,Chicherin:2018yne,Abreu:2018aqd,Gehrmann:2015bfy,Abreu:2018aqd,Abreu:2018jgq,Abreu:2017hqn,Abreu:2018zmy}). This result provides the necessary order to match to our resummed N$^3$LL result. Additionally, many of the perturbative ingredients necessary to extend the factorization to higher orders have appeared. These included the N$^3$LO TMD PDFs and fragmentation functions \cite{Luo:2019szz,Luo:2020epw,Ebert:2020qef,Ebert:2020yqt}, the four loop rapidity anomalous dimension \cite{Moult:2022xzt,Duhr:2022yyp}, and the $2\to 2$ scattering amplitudes at three loops \cite{Bargiela:2021wuy,Caola:2021izf,Caola:2021rqz,Caola:2020dfu}. We therefore believe that it is timely to discuss the higher order structure of hadron collider event shapes.

In addition to its direct phenomenological relevance, we believe that the TEEC provides a particularly clean laboratory for studying factorization violation and rapidity factorization at high perturbative orders.  For the purposes of studying factorization violation, we also highlight a particularly interesting feature of the TEEC, namely that it can be defined not only for dijets, but also for $W/Z/\gamma$+jet events by demanding that one of the energy correlators lies on the $W/Z/\gamma$, as well as for Drell-Yan with the $\gamma/Z$ decaying to leptons, where both correlators are placed on the leptons. Using the results of this paper, the TEEC in all these different final states can be computed at N$^3$LL. We believe that by having the same observable, but with distinct final states, at this level of perturbative accuracy will allow for the detailed study of color flow and factorization violation. In particular, it is known that the TEEC will factorize for the Drell-Yan process (where it is related to the $q_T$ observable which factorizes  \cite{Collins:1981uk,Collins:1981va,Collins:1981ta,Collins:1984kg,Collins:1985ue,Collins:1988ig,Collins:1989gx}), while for the dijet process, it is known that it will not factorize. For $W/Z/\gamma$, the status is unclear. In this sense, we can view the TEEC as a generalization of $q_T$ to final state jets. We hope that this will enable advances in the understanding of QCD in hadronic collisions.

An outline of this paper is as follows. In \Sec{sec:transv-energy-energy} we define the TEEC and describe its relation to the more familiar EEC observable. In \Sec{sec:fact-teec-backw} we discuss the kinematics of the TEEC in the back-to-back limit, and derive the factorization formula describing this limit. We also briefly discuss factorization violation and the effect of underlying event.  
In \Sec{sec:soft} we discuss in detail the soft function appearing in the description of the TEEC in the back-to-back limit, which is the primary new perturbative ingredient required for the TEEC. In \Sec{sec:ingr-resumm-teec} we present a solution to the color space matrix renormalization group equations for the soft function. In \Sec{sec:linear_polarization} we discuss linearly polarized beam and jet functions that first contribute at N$^3$LL. In \Sec{sec:numerics} we verify the singular structure predicted by the factorization formula by comparing with a numerical calculation using the NLO three jet cross section. We also make predictions for the singular behavior of the three jet cross section at NNLO. In \Sec{sec:numerics_resum} we present resummed results for the TEEC at NNLL and N$^3$LL.  We conclude in \Sec{sec:conclusion}.

%%%%%%%%%%%%%%%%%%%%%%%%%%%
\section{The Transverse Energy-Energy Correlator}
\label{sec:transv-energy-energy}
%%%%%%%%%%%%%%%%%%%%%%%%%%%

In this section we define the TEEC, comparing its definition to the more standard EEC. We also provide definitions of the TEEC which are applicable to color singlet production, and for $W/Z/\gamma+$ jet production. We believe that the ability to compute the observable at N$^3$LL for  three distinct final states,  with different color flows will provide an important handle in the study of QCD event shape that has not been available previously.

\begin{figure}
  \centering
\subfloat[]{\label{fig:1a}
\includegraphics[width=5cm]{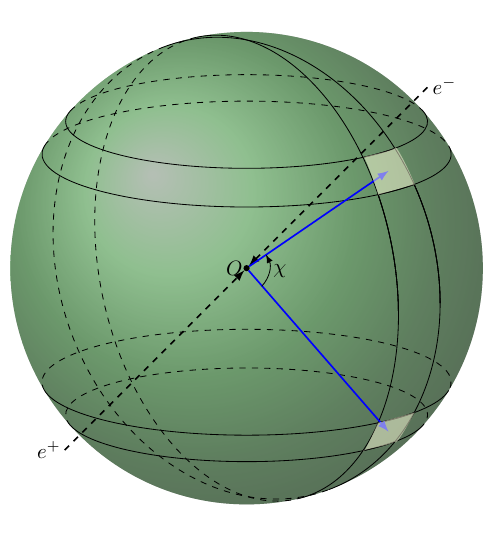}
}  \qquad \qquad
\subfloat[]{\label{fig:1b}
\includegraphics[width=7cm]{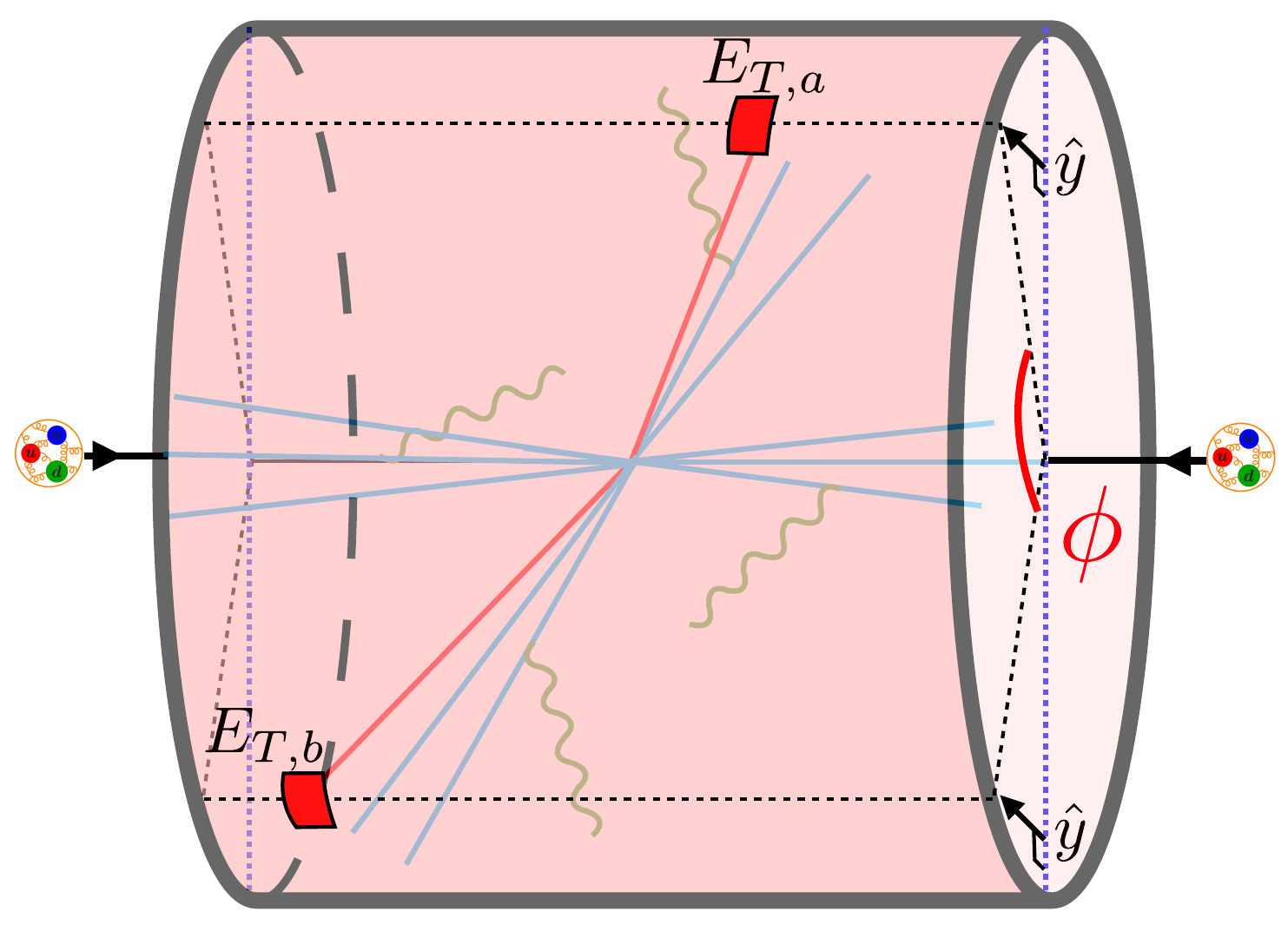}
}  
  \caption{a) The EEC observable in $e^+e^-$ collisions measures the correlation between energy depositions as a function of the angle $\chi$ on the sphere. b)
  The TEEC, which generalizes the EEC to hadronic collisions, measures the correlation between energy depositions as a function of the angle $\phi$ in the plane transverse to the scattering.}
  \label{fig:1}
\end{figure}

The EEC in $e^+e^-$ is defined as \cite{Basham:1979gh,Basham:1978zq,Basham:1978bw,Basham:1977iq}
\begin{align}\label{eq:EEC_intro}
\text{EEC}(\chi)=\sum\limits_{a,b} \int d\sigma_{V\to a+b+X} \frac{2 E_a E_b}{Q^2 \sigma_{\text{tot}}}   \delta(\cos(\theta_{ab}) - \cos(\chi))\,.
\end{align}
Here the sum is over all \emph{different} pairs of hadrons $h_a$ and $h_b$ in the event. It measures the flow of energy in two calorimeters separated by an angle $\chi$, as shown in \Fig{fig:1a}.

The TEEC is the natural extension of the EEC to a hadron collider, and measures the flow of energy in two calorimeters separated by an angle $\phi$ in the plane transverse to the scattering \Fig{fig:1b}. It is defined as  \cite{Basham:1978bw,Ali:1984yp}
\begin{align}\label{eq:TEEC_intro}
\frac{d\sigma}{d\cos \phi}=\sum\limits_{a,b} \int d\sigma_{pp\to a+b+X} \frac{2 E_{T,a} E_{T,b}}{ |\sum_i E_{T,i}|^2 }   \delta(\cos\phi_{ab} - \cos\phi)\,. 
\end{align}
Much like the EEC, the TEEC exhibits singularities at the two extremes of its phase space, which must be resummed to all orders. Much like the EEC, at $\phi=\pi$ we have the back-to-back (dijet) region, which is characterized by Sudakov double logarithms. The $\phi=0$ region is characterized by a collinear limit, which exhibits single collinear logarithms. The focus in this paper will be on the back-to-back region. The resummation of the collinear logarithms can be performed similarly to the case of $e^+e^-$ \cite{Dixon:2019uzg}, which represents an extension of the jet calculus  \cite{Kalinowski:1980wea,Konishi:1979cb,Richards:1982te}.

We can also define extensions of the TEEC for the case of $V+$ jet and Drell-Yan (or more generally arbitrary color singlet production). For the case of $V+$ jet, we define the TEEC as
\begin{align}\label{eq:TEEC_intro_Vjet}
\frac{d\sigma}{d\cos \phi}=\sum\limits_{a} \int d\sigma_{pp\to V+a+X} \frac{ E_{T,a}}{ \sum\limits_{i\in \text{hadrons}} E_{T,i}}   \delta(\cos\phi_{Va} - \cos\phi)\,, 
\end{align}
while for Drell-Yan, we define it as
\begin{align}\label{eq:TEEC_intro_drellyan}
\frac{d\sigma}{d\cos \phi}= \int d\sigma_{pp\to l^++l^-+X}   \delta(\cos\phi_{l^+l^-} - \cos\phi)\,.
\end{align}
Note that in all cases, the definition of the TEEC is chosen such that it obeys a sum rule, namely 
\begin{align}
\int d\cos\phi \frac{d\sigma}{d\cos \phi} =\sigma_{\text{tot}}\,.
\end{align}
The factorization formula that we will derive will trivially also apply to these cases. We believe that having an observable defined for final states with $0$, $1$ and $2$ final state jets, all of which can be computed at N$^3$LL, is particularly interesting from the perspective of studying factorization violation. Another interest in defining the TEEC for these additional final states, in particular $W/Z/\gamma$, is also that they might be the first to which one can match at NNLO. For recent progress towards the $pp \to V+2$ jet amplitudes at NNLO, see \cite{Hartanto:2019uvl}.

We should also note that the TEEC observable is similar to other observables that measure azimuthal correlations in DIS \cite{Banfi:2002vw} in back-to-back jets in $e^+e^-$ or $pp$ \cite{Collins:1981uk}, or between jets and vector bosons \cite{Buffing:2018ggv,Chien:2019gyf,Chien:2022wiq,Chien:2020hzh}.

%%%%%%%%%%%%%%%%%%%%%%%%%%%%
\section{Factorization in the Back-to-Back Limit}
\label{sec:fact-teec-backw}
%%%%%%%%%%%%%%%%%%%%%%%%%%%%

In this section, we derive the factorization formula for the TEEC in the back-to-back limit, which was first presented without derivation in \cite{Gao:2019ojf}. Due to the simplicity of the TEEC observable, we will find that the factorization formula in this region can be expressed in terms of well known functions used in the description of other observables, but combined in a non-trivial way. This is particularly convenient, since it immediately allows us to derive all required anomalous dimensions using existing results in the literature. The factorization we present is expected to be violated at higher orders by Glauber contributions, and we discuss this in \Sec{sec:ingr-resumm-teec-soft-caveat}.

%%%%%%%%%%%%%%%%%%%%%%%%%%%%
\subsection{Kinematics}
\label{sec:kinematics}
%%%%%%%%%%%%%%%%%%%%%%%%%%%%

In this subsection, we identify the relevant kinematical regions for $\var \to 1$. We first review the kinematics for the back-to-back limit of the EEC in $e^+e^-$ colliders. We then generalize this discussion to the case of a hadron collider.

In $e^+e^-$ collisions, it is customary to define a dimensionless variable, $z = (1 - \cos\chi)/2$, where $\chi$ is the angle between two outgoing particles with momenta $k_a$ and $k_b$. In the massless limit, 
\begin{align}
1 - z =  \frac{1 + \cos\chi}{2} = \frac{k_a^0 k_b^0 + \vec k_a \cdot \vec k_b} {2 k_a^0 k_b^0} \,.
\end{align}
In the back-to-back limit, $k_a$ is almost anti-aligned with $k_b$, and we have
\begin{align}
1 - z =  \frac{1 - \cos(\pi - \chi)}{2}  \sim \frac{(\pi - \chi)^2}{4} + \Ord((\pi - \chi)^4)\,.
\end{align}
At leading power, $k_a$ and $k_b$ must come from the splitting of two almost back-to-back jets with momentum $p_1$ and $p_2$. The two jets are not exactly back-to-back due to soft radiation. Let $\chi_J = \vec p_1 \cdot \vec p_2 /(|\vec p_1||\vec p_2|)$ be the angle between the two jets. Its deviation from the exact back-to-back limit, $\chi_J = \pi$, is given by
\begin{align}
\frac{(\pi - \chi_J)^2}{4} =  \frac{\vec{k}_{s\perp}^2}{Q^2} \,,
\end{align}
where $\vec{k}_{s\perp}$ is the transverse momentum of the soft radiation, defined either against $p_1$, or $p_2$, or even the thrust axis of the whole event. Different choices for the axes only lead to power suppressed effects, and are thus irrelevant to the discussion in this paper. $\chi$ differs from $\chi_J$ due to transverse recoil of $k_a$ and $k_b$ against $p_1$ and $p_2$, respectively. Taking this into account, one obtains
\begin{align}
(1 - z) \sim \left( \frac{\vec k_{a\perp}}{\xi_a Q} + \frac{\vec k_{b\perp}}{\xi_b Q} - \frac{\vec{k}_{s\perp}}{Q} \right)^2 + \Ord((1-z)^2)\,,
\label{eq:zdef}
\end{align}
where $\vec k_{a(b)\perp}$ is the transverse momentum of $k_{1(2)}$ against $p_{1(2)}$, and $\xi_{a(b)}$ is the corresponding longitudinal momentum fraction. The fact that $(1-z)$ can be related to the transverse momentum of soft or collinear states leads to enormous simplifications for resummation: 1) many of the ingredients for resummation can be adopted from $q_T$ resummation for Drell-Yan/Higgs production at hadron colliders; 2) the vector sum nature of Eq.~\eqref{eq:zdef} admits a simple factorization of soft and collinear modes in impact parameter space; 3) most importantly, the soft radiation is defined globally, avoiding the need to partition the phase space into different angular directions as in the case of the beam thrust (or $N$-jettiness) soft function. 

Now we would like to apply the above observations to the TEEC in the back-to-back limit. Much like for the EEC, it is convenient to work with the variable
\begin{equation}
  z_\phi=\frac{1-\cos\phi}{2}\,.
\end{equation}
As compared to the case of $e^+e^-$, we use the subscript $\phi$ to emphasize that this variable corresponds to the TEEC.
We consider the LO partonic scattering process, $p_1 + p_2 \to p_3 + p_4$. Generalizing this to the hadronic scattering process will be straightforward in the factorization formula. At LO, the correlation localizes at $\phi \equiv 0$ in the azimuthal plane. The incoming and outgoing momentum $p_i$ span the scattering plane, which we choose to be the $x$-$z$ plane, where the $z$-axis is the beam axis.  Non-trivial $\phi$ dependence is generated through radiation \emph{out of} the scattering plane. In the back-to-back limit where $\phi \to \pi$, hard radiation out of the plane is power suppressed, enforcing that the shape of the event is almost planar. After integrating out the hard virtual contributions, the relevant low energy modes are collinear radiation in the initial and final state, as well as global soft radiation. A simple derivation shows that in this limit we have
\begin{align}
  \label{eq:ydef}
  1 - \var = \sin^2\frac{\pi-\phi_{ab}}{2}=\frac{1}{4 p_T^2}\left(\frac{k_{a,y}}{\xi_a}+\frac{k_{b,y}}{\xi_b}+p_{1,y}+p_{2,y}-k_{s,y}\right)^2+\mathcal{O}((1-\var)^2) \,.
\end{align}
Here $p_1$ and $p_2$ are initial state partons which enter the hard scattering vertex. Initial-state splittings result in non-zero transverse momentum off the scattering plane, which we denote as $p_{1,y}$ and $p_{2,y}$. $k_a$ and $k_b$ are the momenta of final-state particles from $p_3$ and $p_4$ respectively, whose transverse energy correlation is to be measured. They acquire non-zero transverse momentum off the scattering plane due to final-state collinear splitting. $\xi_a$ and $\xi_b$ are their respective longitudinal momentum fraction relative to $p_3$ and $p_4$. $k_{s,y}$ is the transverse momentum of soft radiation off the scattering plane. $p_T$ is the LO transverse momentum of $p_3$ and $p_4$ relative to the beam axis, which plays the role of hard scale in the problem. When discussing the factorization in the back-to-back limit, we will often use
\begin{align}\label{eq:def_tau}
\tau \equiv 1-z_\phi \,,
\end{align}
since it is the appropriate resolution variable for characterizing the back-to-back limit.

Comparing \eqs{zdef}{ydef}, one notices the close similarity between the two. The two-dimensional transverse momentum in Eq.~\eqref{eq:zdef} is replaced by the one-dimension transverse momentum in Eq.~\eqref{eq:ydef}. There is also an additional contribution from initial-state physics in Eq.~\eqref{eq:ydef}. This makes clear that one should view the TEEC in the back-to-back limit as a sort of generalization of the standard $q_T$ observable to dijet final states.

%%%%%%%%%%%%%%%%%%%%
\subsection{Factorization Formula}
\label{sec:fact-form-backw}
%%%%%%%%%%%%%%%%%%%%

We can now derive a factorization formula for the TEEC in the back-to-back limit. This follows closely the derivation for the EEC in \cite{Moult:2018jzp}, which in turn builds on the factorization formula for identified hadron production in the back-to-back limit \cite{Collins:1981uk,Collins:1981va}. In particular, we will use the approach of factorizing a multi-differential cross section in terms of beam and fragmentation functions. We will then show that the integration over this multi-differential cross section to obtain the TEEC allows the use of sum rules to eliminate non-pertubative contributions from the fragmentation functions, as expected for an Infrared and collinear safe observable. Throughout this section, we will suppress Lorentz indices for simplicity. At N$^3$LL, non-trivial spin structures give rise to interesting effects in the form of linearly polarized jet and beam functions. We will discuss these in more detail in \Sec{sec:linear_polarization}. 

To derive the factorization, we start with the expression
\begin{equation}\label{eq:triple_diff}
\frac{d\sigma}{d\var}=\frac{1}{2}\sum_{\text{channels}}\sum_{h,h'}\int\! d\zeta_h\,  d\zeta_{h'}\,\zeta_h \,\zeta_{h'}\,\frac{d^3\sigma}{d\zeta_h d\zeta_{h'} d\var}\,,
\end{equation}
where $h,h'$ are summed over all the final-state hadrons, and
$\frac{d^3\sigma}{d\zeta_h d\zeta_{h'} d\var}$ is the triple differential cross section measuring the transverse energy fraction with respect to the transverse energy of the jet, $\zeta_h=p_{h,T}/p_T$, and the relative angle $\var$. We will denote the momenta of the incoming partons as $p_1^\mu$ and $p_2^\mu$, and $p_3^\mu$ and $p_4^\mu$ are the outgoing parton level quarks or gluons. In the limit that $p_1^\mu$ and $p_2^\mu$ have very small transverse energy, $\vec{p}_1$, $\vec{p}_2$, $\vec{p}_3$, and $\vec{p}_4$ are almost on a plane, which we denote as $zx$-plane, and the transverse momenta of two outgoing partons are nearly the same, which we denote $p_T$. 

In the limit that $h$ and $h'$ are back-to-back, let hadrons $h$ and $h'$ be emitted from $p_3$ and $p_4$ respectively, so that their transverse energy fraction of $\zeta_{h(h')}$ and longitudinal momentum fraction $\xi_{h(h')}$ relative to $p_{3(4)}$ can be used interchangeably. From now on, we will replace $\zeta_{h(h')}$ in \eq{triple_diff} by $\xi_{h(h')}$.
To leading power, we only need to consider recoil effects from soft radiation and collinear fragmentation in the $y$-direction, so that
\begin{equation}
1-z_\phi=\sin^2\frac{\pi-\phi}{2}=\frac{1}{4p_T^2}\left|\frac{k_{h,y}}{\xi_h}+\frac{k_{h',y}}{\xi_{h'}}+p_{1,y}+p_{2,y}-k_{s,y}\right|^2+\mathcal{O}((1-z_\phi)^2).
\end{equation}
Compared with \eq{ydef}, here we use $h$ and $h'$ instead of $a$ and $b$ to emphasize that they are hadrons. 
We can change the variable $\var$ for the $y$-momentum $q_y=k_{h,y}/\xi_h+k_{h',y}/\xi_{h'}+p_{1,y}+p_{2,y}-k_{s,y}$,
\beq
\frac{d^3\sigma}{d\xi_h\, d\xi_{h'}\, d\var}=\int\! dq_y\,\frac{d^3\sigma}{d\xi_h\, d\xi_{h'}\, dq_y}\delta\left(1-\var-\frac
{q_y^2}{4p_T^2}\right).
\eeq
We now proceed to factorize the triple differential distribution,
\begin{align}
  \label{eq:triple}
 &\sum_{h,h'} \frac{d^3 \sigma}{d\xi_h \, d\xi_{h'}\, d\var} = \\
 &\hspace{1cm} \sum_{h,h'}\sum_X\nolimits' \langle P_1P_2 | X \rangle \delta(\xi_h - k_{h,T}/p_T) \delta(\xi_h - k_{h',T}/p_T)\, \delta\biggl(\var - \frac{1}{2} - \frac{\vec{k}_{h\perp} \cdot \vec{k}_{h'\perp}}{2|\vec{k}_{h\perp}||\vec{k}_{h'\perp}|}\biggr) \langle X | P_1P_2 \rangle \,, \nn
\end{align}
where we sum over all hadronic states $X$. Here $| P_{1,2} \rangle$ denote the incoming proton state, and $k_h$ and $k_{h'}$ are two detected particles. $\vec{k}_{h(h')\perp}$ is the transverse momentum of the detected particle against the beam axis. The phase space summation $\sum_X\nolimits'$, is restricted by experimental cuts to select only hard scattering events. For example, the ATLAS measurement for TEEC \cite{ATLAS:2015yaa,Aaboud:2017fml,ATLAS:2023tgo} imposes an average $p_T \geq 250$ GeV for two leading jets, and rapidity $|Y|\leq 2.5$.  Throughout this paper we work in the high energy limit such that the detected hadrons are taken to be massless.

In the limit of $\var \to 1$, the radiation in the event is restricted to lie in a plane, as explained in Sec.~\ref{sec:kinematics}. An illustration of a typical event is depicted in \Fig{fig:1b}, where there are bunches of collinear particles emitted in the beam and jet directions (shown in light blue), and soft particles emitted in all directions (shown in green). To describe the dynamics in this limit, we use the soft collinear effective theory (SCET) \cite{Bauer:2000ew, Bauer:2000yr, Bauer:2001ct, Bauer:2001yt}, which will allow us to provide a factorized description of the dynamics of the soft and collinear radiation. Throughout this section, we do not consider cross talk between the beam remnants and final-state jets, which could potentially invalidate the factorization picture (A brief discussion of factorization violation is given in \Sec{sec:ingr-resumm-teec-soft-caveat}). However, there is evidence that such factorization violation effects exist due to the exchange of Glauber gluons at high orders in perturbation theory. To the accuracy considered in this paper, namely NNLO in fixed-order perturbation and  N$^3$LL in resummed perturbation theory, we strongly believe that the factorization picture is not spoiled. At higher perturbative orders, we believe that our factorization formula can serve as a concrete foundation for a quantitative understanding of factorization violating effects at hadron collider. 

To achieve a factorized description of the TEEC in the back-to-back limit, short distance physics is first integrated out and matched onto a set of SCET hard operators which describe the hard scattering. In our case, these are a set of hard operators that describe the short distance $2\to 2$ scattering processes, e.g., $q\bar{q} \to q'\bar{q}'$, $q\bar{q} \to gg$, $g g \to gg$, etc. We will take $q\bar{q} \to q'\bar{q}'$ as the primary example in our derivation of factorization, but the generalization to other processes is straightforward. The relevant leading power SCET hard operators can be schematically written as
 \begin{align}
{\cal O}_{q\bar{q} q'\bar{q}'} = \sum_I \sum_{\Gamma} {\cal C}_I^{\Gamma} \bar{\chi}_2  \chi_1 \bar{\chi}_3'  \chi_4'  \Gamma t_I\,,
\label{eq:hardO}
 \end{align}
where $\chi_i$ is the gauge invariant collinear quark or anti-quark field in the lightcone direction $n_i$, $\Gamma$ is a basis of Dirac structures, and $t_I$ is a basis of color structures. ${\cal C}_I^{\Gamma}$ is the Wilson coefficient resulted from integrating out the hard modes. We have suppressed the color indices, Lorentz indices, and kinematical dependence in Eq.~\eqref{eq:hardO}.

The leading power SCET Lagrangian describing the dynamics of the TEEC can be written as
\begin{align}
\cL^{(0)}=\cL^{(0)}_{B_1}+\cL^{(0)}_{B_2}+\cL^{(0)}_{J_1}+\cL^{(0)}_{J_2} +\cL^{(0)}_G\,.
\end{align}
Here $\cL^{(0)}_G$ is the Glauber Lagrangian \cite{Rothstein:2016bsq}, which contributes to factorization violation. We will return to this in \Sec{sec:ingr-resumm-teec-soft-caveat}, but for now we set $\cL^{(0)}_G=0$. Once this is done, the dynamics of the different collinear sectors exactly factorizes. This implies that the external state also factorizes into collinear and soft states, 
\begin{align}
  \label{eq:state_fac}
  | X \rangle \to |X_1 \rangle |X_2 \rangle |X_3 \rangle |X_4 \rangle |X_s \rangle \,.
\end{align}
From Sec.~\ref{sec:kinematics} it's convenient to define an auxiliary observable
\begin{align}
  \label{eq:qperp}
    q_y=\frac{k_{h,y}}{\xi_h}+\frac{k_{h',y}}{\xi_{h'}}+p_{1,y}+p_{2,y}-k_{s,y}
\end{align}
where we choose the $y$ component to be the transverse component perpendicular to the scattering plane spanned by the incoming beams and outgoing jets. 
 Writing the triple differential distribution for $d^3\sigma/(d\xi_h\, d\xi_{h'}\, dq_y)$ in terms of SCET hard operators and fields, we obtain 
\begin{align}
 \sum_{h,h'} \frac{d^3 \sigma}{d\xi_h \, d\xi_{h'}\, dq_y} &=\ 2 \sum_{h \in X_3, h' \in X_4}\sum_{X_1,X_2,X_3,X_4,X_s} \langle P_1P_2 | {\cal O}_{q\bar{q} q'\bar{q}'}^\dagger |X_1 \rangle |X_2 \rangle |X_3 \rangle |X_4 \rangle |X_s \rangle 
\nn\\
&\ \times\delta(\xi_h - k_{h,T}/p_T) \delta(\xi_{h'} - k_{h',T}/p_T) \delta\left(q_y - \left( \frac{k_{h,y}}{\xi_h}+\frac{k_{h',y}}{\xi_{h'}}+p_{1,y}+p_{2,y}-k_{s,y}\right) \right) 
\nn\\
&\
\times \langle X_1 | \langle X_2| \langle X_3| \langle X_4| \langle X_s |{\cal O}_{q\bar{q} q'\bar{q}'} | P_1P_2 \rangle + \text{power corrections} \,.
\end{align}
At leading power, $k_h$ and $k_{h'}$ are collinear particles belonging to final-state jets $X_3$ and $X_4$, respectively. They cannot be soft, since the observable is weighted by the energy of the detected particle.  An overall factor of $2$ arises due to restricting $k_h$ to be in jet $X_3$.

With the dynamics factorized, it is now a standard algebraic exercise to manipulate the operators into matrix elements separately describing the dynamics of the different collinear sectors, and the soft sector (for a detailed discussion in the context of jet cross sections at the LHC, see \cite{Stewart:2009yx} ).
We can write our full expression for the leading power TEEC cross section in the back-to-back limit as
\begin{align}\label{eq:full}
\frac{\df\sigma^{(0)}}{\df \var} =\,& \frac{1}{16\pi s^2}\sum\limits_{\text{channels}} \sum\limits_{IJ}\sum\limits_{hh'}\frac{1}{(1 + \delta_{f_3 f_4}) N_{\text{init}}} \int \frac{\df y_3 \df y_4 \df p_T^2}{\xi_1\,\xi_2} 
H_{IJ}^{f_1f_2\to f_3f_4}(p_T,y_3,y_4,\mu)  
 \\\nn
 & \times  
\int\!\df\xi_h \df\xi_{h'} \, \xi_h\, \xi_{h'}    \int \df q_y\,    \delta \left(  1-\var -\frac{q_y^2}{4p_T^2}  \right) 
\int\! \df p_{1,y} \, \df p_{2,y} \, \df k_{h,y} \, \df k_{h',y} \, \df k_{s,y}   \,
 \\\nn
 & \times  
 \delta\left(\frac{k_{h,y}}{\xi_h}+\frac{k_{h',y}}{\xi_{h'}}+p_{1,y}+p_{2,y}-k_{s,y}-q_y\right) S_{JI}(k_{s,y},\mu,\nu) \\
&\times B_{f_1/N_1}(p_{1,y},\,\xi_1,\,\mu,\,\nu)\,B_{f_2/N_2}(p_{2,y},\,\xi_2,\,\mu,\,\nu) D_{h/f_3}(k_{h,y}, \xi_h,\mu,\nu)  D_{h'/f_4}(k_{h',y}, \xi_{h'},\mu,\nu)  \,. \nn
\end{align}
Here, the superscript $(0)$ denotes that this expression describes only the leading power dynamics in the expansion about the back-to-back limit. Summing over channels includes summing over the flavors of partons $f_i$. $N_{\text{init}}$ is the number of initial states averaged over in computing the cross section ($N_{\text{init}}=2^2\times 3^2$ for the $qq$ initial state, $N_{\text{init}}=2^2\times 3\times 8$ for the $qg$ initial state, and $N_{\text{init}}=2^2\times 8^2$ for the $gg$ initial state). $S_{IJ}$ is the soft function, $B$'s are beam functions, $D$'s are fragmentation functions, and the hard function $H_{IJ}$ is defined in terms of the Wilson coefficients $\cC_I^\Gamma$ as
\begin{equation}
H_{IJ}=\sum_\Gamma \cC^\Gamma_I\cC^{\Gamma*}_J\,.
\end{equation}
with $I$ indexing
the different color structures and $\Gamma$ indexing the different spins.
We will describe each of these functions in more detail shortly. 
The variables $y_3$ and $y_4$ are the rapidities of partons $p_3$ and $p_4$, and
$\xi_1$ and $\xi_2$ are the energy fractions of partons $p_1$ and $p_2$ relative to hadrons $P_1$ and $P_2$, which can be expressed as
functions of the Born kinematics $p_T$, $y_3$ and $y_4$,
\begin{align}
\xi_1=\frac{p_T}{\sqrt{s}}\left(e^{y_3}+e^{y_4}\right),~~\xi_2=\frac{p_T}{\sqrt{s}}\left(e^{-y_3}+e^{-y_4}\right)\,.
\end{align}

As currently formulated, this factorization formula is not desirable, since it is expressed in terms of transverse momentum dependent fragmentation functions (TMDFFs), which are intrinsically non-perturbative objects. However, we expect that since the TEEC is an Infrared and collinear safe observable, the only non-perturbative functions appearing in its definition should be the PDFs (or more precisely the TMDPDFs). For perturbative transverse momentum, we can perform an operator product expansion (OPE) to match the TMDFFs onto the standard fragmentation functions, allowing us to use a sum rule to eliminate the fragmentation functions from the result. We therefore briefly review the properties of the TMD fragmentation functions and their OPE, to understand how to convert them into perturbative jet functions.

The standard fragmentation functions (FFs) are defined as \cite{Georgi:1977mg,Ellis:1978ty,Collins:1981uw,Collins:1989gx}
\begin{align}
d_{h/q}(z_h) &
  = \frac{1}{2z_hN_c}
    \sum_{X} \int\!\frac{\df b^+}{4\pi}e^{i k_h^- b^+/(2z_h)}\, \tr_{\rm spin} \Mae{0}{\frac{\slashed\bn}{2}\chi_{n}(b^+)}{h,X}
    \Mae{h,X}{\bar\chi_{n}(0)}{0}
    \,,
    \\
    d_{h/g}(z_h)&
  = -\frac{k_h^-}{(d-2)(N_c^2-1)z_h^2}
    \sum_{X} \int\!\frac{\df b^+}{4\pi}e^{i k_h^- b^+/(2z_h)}\, \Mae{0}{\cB_{n\perp}^\mu(b^+)}{h,X}
    \Mae{h,X}{\cB_{n\perp\mu}(0)}{0}
    \,.
\end{align}
The TMDFFs in position space are defined as \cite{Collins:2011zzd}
\begin{align}\label{eq:TMDFF_def}
D_{h/q}(\vec b_\perp,z_h)&
  = \frac{1}{2z_hN_c}
    \sum_{X} \int\!\frac{\df b^+}{4\pi}e^{i k_h^- b^+/(2z_h)}\,\tr_{\rm spin} \Mae{0}{\frac{\slashed\bn}{2}\chi_{n}(b)}{h,X}
    \Mae{h,X}{\bar\chi_{n}(0)}{0}
    \,,
    \\
D_{h/g}^{\mu\nu}(\vec b_\perp,z_h)&
  = -\frac{k_h^-}{(d-2)(N_c^2-1)z_h^2}
    \sum_{X} \int\!\frac{\df b^+}{4\pi}e^{i k_h^- b^+/(2z_h)}\, \Mae{0}{\cB_{n\perp}^\mu(b)}{h,X}
    \Mae{h,X}{\cB_{n\perp}^\nu(0)}{0}
    \,.
\end{align}
Here we have used the SCET notation, where $\chi_n$ and $\cB_n^\mu$ are the gauge invariant $n$-collinear quark and gluon fields respectively.
The pair of fields are separated by $b^\mu=(b^+,0^-,\vec b_\perp)$,
with $\vec b_\perp$ the conjugate variable to $\vec k_\perp^h$, which is the transverse momentum of $h$ perpendicular to the jet axis $\vec{n}$. 
$D_{h/g}^{\mu\nu}$ can be decomposed as into tensor structures as
\begin{align}
D_{h/g}^{\mu\nu}=\frac{g_\perp^{\mu \nu}}{d-2} D_{h/g} +\left ( \frac{g_\perp^{\mu \nu}}{d-2}+\frac{b_\perp^\mu b_\perp^\nu}{b_T^2}  \right ) D_{h/g}'\,.
\end{align}
Here, $D_{h/g}$ is the unpolarized gluon contribution, while $D_{h/g}'$ is the linearly polarized gluon contribution. We leave the discussion for the linear polarization contribution to \sec{linear_polarization} and only consider $D_{h/g}$ in this section.

The OPE of the TMDFF onto the standard FF, is given in momentum space by
\begin{align}
\label{eq:TMDFF_matching}
D_{h/i}(\vec k_{h\perp},\xi_h) &=\sum \limits_j \int \frac{dz_h}{z_h^3}d_{h/j} (z_h, \mu) \cJ_{ji}\left(\frac{\vec k_{h\perp}}{z_h},\frac{\xi_h}{z_h}\right)  
+ \text{power correction}\,,
\end{align}
where $\cJ_{ij}$ are finite matching coefficients, 
and $d_{h/j}$ are fragmentation functions. 
To convert these TMDFFs as well as their matching coefficients to the ones shown in \eq{full} as functions of the $y$-component momenta, one simply integrates out their $x$-components,
\begin{align}\label{eq:TMDFF_y}
    D_{h/i}(k_{h,y},\xi_h) 
    =\int\!dk_{h,x}\, F_{h/i}(\vec k_{h\perp},\xi_h)
    \,,\qquad \text{and} \qquad
    \cJ_{ji}\left(k_y,\xi\right)=\int\!\df k_x\,\cJ_{ji}\left(\vec k_\perp,\xi\right)\,,
\end{align}
so that we have
\begin{align}
    \label{eq:TMDFF_matching_y}
    D_{h/i}(k_{h,y},\xi_h) 
    % =\int\!dk_{h,x}\, F_{h/i}(\vec k_{h\perp},\xi_h) 
    &=\sum \limits_j \int \frac{dz_h}{z_h^2}d_{h/j} (z_h, \mu)\, \cJ_{ji}\left(\frac{k_{h,y}}{z_h},\frac{\xi_h}{z_h}\right)
    + \text{power correction}\,.
\end{align}

We can now simplify the factorization formula in \eq{full}, by using sum rules to eliminate the dependence on the FF. Inserting \eq{TMDFF_matching_y} into \eq{full}, and changing variables to $\xi_{i}=\xi_{h}/z_{h}$, $\xi_{j}=\xi_{h'}/z_{h'}$, $k_{i,y}=k_{h,y}/z_h$, $k_{j,y}=k_{h',y}/z_{h'}$,
we then find 
\begin{align}\label{eq:fact_tmp}
\frac{d\sigma^{(0)}}{dz_\phi} &= \frac{1}{16\pi s^2}\sum\limits_{\text{channels}} \sum\limits_{IJ}\sum\limits_{ij} \frac{1}{(1 + \delta_{f_3 f_4}) N_{\text{init}}}\int \frac{dy_3dy_4dp_T^2}{\xi_1\xi_2} d\xi_i d\xi_j  \xi_i \xi_j    \int\! d q_y \,   \delta \left(  1-z_\phi -\frac{q_y^2}{4p_T^2}  \right) \nn \\
& \times    \int\! dp_{1,y} dp_{2,y} dk_{s,y} dk_{i,y} dk_{j,y} \, \delta\left(q_y-\left(\frac{k_{i,y}}{\xi_i}+\frac{k_{j,y}}{\xi_j}+p_{1,y}+p_{2,y}-k_{s,y}\right)\right) \nn \\
&\times H_{IJ}^{f_1 f_2 \to f_3 f_4}(p_T,y_3,y_4,\mu) \, B_{f_1/N_1}(p_{1,y},\,\xi_h,\,\mu,\,\nu)\,B_{f_2/N_2}(p_{2,y},\,\xi_2,\,\mu,\,\nu)S_{JI}(k_{s,y},\mu,\nu)\nn\\
&\times  \left[ \sum\limits_h \int dz_h~ z_h~ d_{h/i}(z_h,\mu)  \right]\cJ_{i f_3}\left(\xi_i, k_{i,y}\right)
  \cdot   \left[ \sum\limits_{h'} \int dz_{h'}~ z_{h'}~ d_{h'/j}(z_{h'},\mu)  \right]\cJ_{j f_4}\left(\xi_j, k_{j,y}\right)
\,.
\end{align}
We can now use the momentum-conservation sum rule 
 \begin{align}
   \sum\limits_h \int dz_h~ z_h~ d_{h/j}(z_h,\mu) =1\,,
 \end{align}
to cancel the non-perturbative fragmentation functions, arriving at an expression purely in terms of the perturbative matching coefficients.

Using the Fourier representation of the delta function,
\begin{align}
&\delta\left(q_y-\left(\frac{k_{i,y}}{\xi_i}+\frac{k_{j,y}}{\xi_j}+p_{1,y}+p_{2,y}-k_{s,y}\right)\right)\nn \\
&\hspace{2cm} = \int\! \frac{db_y}{2 \pi} ~ \exp\left[-i b_y q_y + i b_y\left(\frac{k_{i,y}}{\xi_i}+\frac{k_{j,y}}{\xi_j}+p_{1,y}+p_{2,y}-k_{s,y}\right)\right] \,,
\end{align}
we now go to position space, and simply our factorized expression to 
\begin{align}\label{eq:factexp}
\frac{d\sigma^{(0)}}{dz_\phi} =\,& \frac{1}{16\pi s^2}\sum\limits_{\text{channels}} \sum\limits_{IJ}\sum\limits_{ij} \frac{1}{(1 + \delta_{f_3 f_4}) N_{\text{init}}}\int \frac{dy_3dy_4\, p_T dp_T^2}{\xi_1\xi_2} d\xi_i d\xi_j \, \xi_i \xi_j     
\nn\\
&\times\int\frac{db_y}{2\pi}\frac{e^{-2ib_y\sqrt{1-z_\phi}~p_T}}{\sqrt{1-z_\phi}}\, H_{IJ}^{f_1 f_2 \to f_3 f_4}(p_T,y_3,y_4,\mu)\, S_{JI}(b_y,\mu,\nu)
\nn\\
&\times  B_{f_1/N_1}(b_y,\,\xi_1,\,\mu,\,\nu)\,B_{f_2/N_2}(b_y,\,\xi_2,\,\mu,\,\nu)\, \cJ_{i f_3}\left(\frac{b_y}{\xi_i},\xi_i\right)\cJ_{j f_4}\left(\frac{b_y}{\xi_j},\xi_j\right)
\,.
\end{align}
Finally, we define the jet function relevant to the TEEC as
\begin{align}\label{eq:J_NPp}
J_q^\TEEC(b_y) = \sum\limits_i  \int \limits_0^1d\xi~ \xi~ \cJ_{iq}\left(\frac{b_y}{\xi},\xi\right)\,,\nn \\
J_g^\TEEC(b_y) = \sum\limits_i  \int \limits_0^1d\xi~ \xi~ \cJ_{ig}\left(\frac{b_y}{\xi},\xi\right)\,.
\end{align}
This allows us to write the final factorized expression for the TEEC in the back-to-back limit as 
\begin{align}
\frac{d\sigma^{(0)}}{d\tau}
 =&\  \frac{1}{16 \pi s^2 \sqrt{\tau}}\sum\limits_{\text{channels}} \frac{1}{(1 + \delta_{f_3 f_4})N_{\text{init}}}\int \frac{dy_3 dy_4\, p_T dp_T^2}{\xi_1\xi_2} \int_{-\infty}^{\infty}\frac{db_y}{2\pi}e^{-2ib_y\sqrt{\tau} p_T}\nn \\
&\times \mathrm{tr}\big[\mathbf{H}^{f_1 f_2 \to f_3 f_4}(p_T,y^*,\mu) \mathbf{S}(b_y, y^*, \mu,\nu) \big]\nn \\
&\times   B_{f_1/N_1}(b_y,\,\xi_1,\,\mu,\,\nu)\,B_{f_2/N_2}(b_y,\,\xi_2,\,\mu,\,\nu) J^\TEEC_{f_3}\left(b_y,\mu,\nu\right)
  J^\TEEC_{f_4}\left(b_y,\mu,\nu\right). 
\label{eq:master}
\end{align}
Here we have used $y^* = (y_3 - y_4)/2$, to denote the single jet rapidity in the partonic center-of-mass frame.
For simplicity, to make contact with the notation of \cite{Gao:2019ojf}, we have written the factorized result in terms of  $\tau \equiv 1-z_\phi $, as defined in \Eq{eq:def_tau}. 

This provides an expression for the leading power dynamics of the TEEC in the back-to-back limit in terms of a number of remarkably simple functions combined in a highly non-trivial way, and is the primary result of this work. All the functions appearing in this formula are related in some manner to TMD dynamics. The TEEC therefore provides a natural extension of the $q_T$ observable from color singlet production, to both $W/Z/\gamma+$ jet, and dijets.

%%%%%%%%%%%%%%%%%%%%%%%%%%%%
\subsubsection{Summary of Factorized Functions}
\label{sec:sum_pert}
%%%%%%%%%%%%%%%%%%%%%%%%%%%%

For convenience, we now briefly summarize the known functions appearing in the TEEC factorization formula, along with their RG evolution. We postpone a discussion of the soft function, which is a new ingredient of our factorization formula, to \Sec{sec:soft}.

%%%%%%%%%%%%%%%%%%%%%%%%%%%%
\subsubsection*{Hard Functions}
\label{sec:hard}
%%%%%%%%%%%%%%%%%%%%%%%%%%%%

The hard functions  $\mathbf{H}^{f_1 f_2 \to f_3 f_4}$ describe the underlying microscopic scattering for the partonic channel $f_1 f_2 \to f_3 f_4$, and are independent of the TEEC measurement. They are obtained as the infrared finite part of the $2\to 2$ scattering amplitudes (For a precise definition, see e.g. \cite{Moult:2015aoa}). All relevant amplitudes are well known at one-  \cite{Kunszt:1993sd} and two-loops \cite{Anastasiou:2000kg, Anastasiou:2000ue,Glover:2001af, Bern:2002tk, Bern:2003ck, Glover:2003cm,Glover:2004si, DeFreitas:2004kmi}. They have recently been computed at three loops \cite{Bargiela:2021wuy,Caola:2021izf,Caola:2021rqz,Caola:2020dfu}. The NLO hard functions have been extracted in \cite{Kelley:2010fn,Moult:2015aoa}, and the NNLO hard functions are available in a  \textsc{Mathematica} file in \cite{Broggio:2014hoa}. We use the results in \cite{Broggio:2014hoa} in our calculation.

The hard functions satisfy a renormalization group equation
\begin{align}\label{eq:GammaH}
    \frac{d}{d \ln \mu^2}  \mathbf{H} =
    \frac12 \left(\mathbf{\Gamma}_H \ \mathbf{H} + \mathbf{H} \ \mathbf{\Gamma}_H^\dagger\right)\,,
\end{align}
where
\begin{align}
  \label{eq:Gammah}
  \mathbf{\Gamma}_H =& - \sum_{i<j} \mathbf{T}_i \cdot \mathbf{T}_j \gamma_{\rm cusp} \ln\frac{\sigma_{ij}\, \hat s_{ij}- i0 }{\mu^2}
 + \sum_i \gamma_i \mathbf{1} + \boldsymbol{\gamma}_{\rm quad}  \,,
\end{align}
where $\sigma_{ij}=-1$ if both $i$ and $j$ are incoming or outgoing, and $1$ otherwise. 
$s_{ij} = 2p_i\cdot p_j$ is the Mandelstam variables. Here $\gamma_i = \gamma_q$, $\gamma_g$ are the quark or gluon anomalous dimension.
This equation describes a non-trivial matrix evolution in the color space, and its solution will be discussed in detail in \Sec{sec:ingr-resumm-teec}.

%%%%%%%%%%%%%%%%%%%%%%%%%%%%
\subsubsection*{Beam Function}
\label{sec:beam}
%%%%%%%%%%%%%%%%%%%%%%%%%%%%

The beam functions appearing in the factorization formula for the TEEC can be obtained from the standard TMD beam functions. Unlike for the transverse momentum spectrum of color singlet particles, the TEEC only measures the component of the momentum out of the plane (recall that we have taken the Born scattering process to lie in the $xz$-plane, so the momentum component out of the plane is the $y$-component). To obtain the TEEC beam function from the standard TMD beam functions, we must therefore project them onto the $y$-component. We first review the TMD beam functions, and then discuss their projection for the TEEC.

The TMD beam functions are defined in terms of gauge invariant SCET fields as \cite{Stewart:2010qs}
\begin{align} \label{eq:beam_defn}
 B_q(z=\w/P_n^-,\bt) &= \big\langle P(P_n) \big| \bar\chi_n(b_\perp) \frac{\slashed{\bn}}{2} [\delta(\w-\bar\cP_n)\chi_n(0)] \big| P(P_n) \big\rangle
\,,\nn\\
 B_g^{\mu\nu}(z=\w/P_n^-,\bt) &= -\w\,\big\langle P(P_n) \big| \cB_{n\perp}^\mu(b_\perp) [\delta(\w-\bar\cP_n)\cB_{n\perp}^\nu(0)] \big| P(P_n) \big\rangle
\,.\end{align}
Lorentz invariance allows the gluon TMD beam functions to be decomposed as 
\begin{align}
    B_{g}^{\mu \nu}\left(z, \bt  \right)=\frac{g_{\perp}^{\mu \nu}}{2} B_{g}\left(z,  \bt\right)
    +\left(\frac{b_\perp^{\mu}b_\perp^{\nu}}{b^2}-\frac{g_{\perp}^{\mu \nu}}{2}\right) 
    B_{g}^{\prime}\left(z,  \bt \right),
\end{align}
where the second term is referred to as the linearly  polarized contribution. Physically, these linearly polarized terms correspond to contributions that have an azimuthal dependence on the angle out of the hard scattering plane. We will discuss in detail the linearly polarized contributions to the beam and jet functions in \Sec{sec:linear_polarization}.

The TMD beam functions obey the $\mu$ and $\nu$ evolution equations
\begin{align}
 \dlogmu{B_i(z,\bt,\mu,\nu)} &= \gMuB(\mu,\omega,\nu) B_i(z,\bt,\mu,\nu)
\,, \nn\\
 \dlognu{B_i(z,\bt,\mu,\nu)} &= \gNu^B(\bt,\mu) B_i(z,\bt,\mu,\nu)
\,,\end{align}
where $i=q,g$. The anomalous dimensions are given by
\begin{align}
 \gMuB(\mu,\omega,\nu) &= 2 \GammaC[\as(\mu)] \ln\frac{\nu}{\omega} - 2 \gamma_C^i - \frac{\gMuS}{2}\,, \nn \\
 \gNu^B(\bt,\mu) &= - \frac{1}{2} \gNu(\bt,\mu)
\,.\end{align}
Here $\omega = z \Ecm$ is the large lightcone momentum.

The beam functions for the TEEC can now be straightforwardly obtained from the TMD beam functions 
\begin{align}
b_\perp^\mu \to b_y y^\mu\,,
\end{align}
where $y^\mu$ is a unit vector in the $y$-direction, which is orthogonal to the plane of the Born scattering process. This projection is trivial for the standard beam function, but gives rise to an interesting azimuthal dependence for the linearly polarized beam function
\begin{align}
    B_{g}^{\mu \nu}\left(z, b_y y^\mu  \right)=\frac{g_{\perp}^{\mu \nu}}{2} B_{g}\left(z,  b_y y^\mu\right)
    +\left(y^\mu y^\nu-\frac{g_{\perp}^{\mu \nu}}{2}\right) 
    B_{g}^{\prime}\left(z,  b_y y^\mu \right),
\end{align}
However, conveniently, the TEEC beam function can be obtained directly from the TMD beam function, allowing us to use the high loop results available in the literature
 
The beam functions for the TEEC can be matched onto standard PDFs at small but perturbative transverse momentum, 
\begin{align}
B_{i/N}(b_y,\xi,\mu,\nu) =\sum \limits_j \int \frac{dz}{z}\cI_{ij}\left(z,L_b,L_Q\right) f_{j/N} \left(\frac{\xi}{z}, \mu\right)   + \text{power corrections} \,,
\end{align}
where $L_b = \ln (b_y^2 \mu^2/b^2_0)$, $b_0 = 2 e^{-\gamma_E}$, and $L_Q = \ln (Q^2/\nu^2)$, with $Q = 2 p_i^0$, twice the energy of the measured parton energy. The matching coefficients have been derived to two loops in \cite{Gehrmann:2012ze,Gehrmann:2014yya,Echevarria:2016scs,Luebbert:2016itl,Luo:2019hmp,Luo:2019bmw}, and three loops in \cite{Luo:2019szz,Luo:2020epw,Ebert:2020qef,Ebert:2020yqt}.

%%%%%%%%%%%%%%%%%%%%%%%%%%%%
\subsubsection*{Jet Functions}
\label{sec:jet}
%%%%%%%%%%%%%%%%%%%%%%%%%%%%

The jet functions appearing in the TEEC were already discussed in some detail in the derivation of the factorization formula, since it was necessary to perform an OPE of the TMDFFs onto standard FFs to obtain a factorization formula expressed entirely in terms of perturbative functions. 

Definitions for the jet functions appearing in the factorization theorem for the EEC in $e^+e^-$ collisions were first presented in \cite{Moult:2018jzp}, where it was shown that they could be obtained as moments of the TMD matching coefficients $\cJ_{ij}$. The relation between the TEEC jet functions and the EEC jet functions is identical to the relation between the TEEC beam functions and the TMD beam functions, namely one must project them onto the component perpendicular to the hard scattering plane using the substitution $b_\perp^\mu \to b_y y^\mu$. This is convenient, since the EEC jet functions are known at two- \cite{Luo:2019hmp,Luo:2019bmw} and three- \cite{Luo:2020epw,Ebert:2020qef} loops, allowing us to immediately obtain the TEEC jet functions at the same order.

Explicitly, the jet functions appearing in the TEEC factorization formula can be obtained from the EEC jet functions by the substitution, 
\begin{align}\label{eq:J_NP}
J_q^\TEEC(b_y)=J_q^\EEC(b_T=b_y)  = \sum\limits_i  \int \limits_0^1d\xi~ \xi~ \cJ_{iq}\left(\frac{b_y}{\xi},\xi\right)\,,\nn \\
J_g^\TEEC(b_y)=J_g^\EEC(b_T=b_y) = \sum\limits_i  \int \limits_0^1d\xi~ \xi~ \cJ_{ig}\left(\frac{b_y}{\xi},\xi\right)\,,
\end{align}
and are obtained as moments of the (matching coefficients of the) TMD fragmentation functions. Here we have suppressed Lorentz indices. Just as for the beam function, one also has linearly polarized jet functions. We will discuss their structure in \Sec{sec:linear_polarization}. The RG evolution of the jet functions was derived in \cite{Moult:2018jzp} from the known evolution of the TMD fragmentation functions. It is given by 
\begin{align}
  \label{eq:jetRGE}
\mu \frac{dJ_q^\TEEC(b_y, \mu,\nu)}{ d\mu} &=  \left[- \Gamma_{\text{cusp}}(\alpha_s)  \ln \frac{Q^2}{\nu^2} +2\gamma^J_\EEC(\alpha_s)    \right] J_q^\TEEC(b_y,\mu,\nu) \,,
\end{align}
for the $\mu$ evolution, and 
\begin{align}
\nu \frac{d J_q^\TEEC(b_y,  \mu,\nu)}{d\nu}=  \left[  \int\limits_{b_0^2/b_y^2}^{\mu^2} \frac{d\bar{\mu}^2}{\bar{\mu}^2} \Gamma_{\text{cusp}}(\alpha_s(\bar{\mu}))  - \gamma^r_\EEC(\alpha_s(b_0/b_y))  \right] J_q^\TEEC(b_y,\mu,\nu)\,,
\end{align}
for the $\nu$ evolution.

%%%%%%%%%%%%%%%%%%%%%%%%%%%%
\subsection{Underlying Event and Factorization Violation}
\label{sec:ingr-resumm-teec-soft-caveat}
%%%%%%%%%%%%%%%%%%%%%%%%%%%%

\begin{figure}
  \centering
\subfloat[]{\label{fig:spectator}
\includegraphics[width=7cm]{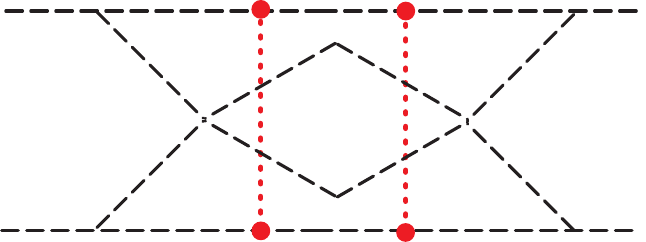}
}  
\subfloat[]{\label{fig:spectator_final}
\includegraphics[width=7cm]{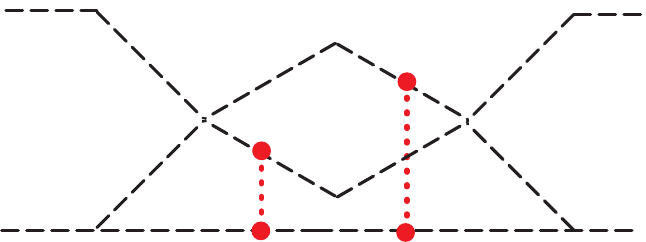}
}  
  \caption{Examples of Glauber diagrams which could violate our factorization formula for the TEEC. The Glaubers, which are illustrated by dashed red lines, couple distinct collinear sectors. For the TEEC, pure spectator graphs, such as shown in a) cancel, while graphs that involve both active and spectator partons, such as shown in b), are not expected to cancel.}
  \label{fig:glauber}
\end{figure}

Before continuing, we must make several comments regarding the validity of our factorization formula. As described earlier, the leading power SCET Lagrangian describing the TEEC in the back-to-back limit can be written as 
\begin{align}
\cL^{(0)}=\cL^{(0)}_{B_1}+\cL^{(0)}_{B_2}+\cL^{(0)}_{J_1}+\cL^{(0)}_{J_2} +\cL^{(0)}_G\,,
\end{align}
where the Glauber Lagrangian  \cite{Rothstein:2016bsq}, $\cL^{(0)}_G$, which was set to zero in our derivation, couples the different collinear directions, leading to potential violations of our formula. Example diagrams involving Glaubers are shown in \Fig{fig:glauber}, where the Glauber gluons are shown as red-dashed lines. In the above derivation, we have assumed that the Glauber Lagrangian does not contribute, which allows us to derive a factorized formula expressed in terms of functions describing the soft and collinear sectors with no connection other than through kinematics. Full proofs of factorization have been given for Drell-Yan production, and the $q_T$ spectrum in Drell-Yan production, in the seminal works of \cite{Collins:1981uk,Collins:1981va,Collins:1981ta,Collins:1984kg,Collins:1985ue,Collins:1988ig,Collins:1989gx}. For the case of the TEEC considered here, it is expected that Glaubers will contribute and factorization will be violated. The TEEC is closely related to momentum imbalance dijet event shapes, for which there is a growing body of evidence that factorization is violated  \cite{Collins:2007nk,Collins:2007jp,Bomhof:2007su,Rogers:2010dm,Buffing:2013dxa,Gaunt:2014ska,Zeng:2015iba,Catani:2011st,Schwartz:2017nmr,Forshaw:2008cq,Forshaw:2006fk,Martinez:2018ffw,Angeles-Martinez:2016dph,Forshaw:2012bi,Angeles-Martinez:2015rna,Schwartz:2018obd,Rothstein:2016bsq}. This ultimately arises due to amplitude-level factorization violation~\cite{Catani:2011st,Forshaw:2012bi,Schwartz:2017nmr}.

There are a number of different effects that can break factorization. First, even in the case of color singlet production, the imposition of a measurement function can block the cancellation of Glauber gluons arising from spectator interactions such as shown in \Fig{fig:spectator}. This can lead to a violation of factorization \cite{Gaunt:2014ska,Zeng:2015iba,Rothstein:2016bsq}, as happens for $E_T$, or beam thrust. This type of factorization violation can lead to large corrections to observables, often characterized by their sensitivity to multi-parton interactions (MPI). This has been studied using a parton shower model for MPI in \cite{Alioli:2016wqt}, and analytic models in \cite{Cacciari:2009dp}. Additionally, when there are jets in the initial and final states, one can have an intrinsic failure of collinear factorization \cite{Catani:2011st,Forshaw:2012bi}. For the case of the TEEC, in principle both types of factorization violation could contribute to the observable.

When discussing factorization violation, one must distinguish two types of factorization violation, namely perturbative, i.e. at a scale $\gg \Lambda_{\text{QCD}}$, and non-perturbative, namely at a scale $\sim\Lambda_{\text{QCD}}$. Non-perturbative factorization violation would imply that one could not factorize into universal PDFs. For color singlet production, this would mean that the two beams are non-perturbatively coupled, while with jets in the final state, it would imply some non-perturbative coupling of the jets and the beams. Perturbative factorization violation, on the other hand, can be explicitly calculated in perturbation theory, and therefore is less of a concern unless it leads to infrared divergence in the cross section.

\begin{figure}
  %\centering
\subfloat[]{\label{fig:UE_a}
\includegraphics[width=8cm]{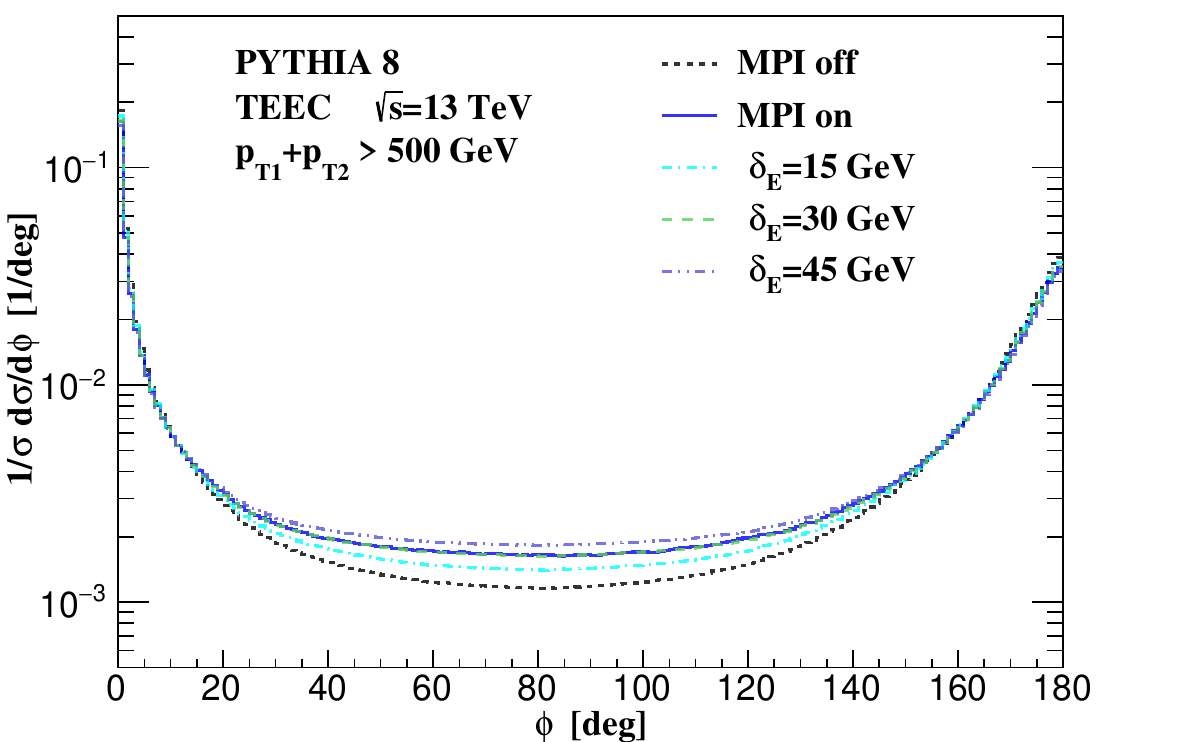}
}  
\subfloat[]{\label{fig:UE_b}
\includegraphics[width=8cm]{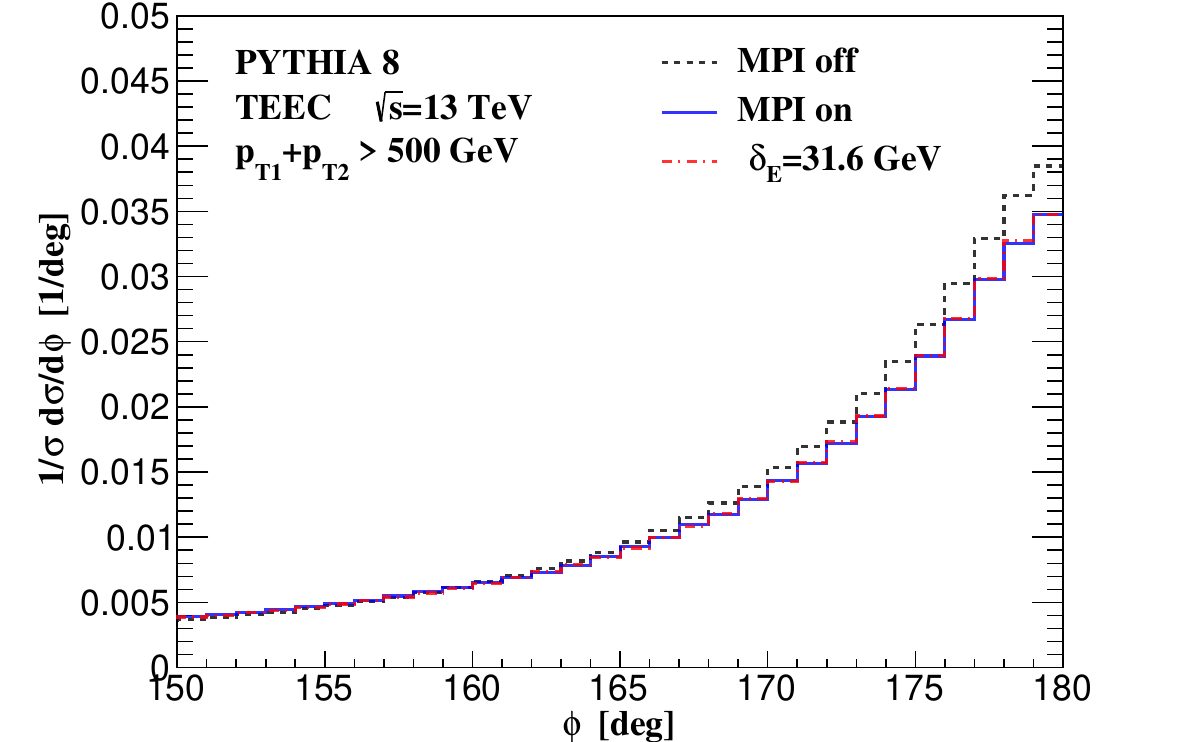}
}  
  \caption{The sensitivity of the TEEC to underlying event as modeled in Pythia. In a) we show the full distribution, while in b) we show a zoomed in version of the back-to-back endpoint region. The effects of the underlying event are small, and consistent with being a power correction. We also show the result of using a model for the underlying event, which adds a uniform energy contribution, as described in the text. We see that this simple model provides an excellent description of the effects of the underlying event throughout the entire distribution.}
  \label{fig:UE}
\end{figure}

When the TEEC scale is perturbative, from the perspective of modes at the scale $\Lambda_{\text{QCD}}$, the TEEC measurement is inclusive over the final state, and we expect a cancellation of the non-perturbative Glaubers. In this case, factorization violation is expected to be a power correction.  However, this argument can be violated if there are many successive spectator interactions leading to a numerically large correction. This is particularly dangerous for observables that are scalar sums. On the other hand, since the TEEC is related to a vector sum, we expect it to be well behaved. In particular, we expect it to be a numerically true statement that non-perturbative factorization violation is a power correction. This can be  tested at a qualitative level in parton shower Monte Carlo using a model for the underlying event.

In \Fig{fig:UE}, we illustrate the sensitivity of the TEEC to underlying event, as modelled in Pythia \cite{Sjostrand:2006za,Sjostrand:2007gs}. In \Fig{fig:UE_a} we show a plot over the whole range of the TEEC, while in \Fig{fig:UE_b}, we have zoomed into the back-to-back region. We see that overall the effects of the underlying event are quite small, except right in the endpoint region where the scale probed by the TEEC reaches  $\Lambda_{\text{QCD}}$. We believe that this minimal sensitivity to the underlying event is quite promising, and ensures that the high perturbative accuracy of our calculation is not destroyed by underlying event. We note that this is a significantly different behavior than for the case of beam thrust, where large contributions from the underlying event are observed (see e.g. \cite{Alioli:2016wqt}).

Interestingly, we can provide an excellent description of the underlying event by adding an energy distribution that is uniform in the azimuthal angle to our perturbative calculation. Inserting a shift in the energies into the formula for the TEEC in \Eq{eq:TEEC_intro}, and expanding, we find that this shifts the TEEC distribution as
\begin{align}\label{eq:mpi}
    \frac{1}{\sigma_{\rm MPI}}\frac{d\sigma_{\rm MPI}}{d\phi} = 
   \frac{Q}{Q+ 2 \delta_E } \left(\frac{1}{\sigma} \frac{d\sigma}{d\phi} +     \frac{2}{\pi} \frac{\delta_E}{Q} \right)\,,
\end{align}
where $\delta_E$ is the total energy added by MPI effects and  $Q=500$ GeV is an approximation to the total transverse energy of the hard scattering that will be used in our numerical results.  The result of this model is also shown in \Fig{fig:UE}. We see that it provides an excellent description of Pythia's model of the underlying event for a value of the parameter $\delta_E\sim 30$ GeV. It would be interesting to measure this in experiment, and also to study its variation with the hard scattering energy. The apparent simplicity of this description of the underlying event may also allow it to be distinguished from hadronization effects. This has been studied for the jet mass in \cite{Stewart:2014nna}, and it would be interesting to perform a similar study for the TEEC.

We can also ask to what extent our factorization formula can be violated perturbatively. For $E_T$ and beam thrust, perturbative factorization violation occurs at N$^4$LL  \cite{Gaunt:2014ska,Zeng:2015iba,Rothstein:2016bsq}, through diagrams such as the one shown in \Fig{fig:spectator}.  While a complete analysis of perturbative Glauber contributions is beyond the scope of this paper, we make several additional comments. First, we note that due to the measurement restriction being a vector sum, following the arguments in \cite{Rothstein:2016bsq}, diagrams with Glaubers exchanged purely between spectators, such as in \Fig{fig:spectator}, are expected to cancel for the TEEC. However, we do believe that factorization will be violated by diagrams such as those shown in \Fig{fig:spectator_final}, which we believe could give at most a constant at three loops.  As has been shown, it is expected that factorization violation should occur when there are two Glaubers, one collinear emission, and one additional emission \cite{Catani:2011st,Forshaw:2012bi}. We strongly expect that this is an N$^4$LL effect. For the case of so called super-leading logarithms \cite{Forshaw:2008cq,Forshaw:2006fk,Martinez:2018ffw,Angeles-Martinez:2016dph,Forshaw:2012bi,Angeles-Martinez:2015rna}, these effects modify the structure of the logarithmic series. However, this arises since there are only single logarithms in the observables of interest in their work. In our case we do not expect this to be the case, however, we have not yet proven this assertion. 

Nevertheless, we believe that pushing to as high perturbative orders as possible is useful, since our factorization formula can be viewed as a baseline, on which factorization violating effects can be added.  We believe that due to the simplicity of the TEEC, it is an ideal observable to attempt to analytically understand the violation of factorization with final state jets using the Glauber Lagrangian of \cite{Rothstein:2016bsq}. We believe that this is particularly true due to the fact that, as discussed in \Sec{sec:transv-energy-energy}, the TEEC can be defined for a color singlet final state, for a final state with a single jet, and for dijets final state. Since it is known that for the case of a color singlet factorization holds to all orders for the TEEC (since it is related to the factorization for transverse momentum of a color singlet boson), and that it is certainly violated for a dijet final state, we believe that this provides an extremely concrete playground to understand theoretically, and also probe experimentally, the effects of factorization violation. We leave this to future work.

%%%%%%%%%%%%%%%%%%%%%%%%%%%%
\subsection{Extension to Drell-Yan and $W/Z/\gamma~+$ Jet}
\label{sec:extend}
%%%%%%%%%%%%%%%%%%%%%%%%%%%%

Although the focus in this paper is on the TEEC for a purely hadronic final state (which at leading power in the back-to-back limit is a dijet final state), as we have discussed previously, we find it interesting that we can also achieve the same level of perturbative accuracy for the TEEC as measured on a leptonic final state in Drell-Yan, or on a $W/Z/\gamma+$ jet final state, and we believe that this will be useful for studying the effects of factorization violation. Therefore, for completeness, we also present the factorization formulas for the TEEC as measured on these final states. All of the relevant factorization formulas can be trivially obtained starting from the factorization formula for the purely hadronic final state, by removing jet functions, altering the Wilson line structure in the soft functions, and using the appropriate hard functions (For recent progress in the calculations of the relevant hard functions, see \cite{Gehrmann:2023jyv,Henn:2023vbd}).  

For the case of Drell-Yan, we can simply eliminate the two jet functions, and obtain
  \begin{align}
  \label{eq:master_drellyan}
\frac{d\sigma^{(0)}}{d\tau}
 =&\  \frac{1}{16 \pi s^2 \sqrt{\tau}}\sum\limits_{\text{channels}} \frac{1}{N_{\text{init}}}\int \frac{dy_{l^+}\, dy_{l^{\prime-}}\, p_T dp_T^2}{\xi_1\xi_2} \int_{-\infty}^{\infty}\frac{db_y}{2\pi}e^{-2ib_y\sqrt{\tau} p_T} \\
 & \times\mathrm{tr}\big[\mathbf{H}^{f_1 f_2 \to l^+l^{\prime-}}(p_T,y^*,\mu) \mathbf{S}(b_y, y^*, \mu,\nu) \big]
 B_{f_1/N_1}(b_y,\,\xi_1,\,\mu,\,\nu)\,B_{f_2/N_2}(b_y,\,\xi_2,\,\mu,\,\nu) \,.  \nn
\end{align}
This is similar to the factorization for the transverse momentum of the vector boson.
For the case of $V+$ jet, we have 
\begin{align}
\label{eq:master_Vjet}
\frac{d\sigma^{(0)}}{d\tau}
 &=\  \frac{1}{16 \pi s^2 \sqrt{\tau}}\sum\limits_{\text{channels}} \frac{1}{N_{\text{init}}}\int \frac{dy_3 dy_V \,p_Tdp_T^2}{\xi_1\xi_2} \int_{-\infty}^{\infty}\frac{db_y}{2\pi}e^{-2ib_y\sqrt{\tau} p_T} \\
 & \times\mathrm{tr}\big[\mathbf{H}^{f_1 f_2 \to f_3 V}(p_T,y^*,\mu) \mathbf{S}(b_y, y^*, \mu,\nu) \big] B_{f_1/N_1}(b_y,\,\xi_1,\,\mu,\,\nu)\,B_{f_2/N_2}(b_y,\,\xi_2,\,\mu,\,\nu) J_{f_3}\left(b_y,\mu,\nu\right).  \nn
\end{align}
Renormalization group consistency holds for all these different factorization formulas. We have used the same notation for the soft functions appearing in all the TEEC factorization formulas, despite the fact that they contain different numbers of Wilson lines. The precise definitions of the soft functions, as well as a discussion of their perturbative structure will be given in  \Sec{sec:soft}.

It would be extremely interesting to study the effects of underlying event and factorization violation for these the TEEC observable on different final states, particularly since factorization has been proven to hold for the TEEC on color singlet final states \cite{Collins:1981uk,Collins:1981va,Collins:1981ta,Collins:1984kg,Collins:1985ue,Collins:1988ig,Collins:1989gx}. This provides a baseline on top of which factorization violation and color flow effects can be studied. We leave detailed phenomenological studies of the different final states to future work.

%%%%%%%%%%%%%%%%%%%%%%%%%%%%
\section{The Transverse Energy-Energy Correlator Soft Function}
\label{sec:soft}
%%%%%%%%%%%%%%%%%%%%%%%%%%%%

The key new perturbative ingredient entering our factorization formula is the TEEC soft function. For multi-jet event shape observables, the soft function is typically the most complicated function entering the factorization formula, since it depends on the directions of all the different jets (For recent progress towards numerical calculations of soft functions at NNLO, see \cite{Bell:2018vaa,Bell:2018oqa,Bell:2018mkk,Bell:2023yso}, and for a semi-numerical calculation of the soft function for $2$-jettiness, see \cite{Jin:2019dho}.). In this section, we highlight the remarkable perturbative simplicity of the TEEC soft function, and we describe how this is related to the particular form of the measurement.

The structure of the TEEC soft function is also of intrinsic theoretical interest, since it provides an example of a rapidity divergent soft function with multiple ($>2)$ Wilson line directions. While the structure of rapidity divergences and their associated renormalization group evolution for soft functions involving two directions is by now quite well understood (the rapidity anomalous dimension is known to four loops in both QCD \cite{Li:2016ctv,Duhr:2022yyp,Moult:2022xzt} and $\cN=4$ super Yang-Mills \cite{Dixon:2017nat}), and has been used to make phenomenological predictions at N$^3$LL \cite{Li:2014bfa}, almost nothing is known about the rapidity anomalous dimension for soft functions involving multiple directions. Part of the understanding of the structure of rapidity divergences comes from the fact that for the particular case of two directions, they are related by conformal symmetry to standard anomalous dimensions \cite{Vladimirov:2017ksc,Vladimirov:2016dll,Moult:2022xzt}. It would therefore be interesting to understand to what extent this holds more generally, and the TEEC provides a concrete example where these questions can be studied.

In \Sec{sec:defn} we define the TEEC soft function and give its renormalization group evolution. In \secs{soft_dipole}{soft_tripole} we present some details of the calculation of the soft function at one- and two-loops. We then summarize our findings and discuss some directions for future study in \sec{discuss}.

\begin{figure}%[ht]
  \centering
\subfloat[]{\label{fig:soft_1}  
  \includegraphics[scale=0.6]{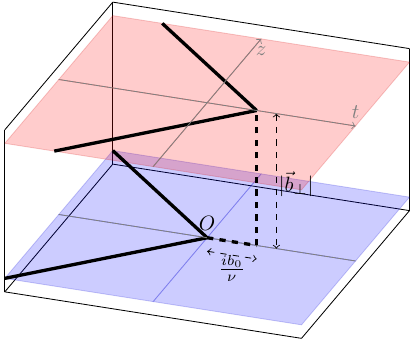} }
  \subfloat[]{\label{fig:soft_2}  
  \includegraphics[scale=0.6]{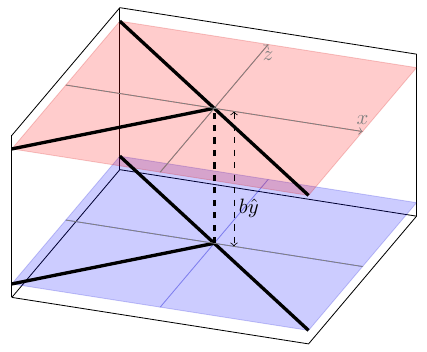} }
    \subfloat[]{\label{fig:soft_3}  
  \includegraphics[scale=0.6]{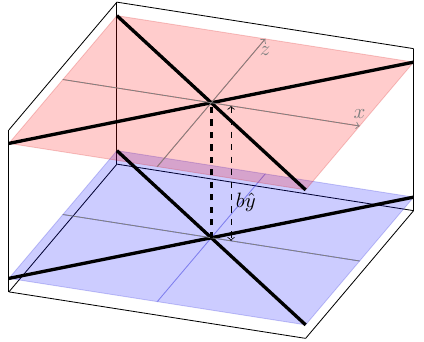} }
  \caption{The TEEC soft functions defined using the exponential rapidity regulator for (a) TEEC for Drell-Yan (or $q_T$), (b) TEEC for $W/Z/\gamma+$ jet, (c) TEEC for dijets.}
  \label{fig:soft}
\end{figure}

%%%%%%%%%%%%%%%%%%%%%%%%%%%%
\subsection{Definition and RG Evolution}
\label{sec:defn}
%%%%%%%%%%%%%%%%%%%%%%%%%%%%

In this section, we will discuss simultaneously the TEEC soft functions for Drell-Yan, $W/Z/\gamma+$ jet and dijets.  The TEEC soft functions are defined, using the exponential regulator of \cite{Li:2016axz}, as
\begin{align}
  \label{eq:soft}
&\text{Drell-Yan}: \qquad \mathbf{S}(b_y,y^*) = \langle 0 |T[\boldsymbol{O}_{n_1 n_2 }(0^\mu)] \overline{T}[\boldsymbol{O}_{n_1 n_2 }^\dagger (b_y^\mu)] | 0 \rangle \,, \nn \\
&\text{$W/Z/\gamma$+jet}: \qquad \mathbf{S}(b,y^*) = \langle 0 |T[\boldsymbol{O}_{n_1 n_2 n_3 }(0^\mu)] \overline{T}[\boldsymbol{O}_{n_1 n_2 n_3}^\dagger (b_y^\mu)] | 0 \rangle \,,  \nn \\
&\text{Dijets}:\qquad \mathbf{S}(b,y^*) = \langle 0 |T[\boldsymbol{O}_{n_1 n_2 n_3 n_4}(0^\mu)] \overline{T}[\boldsymbol{O}_{n_1 n_2 n_3 n_4}^\dagger (b_y^\mu)] | 0 \rangle \,,
\end{align}
as illustrated in \Fig{fig:soft} (There the temporal direction has necessarily been suppressed in (b) and (c)). 
Here $\boldsymbol{O}_{n_1 n_2}(x) = \boldsymbol{Y}_{n_1} \boldsymbol{Y}_{n_2} (x) $, $\boldsymbol{O}_{n_1 n_2 n_3}(x) = \boldsymbol{Y}_{n_1} \boldsymbol{Y}_{n_2} \boldsymbol{Y}_{n_3} (x)$, and $\boldsymbol{O}_{n_1 n_2 n_3 n_4}(x) = \boldsymbol{Y}_{n_1} \boldsymbol{Y}_{n_2} \boldsymbol{Y}_{n_3} \boldsymbol{Y}_{n_4}(x)$, with $\boldsymbol{Y}_{n_i}(x) = \exp[ i \int ds\, n_i \cdot A_s(s n_i + x) \mathbf{T}_i]$ a semi-infinite light-like soft Wilson line, and $n_i^\mu = p_i^\mu/p_i^0$ the light-like direction of the incoming or outgoing parton in the partonic center-of-mass frame. The directions of the Wilson lines are standard and hence suppressed, as are gauge links at infinity.
We have chosen coordinates such that $b_y^\mu = (0^+,0^-,0_x,b_y)$ is in the direction $\hat y$ perpendicular to the scattering plane, $\hat y \cdot n_i = 0$. 

The soft functions defined in Eq.~\eqref{eq:soft} suffer from UV and rapidity divergences. Rapidity divergences are regulated using the exponential regulator of \cite{Li:2016axz}.  The soft function, which is a matrix in color space, satisfies the RG equation
\begin{align}
  \label{eq:softRG}
  \frac{d\mathbf{S}}{d\ln \mu^2} = \frac{1}{2} \left( \mathbf{\Gamma}_S^\dagger \cdot \mathbf{S} + \mathbf{S} \cdot \mathbf{\Gamma}_S \right) \,,
\end{align}
with \cite{Kidonakis:1998bk,Kidonakis:1998nf,Aybat:2006mz,Aybat:2006wq}
\begin{align}
  \label{eq:Gammas}
  \mathbf{\Gamma}_S = \sum_{i<j} \mathbf{T}_i \cdot \mathbf{T}_j \gamma_{\rm cusp} \ln\frac{\sigma_{ij}\nu^2\, n_i \cdot n_j-i 0}{2 \mu^2} - \sum_i \frac{c_i}{2} \gamma_s \mathbf{1} - \boldsymbol{\gamma}_{\rm quad}  \,,
\end{align}
where $\sigma_{ij}=-1$ if both $i$ and $j$ are incoming or outgoing, and $\sigma_{ij}=1$ otherwise.
Here $\nu$ is the rapidity scale, and $c_i = C_F$ or $C_A$ is the Casimir of the parton $i$. 
Here $\gamma_{\rm cusp}$ is the cusp anomalous dimension \cite{Korchemsky:1987wg}, $\gamma_s$ is the threshold soft anomalous dimension \cite{Li:2014afw} and  $\boldsymbol{\gamma}_{\rm quad}$ is the  anomalous dimension for quadrupole color and kinematic entanglement, which first appears at three loops \cite{Almelid:2015jia,Almelid:2017qju} for the case that there are four Wilson lines. Its structure will be discussed in \Sec{sec:ingr-resumm-teec}, where we will also discuss the solution of the RG evolution equation with color mixing.

 The evolution equation associated with the rapidity scale $\nu$ is
\begin{align}
  \label{eq:softnuRG}
  \frac{d \mathbf{S}}{d \ln \nu^2} =
\frac{1}{2} \left( \mathbf{\Gamma}_y^\dagger \cdot \mathbf{S} + \mathbf{S} \cdot \mathbf{\Gamma}_y \right) \,,
\end{align}
with
\begin{align}
\label{eq:nuAD}
\mathbf{\Gamma}_y = &\, \left(
\int_{\mu^2}^{b_0^2/b^2} \frac{d\bar{\mu}^2}{\bar{\mu}^2} \gamma_{\rm cusp} [\alpha_s(\bar{\mu})] + \gamma_r[\alpha_s(b_0/b)] \right) \sum_i c_i  \mathbf{1} 
 + \boldsymbol{\gamma}_X[y^*, \alpha_s(b_0/b)] \,.
\end{align}
This is the generalization of the rapidity RGE~\cite{Chiu:2011qc,Chiu:2012ir} for color singlet production to dijet production at hadron colliders. Here $\gamma_r$ is the rapidity anomalous dimension for the color transverse momentum distribution \cite{Li:2016ctv}, and $b_0=2e^{-\gamma_E}$.

Thanks to the non-abelian exponentiation theorem~\cite{Gatheral:1983cz,Frenkel:1984pz},
we can write the TEEC soft function as an exponential of web diagrams. We further separate web diagrams into dipole, tripole and quadrupole contributions, by which we mean interactions involving two, three, and four Wilson lines respectively,
\begin{align}
  \mathbf{S}=\exp\Bigl[
    \mathbf{S_{\rm dip}} 
    +\mathbf{S}_{\rm tri}+\mathbf{S}_{\rm quad}
    \Bigr]\,.
\end{align}
The dipole contribution starts at $\cO(\as)$, where the calculation is naturally expressed as a sum over dipoles. 
We will show in \sec{soft_dipole} that\footnote{We have assumed here that the TMD soft function $S_{ij}$ takes the same expression no matter $i$ and $j$ being incoming or outgoing. In \cite{Collins:2004nx}, it was argued with contour deformation that the TMD soft function takes the same universal expression for Drell-Yan, $e^+e^-$ and semi-inclusive deep elastic scattering. It would be interesting to use the Glauber SCET~\cite{Rothstein:2016bsq} to prove or disprove this.
}
\begin{align}\label{eq:S_dipole}
  \mathbf{S}_{\rm dip}=-\sum_{i<j} \mathbf{T}_i \cdot \mathbf{T}_j\, S_{ij}
  =-\sum_{i<j} \mathbf{T}_i \cdot \mathbf{T}_j\, S_{\perp}\,\Bigl(L_b,L_\nu+\ln\frac{n_i\cdot n_j}{2}\Bigr)
  \,,
\end{align}
where $S_\perp(L_b,L_\nu)$ is the TMD soft function for color singlet production at hadron colliders (which can be found up to three loops in \cite{Li:2016ctv})

The tripole soft function starts at $\cO(\as^2)$; we will calculate it in \sec{soft_tripole}. The quadrupole contribution starts at three loop: its $\mu$ dependent part is predicted by $\boldsymbol{\gamma}_{\rm quad}$, while the constant piece is beyond the scope of this paper.

We note that if there were no $\mathbf{S}_{\rm tri}$ and $\mathbf{S}_{\rm quad}$, the dipole contribution alone already satisfies the rapidity RGE~\eqref{eq:softnuRG}. Rapidity RG consistency then implies that $\mathbf{S}_{\rm tri}$ and $\mathbf{S}_{\rm quad}$ are both free from rapidity divergences. 
We call the statement that rapidity divergences in the soft function take a dipole form (or equivalently, that $\mathbf{S}_{\rm tri}$ and $\mathbf{S}_{\rm quad}$ are both free from rapidity divergences) the ``rapidity dipole conjecture".
We will show in \sec{soft_tripole} that for $\mathbf{S}_{\rm tri}$ this is indeed the case to two loops. 
However, it is far from obvious that $\mathbf{S}_{\rm tri}$ and $\mathbf{S}_{\rm quad}$ will be free from rapidity divergences at higher loop order.
It would be particularly interesting to prove or to disprove the dipole conjecture.
If the dipole conjecture turns out to be wrong, it implies factorization violation.

While the renormalization group evolution predicts the logarithmic structure of the soft functions, to achieve N$^3$LL accuracy, we also need the soft function constants to two loops (which provide the boundary conditions for the RG evolution).

%%%%%%%%%%%%%%%%%%%%%%%%%%%%
\subsection{Dipole Contribution}
\label{sec:soft_dipole}
%%%%%%%%%%%%%%%%%%%%%%%%%%%%

In this section, we will show that each dipole soft function with general $n_i$ and $n_j$ can be written as $S_\perp$ with $\nu^2$ modified by the $n_i\cdot n_j/2$ factor

We will perform a one loop calculation to demonstrate this.
In the exponential rapidity regulator of \cite{Li:2016axz}, the bare one loop integral for each dipole soft function is given by
\begin{align}
  \label{eq:I_Sij}
  S_{ij} (b_y, \tau) =&\, 2(4\pi)^2\,\left(\frac{\mu^2
  e^{\gae}}{4\pi} \right)^{\frac{4-d}{2}} \! \int\!\frac{d^d k \,
  \delta^+\!(k^2) }{(2\pi)^{d-1}}
  \frac{ n_i \mcdot n_j}{(k\mcdot n_i)( k \mcdot n_j)}
\exp\left( - 2k_0\tau e^{-\gae} + \im b_y
  \mcdot k \right) \,,
\end{align}
where $d = 4 - 2 \e$, $\delta^+(k^2)=\theta(k_0)\delta(k^2)$, and $b_y$ is the impact parameter perpendicular to the scattering plane, $b_y^\mu=(0^+, 0^-, 0_x,b_y)$. 
$\tau$ here is the rapidity regulator: at the end of the calculation, one keeps the leading term of $\tau\to0$ and identify $\nu=\tau^{-1}$.
We work in the $\overline{\textrm{MS}}$ scheme by a redefinition of the bare scale $\mu^2_0
= \mu^2 e^\gae/(4\pi) $.
We then rescale $n_{i,j}$ to $\tilde n_{i,j}$ as
\begin{align}
    \tilde n_i=\frac{n_i}{\sqrt{(n_i\cdot n_j)/2}}
    \,,\qquad
    \tilde n_j=\frac{n_j}{\sqrt{(n_i\cdot n_j)/2}}
    \,,
\end{align}
such that $\tilde n_i\cdot \tilde n_j=2$, and work in the lightcone frame in terms of $\tilde n_i$ and $\tilde n_j$, i.e., we decompose momentum $k$ as,
\begin{align}
    k^\mu=\frac{k^+}{2} \tilde n_j^\mu +\frac{k^-}{2} \tilde n_i^\mu +  k_\perp^\mu\,,
\end{align}
where $k^+=\tilde n_i\cdot k$, $k^-=\tilde n_j\cdot k$. 
The one loop bare soft function now becomes
\begin{align}
  \label{eq:I_Sij_lc}
  S_{ij} (b_y, \tau) =&\, 2(4\pi)^2 \, \left(\frac{\mu^2
  e^{\gae}}{4\pi} \right)^{2-d/2} \int \frac{dk^-dk^+d^{d-2} k_\perp}{2(2\pi)^{d-1}}
  \delta^+(k^2)  \frac{2}{k^+ k^-}
\brk \qquad
\times \exp\left( -\Bigl[\sqrt{\frac{n_i\cdot n_j}{2}}\,(k^++k^-)+v_\perp\cdot k\Bigr]\tau e^{-\gae} + \im b_y
  \mcdot k \right)\,.
\end{align}
Comparing this integral with the one-loop calculation in \cite{Li:2016axz}, we see that the only difference is that $2k_0$ is replaced by the term in the square bracket instead of simply $k^++k^-$. Here $v_\perp$ is a perpendicular vector. 
The $v_\perp\cdot k$ part does not matter when we take the $\tau\to0$ limit since its contribution is $\cO(\tau)$ suppressed compared with $\im b_y\cdot k$. The $\sqrt{\frac{n_i\cdot n_j}{2}}$ is a multiplicative factor to $\tau$, which amounts to dividing $\nu^2$ by $\frac{n_i\cdot n_j}{2}$ in the final result.

We expect that a similar argument can be generalized to all loop order since there is only single logarithm for $\nu$ (or single pole in $\eta$ for the $\eta$ regulator~\cite{Chiu:2011qc,Chiu:2012ir}) in the exponential of the soft function, and thus the subleading terms in $\tau\to0$ do not matter.
 
Note that for $2\to 2$ kinematics, we have the following kinematic relations
\begin{align}
n_1\cdot n_2&=n_3\cdot n_4=2\,,\nn\\
n_1\cdot n_3&=n_2\cdot n_4=-\frac{2\hat{t}}{\hat{s}}\,,\nn\\
n_1\cdot n_4&=n_2\cdot n_3=-\frac{2\hat{u}}{\hat{s}}\,,
\end{align}
where $\hat s=(p_1+p_2)^2$, $\hat t=(p_1-p_3)^2$ and $\hat u=(p_2-p_3)^2$ are the partonic Mandelstam variables.

%%%%%%%%%%%%%%%%%%%%%%%%%%%%
\subsection{Tripole Contribution at Two Loops}
\label{sec:soft_tripole}
%%%%%%%%%%%%%%%%%%%%%%%%%%%%

In this section, we calculate the ``tripole" contribution that arises at two loops, which correlates three lines in the soft function. We will show that it is purely imaginary (antisymmetric and thus Hermitian) at this order.
Since the tree-level hard functions are real, the imaginary constant piece of the soft function at two loops is not relevant at the accuracy with which we work in this paper. However, the scale dependent part does contribute and is already predicted in the imaginary part of the soft anomalous dimension in \Eq{eq:Gammas}.

According to \cite{Catani:1999ss}, the double real contribution is dipole-like and is completely incorporated in $\mathbf{S_{\rm dip}}$. Therefore, we only need to calculate the real-virtual contribution.
To this end, we make use of the tree-level and one-loop soft-gluon currents in \cite{Catani:2000pi},
\begin{align}
  \mathbf J^{(0)\mu}_a(q)&=\sum_{i} \mathbf{T}_i^a \frac{p_i^\mu}{p_i\cdot q}\\
    \mathbf J_{a}^{(1)\mu}(q, \eps) &= -\frac{1}{16\pi^2}\; \frac{1}{\eps^2} \;\frac{\Gamma^3(1-\eps)
    \,\Gamma^2(1+\eps)}{\Gamma(1-2\eps)} \nn \\
    &\,\times \;\; i\, f^{abc}\sum_{i\neq j}\mathbf{T}_i^b\;\mathbf{T}_j^c
    \left(\frac{p_{i}^{\mu}}{p_i\cdot q}-\frac{p_{j}^{\mu}}{p_j\cdot q}\right)
    \left[\frac{4\pi\, p_i\cdot p_j \,e^{-i\pi\lambda_{ij}}}{2 (p_i\cdot q)\, (p_j\cdot q) \, 
    e^{-i\pi\lambda_{iq}} \,e^{-i\pi\lambda_{jq}}} \right]^\eps\, 
\end{align}
where $\lambda_{AB} = +1$ if $A$ and $B$ are both incoming or outgoing, and $\lambda_{AB} = 0$ otherwise. The real-virtual integrand is~\cite{Catani:2000pi},
\begin{align} \label{eq:vrInt}
&\mathbf J_{\mu}^{(0)}(q) \cdot \mathbf J^{(1)\mu}(q, \epsilon)+\text{h.c.}
\nn\\ 
=\,&  -\frac{1}{4 \pi^{2}} \frac{(4 \pi)^{\epsilon}}{\epsilon^{2}} \frac{\Gamma^{3}(1-\epsilon) \Gamma^{2}(1+\epsilon)}{\Gamma(1-2 \epsilon)}
\biggl\{C_{A} \cos (\pi \epsilon) \sum_{i, j}~\!\!\!^{\prime} \left[\mathcal{S}_{i j}(q)\right]^{1+\epsilon} (\mathbf{T}_i\cdot \mathbf{T}_j)
\nn  \\ & \qquad\quad
-2i \sin (\pi \epsilon) \sum_{i, j, k}~\!\!\!^{\prime} \mathcal{S}_{k i}(q)\left[\mathcal{S}_{i j}(q)\right]^{\epsilon}\left(\lambda_{i j}-\lambda_{i q}-\lambda_{j q}\right)
(if^{abc})\,\mathbf{T}_k^{a} \mathbf{T}_i^{b} \mathbf{T}_j^{c}
 \biggr\} 
 \,.
\end{align}
The dependence on whether partons are incoming or outgoing is encoded in $\lambda$, 
\begin{align}
\lambda_{i j}-\lambda_{i q}-\lambda_{j q}  & = 1 \text{ ($i$, $j$ both incoming)},\qquad -1 \text{ (otherwise)}.
\end{align}
The summation $\sum~\!\!\!^{\prime} $ stands for the sum over the different values of the indices. 
The soft eikonal function  is defined as 
\begin{align}
\mathcal{S}_{i j}(q)=\frac{p_{i} \cdot p_{j}}{2\left(p_{i} \cdot q\right)\left(p_{j} \cdot q\right)}=\frac{s_{i j}}{s_{i q} s_{j q}}\,.
\end{align}
As discussed above, here we will focus on the three Wilson-line contribution  which is represented as the last line of eq.~(\ref{eq:vrInt}). 

Notice that only the imaginary part of the one-loop soft-gluon current survives in the tripole contribution (last line of \eq{vrInt}). According to \cite{Rothstein:2016bsq}, this imaginary part is exactly the Glauber contribution.
Therefore, we see that the tripole contribution at $\cO(\as^2)$ in the soft function is purely from Glaubers. This is quite interesting since it is very different from the case for the dipole soft function, where Glauber effects only start to contribute at $\cO(\as^3)$.

We now perform the calculation for the tripole contribution to the TEEC soft function with four Wilson lines. Making use of color conservation
\begin{align}
   \mathbf{T}_1 + \mathbf{T}_2 + \mathbf{T}_3 + \mathbf{T}_4  = 0 ~,
\end{align}
the color structure of the three-parton correlation can be reduced to $f^{abc}\mathbf{T}_1^a \mathbf{T}_2^b\mathbf{T}_3^c $. For example, 
\begin{align}
   f^{abc} \mathbf{T}_1^a \mathbf{T}_2^b\mathbf{T}_4^c =& -f^{abc} \mathbf{T}_1^a \mathbf{T}_2^b\mathbf{T}_3^c -f^{abc} \mathbf{T}_1^a \mathbf{T}_2^b \mathbf{T}_1^c -f^{abc} \mathbf{T}_1^a \mathbf{T}_2^b\mathbf{T}_2^c 
   \nonumber \\
   =&  -f^{abc} \mathbf{T}_1^a \mathbf{T}_2^b\mathbf{T}_3^c \,,
\end{align} 
where we have used the relation 
\begin{align}
    [\mathbf{T}_k^a, \mathbf{T}_j^b] = i \delta_{kj} f^{abc} \mathbf{T}_k^c\,.
\end{align}
We define 
\begin{align}
    I_{ijk} = \mathcal{S}_{k i}(q)\left[\mathcal{S}_{i j}(q)\right]^{\epsilon}~.
\end{align}
The soft integrand becomes 
\begin{align}
 f_{a b c} &\sum_{i, j, k}~\!\!\!^{\prime} \mathbf{T}_{k}^{a} \mathbf{T}_{i}^{b} \mathbf{T}_{j}^{c}\   \mathcal{S}_{k i}(q)\left[\mathcal{S}_{i j}(q)\right]^{\epsilon}\left(\lambda_{i j}-\lambda_{i q}-\lambda_{j q}\right) = 
 \nonumber \\ 
    &  f^{abc} \mathbf{T}_1^a \mathbf{T}_2^b\mathbf{T}_3^c  
    \big\{I_{123}-I_{124}+I_{132}-I_{134}-I_{142}+I_{143} -I_{213}+I_{214}
    \nonumber \\ & 
    -I_{231}+I_{234}+I_{241}-I_{243}-I_{312}+I_{314}+I_{321}-I_{324}-I_{341}+I_{342}
    \nonumber \\ & 
   +I_{412}-I_{413}-I_{421}+I_{423}
   +I_{431}-I_{432}
    \big\}\,.
\end{align}
The rapidity divergences cancel separately in each term of the form $I_{ijk}-I_{jik}$, giving rise to a rapidity finite result.   The phase space integrals can be performed straightforwardly. The final result is
\begin{align} \label{eq:vr}
    \mathbf{S}_{\rm tri}^{bare}= 
    i f^{abc} \mathbf{T}_1^a \mathbf{T}_2^b\mathbf{T}_3^c\, S_{\rm tri}^{bare}
    =\left(\frac{\alpha_s}{4\pi}\right)^2 f^{abc} \mathbf{T}_1^a \mathbf{T}_2^b\mathbf{T}_3^c \, \ln\frac{\hat t}{\hat u}  \left(\frac{b^2\mu^2}{4 e^{-2\gamma_E}}\right)^{2\epsilon} \frac{8\pi}{3}\left( \frac{6}{\epsilon^2}+\pi^2  + \mathcal{O}(\epsilon)\right)\,.
\end{align}
Notice that the color factor $f^{abc} \mathbf{T}_1^a \mathbf{T}_2^b\mathbf{T}_3^c$ is a purely imaginary matrix once the color basis is specified. The divergent terms appearing in this result can be predicted by the RG equation with the imaginary part of the anomalous dimensions in \Eq{eq:Gammas}.

We can also obtain the result for the TEEC soft function with three Wilson lines, which is relevant for $W/Z/\gamma+$jet. 
In that case, using the color conservation identity
\begin{align}
\mathbf{T}_1+\mathbf{T}_2+\mathbf{T}_3=0\,,
\end{align}
we clearly see that the tripole contribution vanishes due to the antisymmetry of $f^{abc}$,
\begin{align}
\left.\mathbf{S}_{\rm tri}^{bare}
\right|_{\text{Three Wilson Lines}}=0\,.
\end{align}
It would be interesting to understand whether or not a tripole contribution can contribute at higher perturbative orders.

In summary, this calculation explicitly shows that to two loops, the soft function is purely dipole for the two and three Wilson line soft functions, and for the four Wilson line soft function, there is a purely imaginary contribution.

%%%%%%%%%%%%%%%%%%%%%%%%%%%%
\subsection{Summary and Discussion}
\label{sec:discuss}
%%%%%%%%%%%%%%%%%%%%%%%%%%%%

We can summarize our one and two loop calculations as follows.  The TEEC soft function has the perturbative expansion
\begin{align}
\mathbf{S}(b,y^*, \mu,\nu) = \mathbf{1} + \frac{\alpha_s}{4 \pi} \mathbf{S}^{(1)}(y^*, L_b, L_\nu) + 
\left(\frac{\alpha_s}{4 \pi} \right)^2 \mathbf{S}^{(2)}(y^*, L_b, L_\nu) + \cO(\alpha_s^3) \,,
\end{align}
with
\begin{align}
  \mathbf{S}^{(1)}(y^*, L_b, L_\nu) =&  - \sum_{i<j} \left(\mathbf{T}_i \cdot \mathbf{T}_j \right) S_\perp^{(1)}\left(L_b, L_\nu + \ln \frac{n_i \cdot n_j}{2} \right) \,, \nonumber \\
\mathbf{S}^{(2)}(y^*, L_b, L_\nu) = &  - \sum_{i<j} \left(\mathbf{T}_i \cdot \mathbf{T}_j \right) S_\perp^{(2)}\left(L_b, L_\nu + \ln \frac{n_i \cdot n_j}{2} \right)  +\frac{1}{2!} \left( \mathbf{S}^{(1)}(y^*, L_b, L_\nu)\right)^2
\nn\\
&+if^{abc}\mathbf{T}_1^a \mathbf{T}_2^b \mathbf{T}_3^c\, S_{\rm tri}(y^*, L_b)
\,.
\end{align}
Here $S_\perp^{(n)}(L_b, L_\nu)$ is the $n$-loop TMD soft function for color-singlet production at hadron colliders,
and $S_{\rm tri}(y^*,L_b)$ is the purely imaginary tripole contribution calculated in \eq{vr}.

Our calculation also explicitly shows that at least to two loops, the color non-diagonal rapidity anomalous dimension, $\boldsymbol{\gamma}_X$, vanishes. This is guaranteed by rescaling invariance, $n_i \to e^{\lambda_i} n_i$, but is a non-trivial check on our calculation. We note that the consistency of the factorization formula, which is derived from demanding the rapidity scale independence of the cross section implies that $\boldsymbol{\gamma}_X=0$ at all orders. This is a highly non-trivial statement, since at three loops there is a scaling invariant cross ratio $n_1 \cdot n_3\, n_2 \cdot n_4 /(n_1 \cdot n_2\, n_3 \cdot n_4) = (1 - \tanh y^*)^2/4$. If it is indeed true that $\boldsymbol{\gamma}_X=0$, then we believe that there should be some argument for this fact purely at the level of the soft function definition. On the other hand, if $\boldsymbol{\gamma}_X \neq 0$, then this would indicate an explicit violation of factorization.

We note that one must be careful in what is meant by the soft function when discussing potential violations of factorization. In particular, it is normally assumed that Glauber contributions can simply be absorbed into the directions of Wilson lines (or proven in certain cases such as $q_T$). In the present case of the TEEC, where this is not expected to be true, one should probably work with the true soft function, defined by removing the Glauber zero-bin
\begin{align}
S=\tilde S-S^{(G)}\,,
\end{align}
where $\tilde S$ denotes the naive soft function, and $S^{(G)}$ denotes is Glauber zero-bin \cite{Rothstein:2016bsq}.
Once this Glauber zero-bin contribution is removed, we expect that the naive factorization formula is not-violated. We are therefore led to the following conjecture

\vspace{0.25cm}
{\bf Rapidity Dipole Conjecture}: The rapidity anomalous dimension for a soft function with Wilson lines in distinct directions $n_i$, is dipole to all loop order once the Glauber zero-bin is performed.
\vspace{0.25cm}

It would be extremely interesting to prove or disprove this statement, and we believe that it would improve our understanding of rapidity factorization with multiple collinear directions, for which little is known. This motivates a direct calculation of the TEEC soft function at three loops, as well as a better understanding of rapidity regularization for multi-Wilson line soft functions.

Finally, along the lines of \cite{Vladimirov:2017ksc,Vladimirov:2016dll,Moult:2022xzt}, it will be important to better understand the all orders structure of rapidity anomalous dimensions for multi-Wilson line soft functions, perhaps by relating them to standard virtuality ($\mu$) anomalous dimensions.  (For other recent work understanding relations between anomalous dimensions defined by Wilson lines, see \cite{Falcioni:2019nxk}). It would be interesting to obtain the rapidity anomalous dimension for multi-Wilson line soft functions in a similar manner, perhaps from some self crossing limit of a more complicated Wilson loop structure (see e.g.  \cite{Dixon:2016epj}).

%%%%%%%%%%%%%%%%%%%%%%%%%%%%
\section{Color Evolution at N$^3$LL}
\label{sec:ingr-resumm-teec}
%%%%%%%%%%%%%%%%%%%%%%%%%%%%

The beam and jet functions are color singlets, and their renormalization group evolution structure is standard. The most complicated aspect of the renormalization at N$^3$LL is the non-trivial color evolution of the hard and soft functions. Since this is applicable to any dijet soft function, and has not previously been presented at N$^3$LL accuracy, we discuss it in some detail with the hope that it will be useful more generally.

%%%%%%%%%%%%%%%%%%%%%%%%%%%%
\subsection{Hard and Soft Function Anomalous Dimensions}
\label{sec:ingr-resumm-teec-hard}
%%%%%%%%%%%%%%%%%%%%%%%%%%%%

 The soft function is a matrix in color space, and satisfies the RG equation
\begin{align}
  \label{eq:softRG_p}
  \frac{d\mathbf{S}}{d\ln \mu^2} = \frac{1}{2} \left( \mathbf{\Gamma}_S^\dagger \cdot \mathbf{S} + \mathbf{S} \cdot \mathbf{\Gamma}_S \right) \,.
\end{align}
The anomalous dimension $\mathbf{\Gamma}_S$ takes the form given in \eq{Gammas}, which we repeat here for convenience \cite{Kidonakis:1998bk,Kidonakis:1998nf,Aybat:2006mz,Aybat:2006wq} 
\begin{align}
  \label{eq:Gammas'}
  \mathbf{\Gamma}_S = \sum_{i<j} \mathbf{T}_i \cdot \mathbf{T}_j \gamma_{\rm cusp} \ln\frac{\sigma_{ij}\nu^2\, n_i \cdot n_j-i 0}{2 \mu^2} - \sum_i \frac{c_i}{2} \gamma_s \mathbf{1} - \boldsymbol{\gamma}_{\rm quad}  \,.
\end{align}
Here $\sigma_{ij}=-1$ if both $i$ and $j$ are incoming or outgoing, and $\sigma_{ij}=1$ otherwise.
$\nu$ is the rapidity scale, and $c_i = C_F$ or $C_A$ is the Casimir of the parton $i$. Here $\gamma_{\rm cusp}$ is the cusp anomalous dimension \cite{Korchemsky:1987wg}, $\gamma_s$ is the threshold soft anomalous dimension \cite{Li:2014afw} and  $\boldsymbol{\gamma}_{\rm quad}$ is the  anomalous dimension for quadrupole color and kinematic entanglement, which first appears at three loops \cite{Almelid:2015jia,Almelid:2017qju}. Here we use the color space notation of \cite{Catani:1996vz}. The quadrupole anomalous dimension is universal (matter independent) \cite{Dixon:2009gx}, and can be written as a function of the conformal cross ratios. For extensive earlier work on its structure, see \cite{Kidonakis:1998nf,Aybat:2006wq,Aybat:2006mz,Dixon:2008gr,Dixon:2009gx,Gardi:2009zv, Gardi:2009qi, Dixon:2009ur, Becher:2009cu, Becher:2009qa,Gardi:2010rn,DelDuca:2011ae,Gardi:2011wa,Gardi:2011yz,Bret:2011xm, Ahrens:2012qz,DelDuca:2013ara, Caron-Huot:2013fea,Gardi:2013saa,Gardi:2013ita,Falcioni:2014pka}, and for a detailed discussion summarizing the complete set of known results and their consistency, see \cite{Becher:2019avh}.

To our knowledge, the quadrupole term in the anomalous dimension matrix has not yet entered into a physical observable. The quadrupole part of the soft anomalous dimension can be written as a sum of two terms 
\begin{align}
\Delta=16(\bold{\Delta}_4^{(3)}+\bold{\Delta}_3^{(3)})\,.
\end{align}
In \cite{Almelid:2017qju} it was shown in detail how to analytically continue the functions appearing in the quadrupole anomalous dimension to the physical region. However, in \cite{Henn:2016jdu}, an analytically continued form was given for $13\to24$ kinematics ($u=-s-t>0$), which is sufficient for our purposes. Here we use the results of \cite{Henn:2016jdu}.

Restricting the general form of the quadrupole anomalous dimension to four external partons, for $\bold{\Delta}_3^{(3)}$, we have
\begin{align}\label{eq:quadrupole}
\bold{\Delta}_3^{(3)}&=-  C \,f_{abe}f_{cde}  \sum_{\substack{{i=1 \ldots 4}\\{1\leq j<k\leq 4}\\ j,k\neq i}}\left\{{\rm \bf T}_i^a,  {\rm \bf T}_i^d\right\}   {\rm \bf T}_j^b {\rm \bf T}_k^c\,,
\end{align}
or very explicitly
\begin{align}
\bold{\Delta}_3^{(3)}=-  C \,f_{abe}f_{cde}\Big [ & \left\{{\rm \bf T}_1^a,  {\rm \bf T}_1^d\right\}  ( {\rm \bf T}_2^b {\rm \bf T}_3^c +  {\rm \bf T}_2^b {\rm \bf T}_4^c+  {\rm \bf T}_3^b {\rm \bf T}_4^c ) \nn \\
+&\left\{{\rm \bf T}_2^a,  {\rm \bf T}_2^d\right\}  ( {\rm \bf T}_1^b {\rm \bf T}_3^c +  {\rm \bf T}_1^b {\rm \bf T}_4^c+  {\rm \bf T}_3^b {\rm \bf T}_4^c ) \nn \\
+&\left\{{\rm \bf T}_3^a,  {\rm \bf T}_3^d\right\}  ( {\rm \bf T}_1^b {\rm \bf T}_2^c +  {\rm \bf T}_1^b {\rm \bf T}_4^c+  {\rm \bf T}_2^b {\rm \bf T}_4^c )\nn \\
+&\left\{{\rm \bf T}_4^a,  {\rm \bf T}_4^d\right\}  ( {\rm \bf T}_1^b {\rm \bf T}_2^c +  {\rm \bf T}_1^b {\rm \bf T}_3^c+  {\rm \bf T}_2^b {\rm \bf T}_3^c ) \Big ]\,,
\end{align}
where the constant $C$ is given by
\begin{align}
C=\zeta_5+2\zeta_3\zeta_2\,.
\end{align}
For the analytic continuation of $\bold{\Delta}_4^{(3)}$, we use the form given in \cite{Henn:2016jdu}
\begin{align}
\bold{\Delta}_4^{(3)} &=  \frac{1}{4} \, f_{abe}f_{cde} \Big[ 
 {\rm \bf T}_1^a  {\rm \bf T}_2^b   {\rm \bf T}_3^c {\rm \bf T}_4^d   \, \mathcal{S}(x) 
  +{\rm \bf T}_4^a   {\rm \bf T}_1^b  {\rm \bf T}_2^c    {\rm \bf T}_3^d \, \mathcal{S}(1/x) \Big], 
\end{align}
where 
\begin{align}
\mathcal{S}(x) &=    2 H_{-3,-2}+2 H_{-2,-3}-2 H_{-3,-1,-1}+2 H_{-3,-1,0} 
  -2 H_{-2,-2,-1}    +2 H_{-2,-2,0}-2
   H_{-2,-1,-2} \nn \\
   &-H_{-1,-2,-2} 
-H_{-1,-1,-3}+4 H_{-2,-1,-1,-1}  -2
   H_{-2,-1,-1,0} 
   -H_{-1,-2,-1,0}-H_{-1,-1,-2,0} \nn \\
   &    +\zeta _3 H_{-1,-1}
  +4 \zeta _3 \zeta _2-\zeta _5  
+ \zeta _2 ( 6  H_{-3}   -10 H_{-2,-1}+6  H_{-2,0}- H_{-1,-2} - H_{-1,-1,0}) \nonumber\\ 
   &+ i \pi \Big[
2 H_{-3,-1}-2 H_{-3,0}+2 H_{-2,-2}-4 H_{-2,-1,-1} 
 +2 H_{-2,-1,0}-2
   H_{-2,0,0}+H_{-1,-2,0}\nn \\
   &+H_{-1,-1,0,0}+   \zeta_2 ( 3   H_{-1,-1}-4  H_{-2}) -\zeta _3 H_{-1} \Big] \,.
\end{align}
Here $H$ are harmonic polylogarithms (HPLs), and the standard convention for the weights has been followed. The argument of the HPLs has been suppressed, and is $x=t/s$. This result contains explicit factors of $i\pi$ that are generated by the analytic continuation. While they do not contribute to the cross section at the order that we work, they would be interesting to understand from the perspective of Glaubers. We leave this to future work.

At N$^3$LL, we also need the cusp anomalous dimension at four loops \cite{Henn:2019swt}. Its value, as well as the value of all other anomalous dimensions required to derive the N$^3$LL result for the TEEC, are provided in \App{sec:anomalous-dimensions}.

%%%%%%%%%%%%%%%%%%%%%%%%%%%%
\subsection{Solving Color Evolution Equations to N$^3$LL}
\label{sec:color}
%%%%%%%%%%%%%%%%%%%%%%%%%%%%

The anomalous dimensions of the hard function in Eq.~(\ref{eq:Gammah}) can be decomposed as the sum of a diagonal matrix and a non-diagonal matrix,  
\begin{align}\label{eq:gammahMatrix}
\boldsymbol{\Gamma}_H = \Gamma^{ D}_h(\alpha_s, \mu)   \boldsymbol{1} + \boldsymbol{\gamma}_h(\alpha_s)\,.    
\end{align}
where  $\boldsymbol{\gamma}_h$ is the non-diagonal matrix contribution, which can be written as  
\begin{align}
    \boldsymbol{\gamma}_h = \frac{\alpha_s}{4\pi}\boldsymbol{\gamma}^h_0  + \left( \frac{\alpha_s}{4\pi} \right)^2 \boldsymbol{\gamma}^h_1 + \left( \frac{\alpha_s}{4\pi} \right)^3\boldsymbol{\gamma}^h_2 + \cdots\,.
\end{align}
The solution to the hard function RG equation is 
\begin{align}
    \boldsymbol{H}(\mu) = \boldsymbol{U}(\mu_h, \mu)\  \boldsymbol{H}(\mu_h)  \ \boldsymbol{U}^\dagger(\mu_h, \mu)\,,
\end{align}
with  
\begin{align}
    \boldsymbol{U}(\mu_h, \mu) = \exp\left[ \int_{\mu_h}^\mu \frac{d\bar{\mu}}{\bar{\mu}} \Gamma^{ D}_h(\alpha_s(\bar{\mu}), \bar{\mu})  \right]  \boldsymbol{u}(\mu_h,\mu).
\end{align}
The factor $\int_{\mu_h}^\mu \frac{d\bar{\mu}}{\bar{\mu}} \Gamma^{ D}_h(\alpha_s(\bar{\mu}), \bar{\mu})$ includes the evolution similar to  a color singlet state. The non-trivial color evolution is included in $\boldsymbol{u}$ which obeys the differential equations
\begin{align} \label{eq:rgu}
    \frac{d}{d\ln \mu }\boldsymbol{u}(\mu_h,\mu) =& \boldsymbol{\gamma}_h(\alpha_s(\mu)) \boldsymbol{u}(\mu_h,\mu)\,,
    \nn \\ 
    \frac{d}{d\ln \mu_h }\boldsymbol{u}(\mu_h,\mu) =& -\boldsymbol{u}(\mu_h,\mu) \boldsymbol{\gamma}_h(\alpha_s(\mu_h))\,.
\end{align}
The solution to these equations is
\begin{align}
    \boldsymbol{u}(\mu_h,\mu) = \mathcal{P} \exp\left[ \int_{\alpha_s(\mu_h)}^{\alpha_s(\mu)} \frac{d\alpha}{\beta(\alpha)} \boldsymbol{\gamma}_h(\alpha)  \right]\,,
\end{align}
where $\mathcal{P}$ denotes the ordering in the coupling constant with the scale in the coupling increasing from left to right. The ordering operator is necessary when $[\boldsymbol{\gamma}_h(\alpha_1), \boldsymbol{\gamma}_h(\alpha_2)] \neq 0 $\,. Here we will give explicit expressions for the $\mathbf{u}$ matrix at NLL, NNLL and N$^3$LL.

At NLL  the $\mathbf{u}$  matrix is  given by
\begin{align}
    \boldsymbol{u}_{\rm NLL}(\mu_h,\mu) = \boldsymbol{V}\ \left( \frac{\alpha_s(\mu)}{\alpha_s(\mu_h)}\right)^{-\frac{\mathrm{\gamma^0_D}}{2\beta_0}} \boldsymbol{V}^{-1}\,,
\end{align}
where $\boldsymbol{V}$ is the  matrix that diagonalizes the LO anomalous dimension
\begin{align}\label{eq:diag}
    \boldsymbol{\gamma}^0_D = \boldsymbol{V}^{-1} \ \boldsymbol{\gamma}^h_0 \ \boldsymbol{V}\,. 
\end{align}
Higher order QCD corrections can be included by expressing them in terms of $\mathbf{u}_{\rm NLL}$ as~\cite{Buras:1991jm}
\begin{align}
    \boldsymbol{u}(\mu_h,\mu) = \boldsymbol{K}(\mu)\boldsymbol{u}_{\rm NLL}(\mu_h,\mu)\boldsymbol{K}^{-1}(\mu_h)\,.
\end{align}
From Eq.~(\ref{eq:rgu}), the differential equation for the matrix $\mathbf{K}$ up to NNLO is
\begin{align}
   \beta(\alpha_s(\mu))  \frac{d}{d\alpha_s(\mu)} \boldsymbol{K}(\alpha_s(\mu)) - \frac{\beta(\alpha_s(\mu))}{2\alpha_s(\mu)\beta_0} \boldsymbol{K}(\alpha_s(\mu))  \boldsymbol{\gamma}^h_0 &
   \nonumber \\
   & \hspace{-6cm} =\frac{\alpha_s(\mu)}{4\pi} \boldsymbol{\gamma}^h_0 \boldsymbol{K}(\mu) +\left(\frac{\alpha_s(\mu)}{4\pi}\right)^2 \boldsymbol{\gamma}^h_1\boldsymbol{K}(\mu)  + \left(\frac{\alpha_s(\mu)}{4\pi}\right)^3 \boldsymbol{\gamma}^h_2\boldsymbol{K}(\mu)\,.
\end{align}
Defining the perturbative expansion of  $\boldsymbol{K}$ as
\begin{align}
     \boldsymbol{K}  = \boldsymbol{1} + \frac{\alpha_s(\mu)} {4\pi}  \boldsymbol{K}_0 +  \left(\frac{\alpha_s(\mu)}{4\pi}\right)^2 \boldsymbol{K}_1  + \cdots \,,
\end{align}
we have 
\begin{align} \label{eq:k0}
    \boldsymbol{K}_0 + \frac{1}{2\beta_0} [\boldsymbol{\gamma}^h_0, \boldsymbol{K}_0] =& \frac{\beta_1 \boldsymbol{\gamma}^h_0}{2\beta_0^2} -\frac{\boldsymbol{\gamma}^h_1}{2\beta_0}\,.
    \nonumber \\
    \boldsymbol{K}_1 + \frac{1}{4 \beta_0} [ \boldsymbol{\gamma}^h_0, \boldsymbol{K}_1] =& \frac{1}{4\beta_0}\left(
    \frac{\beta_1}{\beta_0} \boldsymbol{\gamma}^h_0 \boldsymbol{K}_0 -\boldsymbol{\gamma}^h_1 \boldsymbol{K}_0-\frac{\beta_1^2}{\beta_0^2} \boldsymbol{\gamma}^h_0+\frac{\beta_1}{\beta_0} \boldsymbol{\gamma}^h_1+\frac{\beta_2}{\beta_0} \boldsymbol{\gamma}^h_1 -\boldsymbol{\gamma}^h_2
    \right)~. 
\end{align}
Using Eq.~(\ref{eq:diag}) we transform $\boldsymbol{K}$ into  $\boldsymbol{S} = \boldsymbol{V}^{-1}\boldsymbol{K} \boldsymbol{V}$, and perturbatively expand $\boldsymbol{S}_i$ as
\begin{align}
     \boldsymbol{S}  = \boldsymbol{1} + \frac{\alpha_s(\mu)} {4\pi}  \boldsymbol{S}_0 +  \left(\frac{\alpha_s(\mu)}{4\pi}\right)^2 \boldsymbol{S}_1  + \cdots \,.
\end{align}
The solution of Eq.~(\ref{eq:k0}) is then given by
\begin{align}
    \boldsymbol{S}_{0, IJ} =& \delta_{IJ} \frac{\beta_1}{ 2\beta_0^2} (\gamma_D^0)_{JJ} - \frac{(\boldsymbol{V}^{-1} \boldsymbol{\gamma}^h_1 \boldsymbol{V})_{IJ}}{2\beta_0 +(\gamma_D^0)_{II}-(\gamma_D^0)_{JJ} }\,,
    \nn  \\
    \boldsymbol{S}_{1, IJ} =& \frac{\delta_{IJ}}{4\beta_0} \left( \frac{\beta_2}{\beta_0} - \frac{\beta_1^2}{\beta_2^2}\right)(\gamma_D^0)_{JJ}
    \nn \\ &
    +\frac{
    1}{4\beta_0 +(\gamma_D^0)_{II}-(\gamma_D^0)_{JJ} }
    \left(\boldsymbol{V}^{-1}\left( \frac{\beta_1}{\beta_0} \boldsymbol{\gamma}^h_0 \boldsymbol{K}_0 - \boldsymbol{\gamma}^h_1 \boldsymbol{K}_0+\frac{\beta_1}{\beta_0}  \boldsymbol{\gamma}^h_1 - \boldsymbol{\gamma}^h_2  \right)\boldsymbol{V} \right)_{IJ}\,.
\end{align}
The expressions for the $\boldsymbol{u}$ matrices at NNLL and N$^3$LL are
\begin{align}
    \boldsymbol{u}_{\rm NNLL}(\mu_h,\mu) = & \boldsymbol{V}\ \left( 1+\frac{\alpha_s(\mu)}{4\pi } \boldsymbol{S}_0  \right)\  \left( \frac{\alpha_s(\mu)}{\alpha_s(\mu_h)}\right)^{-\frac{\mathrm{\gamma^0_D}}{2\beta_0}} \left( 1+\frac{\alpha_s(\mu_h)}{4\pi } \boldsymbol{S}_0  \right)^{-1}  \boldsymbol{V}^{-1}\,,
    \nn  \\ 
    \boldsymbol{u}_{\rm N^3LL}(\mu_h,\mu) =  &  \boldsymbol{V}\ \left( 1+\frac{\alpha_s(\mu)}{4\pi } \boldsymbol{S}_0 +\left(\frac{\alpha_s(\mu)}{4\pi }\right)^2 \boldsymbol{S}_1 \right)\  \left( \frac{\alpha_s(\mu)}{\alpha_s(\mu_h)}\right)^{-\frac{\mathrm{\gamma^0_D}}{2\beta_0}}     
    \nn \\ & \times 
     \left( 1+\frac{\alpha_s(\mu_h)}{4\pi } \boldsymbol{S}_0 +\left(\frac{\alpha_s(\mu_h)}{4\pi }\right)^2 \boldsymbol{S}_1 \right)^{-1}  \boldsymbol{V}^{-1}\,. 
\end{align}
These matrices allow for the resummation up to N$^3$LL for generic color mixing matrices, and we believe that they will prove useful in many future studies of event shapes at hadron colliders, or multi-jet event shapes in $e^+e^-$ colliders.

%%%%%%%%%%%%%%%%%%%%%%
\section{Linearly Polarized Beam and Jet Functions}
\label{sec:linear_polarization}
%%%%%%%%%%%%%%%%%%%%%%

Another interesting feature of the TEEC which first appears at N$^3$LL, is the presence of linearly polarized jet and beam functions. Since the TEEC is measuring radiation perpendicular to the scattering plane formed by the hard process, the helicity of gluons in the jets or beams can lead to terms which have an azimuthal dependence as they are rotated through this plane. These are described by the linearly polarized beam and jet functions. Similar polarization effects in the collinear limit of the energy correlators were discussed in \cite{Chen:2020adz,Chen:2021gdk,Li:2023gkh}. There the effect came from the presence of other detectors, instead of the plane of the hard scattering process. The matching coefficients for the gluonic TMD beam and fragmentation functions can be decomposed into tensor structures as
\begin{align}
\cI_{gi}^{\mu \nu}(\xi,b_\perp)=\frac{g_\perp^{\mu \nu}}{d-2} \cI_{gi}(\xi,b_\perp) +\left ( \frac{g_\perp^{\mu \nu}}{d-2}+\frac{b_\perp^\mu b_\perp^\nu}{b_T^2}  \right ) \cI'_{gi}(\xi,b_\perp)\,.
\end{align}
For a detailed discussion and NNLO calculation for both TMD beam functions and fragmentation functions in our regulator, see \cite{Luo:2019bmw}. The one loop matching coefficients and the one-loop beam and jet functions for  linearly polarized gluons are given in \App{app:pert}. Importantly, they are first non-zero at one-loop, where they give finite results. 

An interesting feature of the TEEC on dijets, is that the tree level hard scattering matrix elements are maximal helicity violating (MHV). Therefore, when contracted with a single linearly polarized jet or beam function, the result vanishes. One must either have two linearly polarized jet or beam functions, or a one-loop correction to the hard scattering matrix element to get a non-vanishing NMHV amplitude combined with a single linearly polarized beam or jet function. For a detailed discussion of the helicity structure of SCET hard matching coefficients, and explicit results, see \cite{Moult:2015aoa}. This shows that for the TEEC on dijets, effects from linearly polarized beam and jet functions first enter as a constant at $\cO(\alpha_s^2)$, namely at N$^3$LL. For this reason, they were not needed in our NNLL calculation \cite{Gao:2019ojf}. This is in distinction to the case of the TEEC on $V+$ jets, or other related $V+$ jet observables \cite{Chien:2020hzh,Chien:2022wiq}, where linearly polarized jet and beam functions enter at NNLL, since the hard function does not have an MHV structure.

It would be interesting to study the phenomenological impact of these linearly polarized terms in more detail. However, since the focus of this paper has been on the derivation of the factorization formula, we leave such studies to future work.

%%%%%%%%%%%%%%%%%%%%%%
\section{Fixed Order Singular Behavior}
\label{sec:numerics}
%%%%%%%%%%%%%%%%%%%%%%

\begin{figure}%[h]
  \centering
  \includegraphics[width=0.9\linewidth]{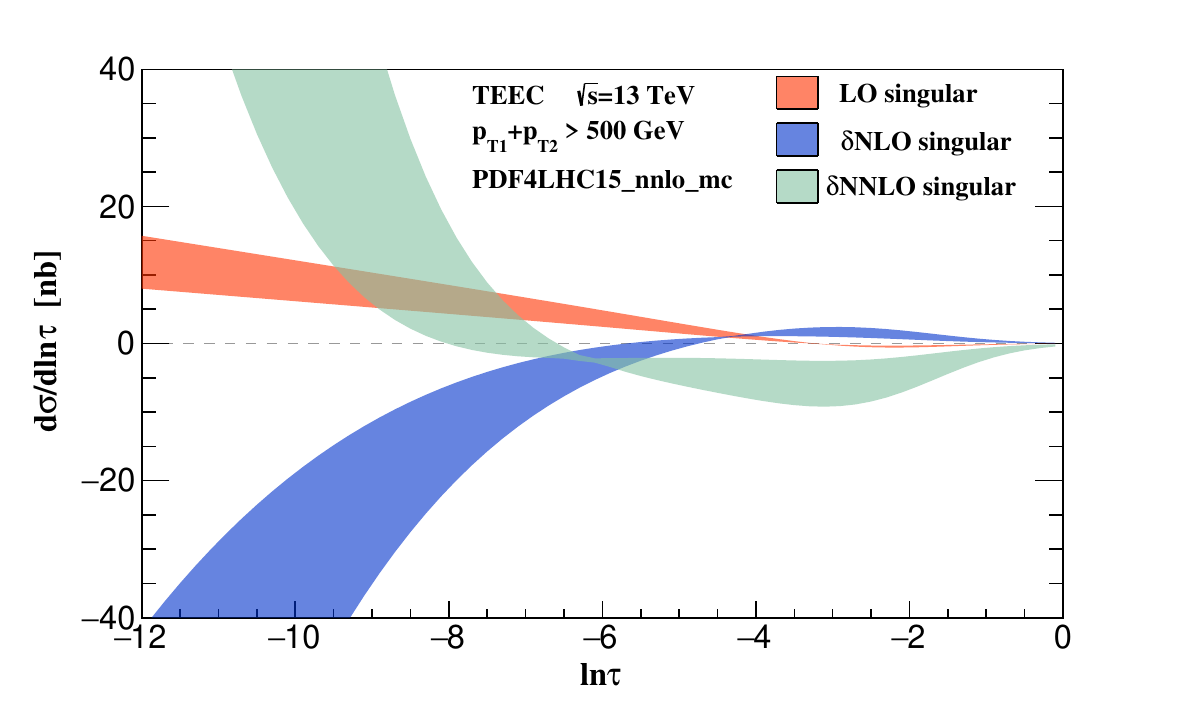}
  \caption{The fixed order singular behavior of the TEEC cross section at LO, NLO and NNLO derived using our factorization formula.}
  \label{fig:NNNLO_singular_2}
\end{figure}

In \cite{Gao:2019ojf}, we verified that our factorization formula correctly reproduced the singular behavior of the TEEC observable by comparing with numerical results obtained using \textsc{Nlojet++} \cite{Nagy:2001fj,Nagy:2003tz} for all partonic channels.  Here we can use our factorization formula, expanded to fixed order, to predict the NNLO singular behavior of the $pp \to 3$ jet cross section in the $\tau \to 0$ limit. For simplicity, we neglect the linearly polarized terms.

For our numerical results we consider the conditions of the LHC at $\sqrt{s} = 13\,$TeV.  We select events with two leading jets having averaged jet $P_T \geq 250\,$GeV and individual jet rapidity $|Y|<2.5$, where the jets are defined using the anti-$k_T$ algorithm \cite{Cacciari:2008gp} with $R =0.4$. The TEEC is then computed on all particles with rapidity $|y|<2.5$. For PDFs, we use the PDF4LHC15$\_$nnlo$\_$mc \cite{Butterworth:2015oua} parton distribution functions. We take $\alpha_s(M_Z) = 0.118$.

\begin{figure}%[h]
  \centering
  \includegraphics[width=\linewidth]{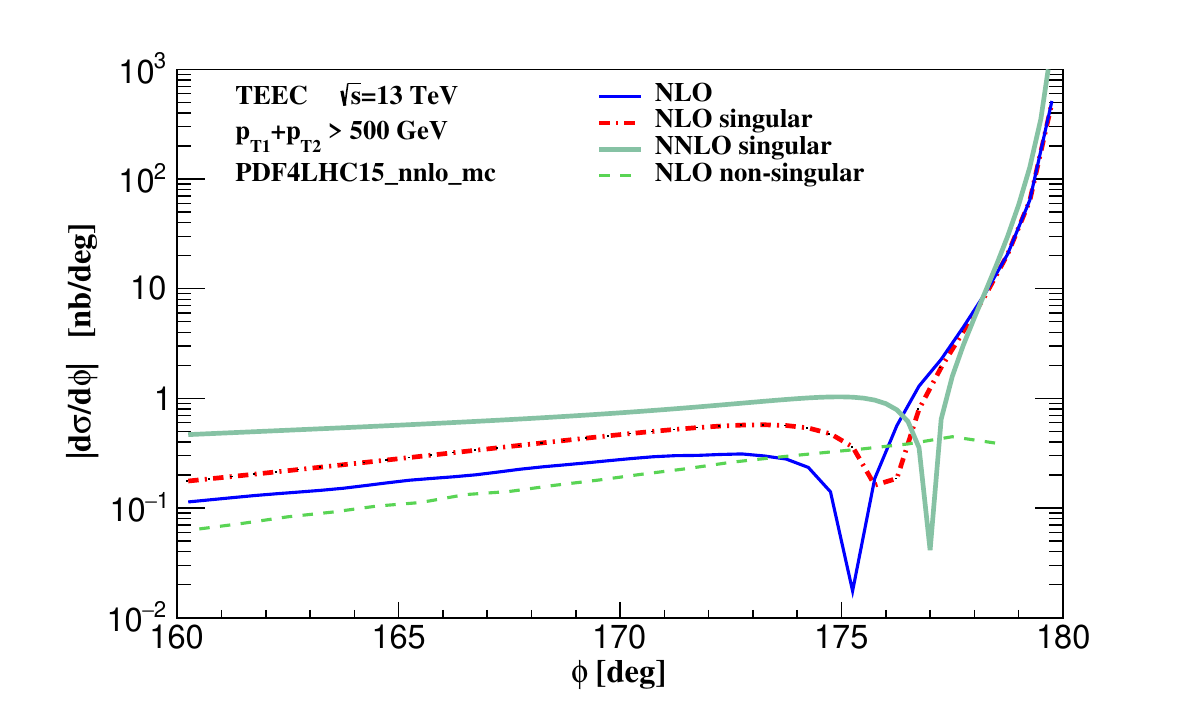}
  \caption{The singular behavior of the TEEC cross section at NLO and NNLO. At NLO, we also show the full result, as computed numerically using \textsc{Nlojet++}, as well as the non-singular contribution. It would be interesting to compare to the recent calculation of the TEEC at NNLO \cite{Czakon:2021mjy,Alvarez:2023fhi}, but this is beyond the scope of this paper.}
  \label{fig:NNNLO_singular}
\end{figure}

In \Fig{fig:NNNLO_singular_2} we show the singular structure at LO, NLO, and NNLO on a logarithmic scale, and in \Fig{fig:NNNLO_singular} on a linear scale. In \Fig{fig:NNNLO_singular}, we also show the non-singular contributions (power corrections) at NLO. There has recently been progress in understanding the structure of the power corrections for $q_T$ in color singlet production \cite{Ebert:2018gsn} and the EEC \cite{Moult:2019vou,Chen:2023wah}, and it would be interesting to extend this to the case of the TEEC. As mentioned above, the NNLO result is obtained under the assumption that there is no factorization violation at this order, namely that our factorization formula predicts the complete singular structure. While this remains to be proven, it is strongly suggested by previous work \cite{Catani:2011st,Forshaw:2012bi}.

%%%%%%%%%%%%%%%%%%%%%%
\section{Resummed Results at N$^3$LL}
\label{sec:numerics_resum}
%%%%%%%%%%%%%%%%%%%%%%

Although the main focus of this paper has been on the derivation of the factorization formula, and the calculation of the relevant ingredients necessary for resummation at N$^3$LL level, here we provide illustrative numerical results to study the perturbative convergence. Ultimately, to provide a complete description of the TEEC, one should match to fixed order perturbative results. At N$^3$LL, one should match to the NNLO calculation of the three jet cross section. Remarkably, this has recently been achieved, and the NNLO calculation of the three jet cross section \cite{Czakon:2021mjy}, has enabled the NNLO calculation of hadron collider dijet event shapes \cite{Alvarez:2023fhi}. This provides the perturbative accuracy necessary to match our resummed calculation to fixed order perturbation theory. Unfortunatley, performing this matching is beyond the scope of the current paper, since these results just became available. Therefore, we settle for illustrating the resummed singular results. Since we are not performing the full matching, for simplicity, we do not include the linearly polarized contributions. These are straightforward to include, and will be incorporated in phenomenological results in future work.

In  \Fig{fig:NNNLO_singular} we present the resummed results at NLL, NNLL and N$^3$LL, extending the results of \cite{Gao:2019ojf}. Good perturbative convergence is observed from NNLL to N$^3$LL.  The fact that the cross section goes negative as $\tau\to 1$ is unphysical, and is a result of the fact that one should match to the full fixed order result. This represents the first example of a dijet event shape (or any event shape) resummed to this accuracy at the LHC.

\begin{figure}%[h]
  \centering
  \includegraphics[width=\linewidth]{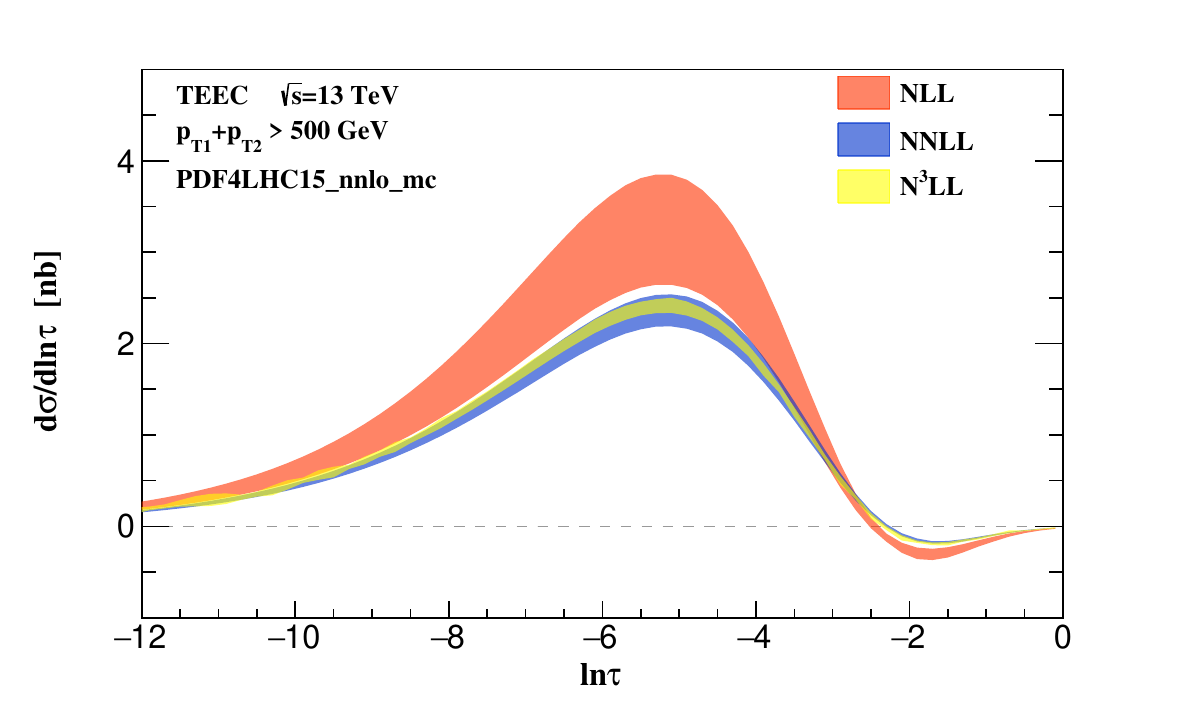}
  \caption{Resummed results for the TEEC at NLL, NNLL and N$^3$LL. Good perturbative convergence is observed. It will be particularly interesting to match the N$^3$LL result to the NNLO $pp \to 3$ jet fixed order calculation \cite{Czakon:2021mjy}.}
  \label{fig:resummed}
\end{figure}

%%%%%%%%%%%%%%%%%%%%%%
\section{Conclusions}
\label{sec:conclusion}
%%%%%%%%%%%%%%%%%%%%%%

Despite their interest for understanding QCD, hadron collider event shapes with final state jets are notoriously difficult to calculate, and have so far resisted higher order resummation. In this paper we have derived a factorization formula for the TEEC in the back-to-back limit, and shown that it exhibits a remarkable theoretical simplicity. This simplicity allowed us to perform its resummation to N$^3$LL, greatly extending the previous highest order NLL resummation. Of particular interest, at N$^3$LL, we first encounter a contribution from the quadrupole anomalous dimension \cite{Almelid:2015jia,Almelid:2017qju}, where the colors of all jets are entangled. This represents, to our knowledge, the first time it has appeared in a physical event shape observable. We also observe interesting contributions from linearly polarized gluons in both beam and jet functions.

We also introduced definitions of the TEEC observable for Drell-Yan, and $W/Z/\gamma+$ jet processes, and showed that they all fit into the same framework and can all be computed to N$^3$LL accuracy. The TEEC therefore provides a natural generalization of transverse momentum observable to states with jets, making it a valuable probe of TMD dynamics at hadron colliders.

A complete description of the TEEC description requires a number of further ingredients, beyond the resummation of singular contributions presented here. First, we would like to be able to match to the fixed order $2\to 3$ jet amplitudes at NNLO, which have recently become available \cite{Czakon:2021mjy,Alvarez:2023fhi,Agarwal:2023suw,DeLaurentis:2023nss,DeLaurentis:2023izi}. Second, it will be important to understand the structure of non-perturbative corrections to the TEEC in the back-to-back limit. Non-perturbative corrections have been studied for the EEC \cite{Dokshitzer:1999sh,Fiore:1992sa,deFlorian:2004mp,Tulipant:2017ybb,Schindler:2023cww}, and we believe that using our operator based definition, we can extend these studies to the TEEC. 

It would be particularly interesting to measure the back-to-back limit of the TEEC precisely to study the resummation effects calculated in this paper in data. Recent measurements of the TEEC have used jets instead of hadrons \cite{ATLAS:2023tgo} to achieve increased precision. It would be interesting to understand if this measurement could be done precisely on hadrons. On the theory side, there has been significant progress in understand the incorporation of tracking information \cite{Chang:2013rca,Chang:2013iba,Li:2021zcf,Jaarsma:2022kdd,Chen:2022pdu,Chen:2022muj} in perturbative calculations. This allows the calculation of the track based TEEC, which perhaps could be measured more precisely.

Finally, it will be important to understand potential factorization violating effects. We have shown that for the TEEC the contributions of the underlying event are quite small, and are easily accounted for by adding an energy distribution that is uniform in the azimuthal angle. We believe that perturbative factorization violation will occur at N$^4$LL, and it would be interesting to prove this. In particular, it will be interesting to compute the TEEC soft function at N$^3$LO to see if it has a quadrupole rapidity anomalous dimension, which would provide a concrete illustration of the violation of factorization. We are also optimistic that due to the simple perturbative structure of the TEEC observable, it can provide a playground for understanding perturbative factorization violation, and we have highlighted how the fact that the TEEC can be defined on a number of distinct final states may facilitate this understanding. We intend to pursue these directions in future work.

%%%%%%%%%%%%%%%%%%%%%%%%%%%%%%%%%%%%%%%%%%
\begin{acknowledgments}
%%%%%%%%%%%%%%%%%%%%%%%%%%%%%%%%%%%%%%%%%%
	
We would like to thank Thomas Gehrmann, Xuan Chen, Duff Neill, Wouter Waalewijn, Matt LeBlanc, Jennifer Roloff, Ben Nachman and Iain Stewart for useful discussions.
I.M and H.X.Z would like to thank the MITP for hospitality while portions of this work were performed.
A.G. was supported by the U.S. DOE under
contract number DE-SC0011090.
H.T.L. is supported by the
National Science Foundation of China under grant No.12275156. I.M was supported by start-up funds from Yale University. H.X.Z was supported by the National Science Foundation of China under
grant No.11975200 and the Asian Young Scientist Fellowship.

%%%%%%%%%%%%%%%%%%%%%%%%%%%%%%%%%%%%%%%%%%
\end{acknowledgments}
%%%%%%%%%%%%%%%%%%%%%%%%%%%%%%%%%%%%%%%%%%

%%%%%%%%%%%%%%%%%%%%%%%%%%%%%%%%%%%%%%%%%%%%%%%%%%%%%%%%%%%%%%%%%%%%%%%%%%%%%%%%
\appendix
%%%%%%%%%%%%%%%%%%%%%%%%%%%%%%%%%%%%%%%%%%%%%%%%%%%%%%%%%%%%%%%%%%%%%%%%%%%%%%%%

%%%%%%%%%%%%%%%%%%%%%%%%%%%%%%%%%%%%%%%%%%%%%%%%%%%%%%%%%%%%%%%%%%%%%%%%%%%%%%%%
\section{Summary of Perturbative Ingredients}\label{app:pert}
%%%%%%%%%%%%%%%%%%%%%%%%%%%%%%%%%%%%%%%%%%%%%%%%%%%%%%%%%%%%%%%%%%%%%%%%%%%%%%%%

In this Appendix, we summarize the perturbative ingredients entering our calculation.

%%%%%%%%%%%%%%%%%%%%%
\subsection{Anomalous Dimensions}
\label{sec:anomalous-dimensions}
%%%%%%%%%%%%%%%%%%%%%

We denote our anomalous dimensions as $\gamma[\alpha_s, \ldots]$, with the dots representing potential dependence on kinematical variables. We expand them perturbatively in $\alpha_s$ as,
\begin{align}
\gamma[\alpha_s,\ldots] = \sum_{n=0}^\infty \left(\frac{\alpha_s}{4 \pi} \right)^{n+1} \gamma_n[\ldots]   \,.
\end{align}
The QCD beta function $\beta[\alpha_s] = -2 \alpha_s \sum_{n=0} (\alpha_s/(4 \pi))^{n+1} \beta_n$ through to three loops is given by~\cite{Tarasov:1980au, Larin:1993tp}
\begin{align}
  \beta_0 = &\, \frac{11 C_A}{3}-\frac{2 n_f}{3} \,,
\nbrk
\beta_1 = &\, \frac{34 C_A^2}{3}-\frac{10 C_A n_f}{3}-2 C_F n_f \,,
\nbrk
\beta_2 = &\, \frac{2857 C_A^2}{54} + C_F^2 n_f - \frac{205 C_F C_A n_f}{18} - \frac{1415 C_A^2 n_f}{54} + \frac{11 C_F n_f^2}{9} + \frac{79 C_A n_f^2}{54} \,,
\nbrk
\beta_3 = &\, \frac{149753}{6}+ 3564 \zeta _3 -\left(\frac{6508 \zeta
   _3}{27}+\frac{1078361}{162}\right) n_f+\left(\frac{6472 \zeta
   _3}{81}+\frac{50065}{162}\right) n_f^2+\frac{1093 n_f^3}{729} \,.
\end{align}
The cusp anomalous dimension is given through to four loops by~\cite{Korchemsky:1987wg, Moch:2004pa,Moch:2017uml, Moch:2018wjh, Davies:2016jie,Henn:2019swt}
\begin{align}
\label{eq:c1}
  \gamma_{0}^{\rm cusp} = & \, 4 \,,
\nbrk
  \gamma_{1}^{\rm cusp} = & \, C_A  \left(\frac{268}{9}-8 \zeta_2\right)-\frac{40  n_f}{9} \,,
\nbrk
  \gamma_{2}^{\rm cusp} = & \, C_A^2  \left(-\frac{1072 \zeta_2}{9}+\frac{88
      \zeta_3}{3}+88 \zeta_4+\frac{490}{3}\right)
+C_A  n_f \left(\frac{160 \zeta_2}{9}-\frac{112
    \zeta_3}{3}-\frac{836}{27}\right)
\nbrk &
+ C_F n_f \left(32 \zeta_3-\frac{110}{3}\right)-\frac{16
                 n_f^2}{27} \,,
\nbrk
  \gamma_{3,q}^{\rm cusp} = &\,   15526.5-3878.93\, n_f +146.683 \, n_f^2+ 2.454\,  n_f^3 \,,
\nbrk
  \gamma_{3,g}^{\rm cusp} = & \,  13626.7-3904.67 \, n_f+146.683 \, n_f^2+ 2.454 \, n_f^3 \,. 
\nn
\end{align}
The full analytic result can be found in \cite{Henn:2019swt}.
Note that at four-loops the Casimir scaling between quark and gluon cusp anomalous dimensions is broken. 
The quark and gluon anomalous dimensions through to three loops are \cite{Moch:2005id,Moch:2005tm,Idilbi:2005ni,Idilbi:2006dg,Becher:2006mr}
\begin{align}
  \gamma^{q}_0 = & \, -3 C_F \,,
\nbrk 
  \gamma^{q}_1 = & \, C_A C_F \left(-11 \zeta_2+26
    \zeta_3-\frac{961}{54}\right)
+C_F^2 \left(12 \zeta_2-24 \zeta_3-\frac{3}{2}\right)+C_F n_f \left(2 \zeta_2+\frac{65}{27}\right) \,,
\nbrk
  \gamma_2^{q} = & \,   
     -\frac{4880 \pi ^2 \zeta _3}{81}+\frac{82072 \zeta _3}{27}-\frac{15328 \zeta
   _5}{9}-\frac{2066 \pi ^4}{405}-\frac{5062 \pi
   ^2}{81}-\frac{196621}{243}
   \nbrk &
   +\left(-\frac{7472 \zeta _3}{81}+\frac{68 \pi ^4}{1215}+\frac{4564 \pi
   ^2}{243}+\frac{36236}{729}\right) n_f+\left(-\frac{32 \zeta _3}{81}-\frac{40 \pi ^2}{81}+\frac{9668}{2187}\right)
   n_f^2 \,,
\nbrk
  \gamma_0^{g} = &\, - \beta_0 \,,
\nbrk
  \gamma_1^{g} = &\, \,C_A^2 \left(\frac{11 \zeta_2}{3}+2 \zeta_3-\frac{692}{27}\right)+C_A n_f \left(\frac{128}{27}-\frac{2 \zeta_2}{3}\right)+2 C_F n_f \,,
  \nbrk
  \gamma_2^{g} = &\,  -60 \pi ^2 \zeta _3+1098 \zeta _3-432 \zeta
   _5-\frac{319 \pi ^4}{10}+\frac{6109 \pi ^2}{18}-\frac{97186}{27}
   \nbrk &
  +\left(\frac{460 \zeta _3}{9}+\frac{107 \pi ^4}{45}-\frac{635 \pi
   ^2}{27}+\frac{59635}{162}\right) n_f+\left(-\frac{56 \zeta _3}{9}+\frac{10 \pi ^2}{27}-\frac{1061}{486}\right)
   n_f^2 \,.
\end{align}

The soft anomalous dimension up to three-loops is~\cite{Li:2014afw}
\begin{align}
    \gsoft_{0} &=  \, 0\,,
\nbrk
  \gsoft_{1} &= \, C_A C_F \left(-\frac{808}{27}+\frac{22}{3}\zeta_2+28 \zeta_3\right)+C_F n_f \left(\frac{112}{27}-\frac{4}{3}\zeta_2\right)\,,
\nbrk
  \gsoft_{2} &=  \, C_A^2 C_F \left(-\frac{136781}{729}+\frac{12650}{81}\zeta_2+\frac{1316}{3}\zeta_3-176\zeta_4-192 \zeta_5-\frac{176}{3} \zeta_3\zeta_2\right)
\brk
+C_A C_F n_f \left(\frac{11842}{729}-\frac{2828}{81}\zeta_2-\frac{728}{27}\zeta_3+48 \zeta_4\right) \brk
+C_F^2 n_f \left(\frac{1711}{27}-4\zeta_2-\frac{304}{9}\zeta_3-16 \zeta_4\right)
+C_F n_f^2 \left(\frac{2080}{729}+\frac{40}{27}\zeta_2-\frac{112}{27}\zeta_3\right)\,.
\label{eq:c2}
\end{align}
The rapidity anomalous dimension up to three-loops is \cite{Li:2016ctv}
\begin{align}
  \label{eq:8}
  \ugr_0 &=  \, 0\,,
\nbrk
\ugr_1 &= \, C_F C_A  \left(-\frac{808}{27}+28 \zeta_3\right)+C_F n_f\frac{112}{27}\,,
\nbrk
\ugr_2 &= \, C_F C_A^2 \left(-\frac{297029}{729} +\frac{6392}{81}\zeta_2+\frac{12328}{27} \zeta_3+\frac{154}{3}\zeta_4-192\zeta_5-\frac{176}{3} \zeta_3\zeta_2 \right) \brk
+ C_F C_A n_f \left(\frac{62626}{729}-\frac{824}{81}\zeta_2-\frac{904}{27}\zeta_3+\frac{20}{3}\zeta_4
\right) 
+ C_F n_f^2 \left(-\frac{1856}{729}-\frac{32}{9} \zeta_3 \right)
\brk
+ C_F^2 n_f \left(\frac{1711}{27} -\frac{304}{9}\zeta_3 -16 \zeta_4 \right)\,.
\end{align}

%%%%%%%%%%%%%%%%%%%%%
\subsection{Beam Functions}
%%%%%%%%%%%%%%%%%%%%%

The unpolarized TMD beam functions at one loop are given by
\begin{align}
  \label{eq:15}
 \mathcal{I}_{qq}(z,L_b,L_Q) = \, & \delta(1-z) + \left(\frac{\alpha_s}{4 \pi}\right) \Big[ C_F \left( -2 L_b L_Q + 3 L_b\right)
                       \delta(1-z) - P_{0,\,qq}(z)L_b  + 2C_F
                       (1-z) \Big] + \cO(\alpha_s^2)\,,\nbrk
\mathcal{I}_{qg}(z,L_b,L_Q) = \, & \left(\frac{\alpha_s}{4 \pi}\right) \Big[ 2 z (  1 - z ) -P_{0,qg}(z) L_b \Big] + \cO(\alpha_s^2)\,,\nn\\
\mathcal{I}_{gq}(z,L_b,L_Q) =\,& \left(\frac{\alpha_s}{4 \pi}\right) \Big[ - P_{0,\,gq}(z)L_b  + 2C_F\, z  \Big] + \cO(\alpha_s^2) \,,\\
\mathcal{I}_{gg}(z,L_b,L_Q) =\, & \delta(1-z) + \left(\frac{\alpha_s}{4 \pi}\right) \Big[ \left( -2C_A L_b L_Q + \beta_0 L_b\right)
                       \delta(1-z) - P_{0,\,gg}(z)L_b \Big] + \cO(\alpha_s^2)  \,. \nn
\end{align}
where $P_{0,ij}(z)$ are the LO splitting functions
\begin{align}
  \label{eq:17}
  P_{0,qq}(z) =\,& C_F \left[ 3  \delta(1-z) + \frac{4}{\left[1-z\right]}_+
                    -2  (1+z) \right]\,,\nbrk
  P_{0,qg}(z) = \, & 1 - 2 z + 2 z^2\,,\nbrk
  P_{0,gq}(z) = \, & 2 C_F \left[\frac{1+(1-z)^2}{z}\right]\,,\nbrk
  P_{0,gg}(z) = \, & 4 C_A \left[\frac{z}{\left[1-z\right]}_++\frac{1-z}{z}+z(1-z)\right]+\beta_0\delta(1-z)\,.
\end{align}
The complete two-loop results using the exponential regulator used in this paper can be found in \cite{Luo:2019hmp,Luo:2019bmw}, and the three-loop results can be found in \cite{Luo:2019szz,Luo:2020epw,Ebert:2020qef,Ebert:2020yqt}.

The linearly polarized beam functions at one loop are finite, and are given by
\begin{align}
\cI'_{gq}&=\left(\frac{\alpha_s}{4 \pi}\right)4C_F \frac{1-x}{x} + \cO(\alpha_s^2)\nn \,,\\
\cI'_{gg}&=\left(\frac{\alpha_s}{4 \pi}\right)4C_A \frac{1-x}{x} + \cO(\alpha_s^2)\,.
\end{align}
A detailed discussion, and results to two loops can be found in \cite{Luo:2019bmw}.

%%%%%%%%%%%%%%%%%%%%%
\subsection{Jet Functions}
%%%%%%%%%%%%%%%%%%%%%

The TEEC jet functions are given at one loop by \cite{Moult:2018jzp}
\begin{align}
  J_q(b,\mu,\nu)=\, & J_{\bar{q}}(b,\mu,\nu)=  1+\left(\frac{\alpha_s}{4\pi}\right) C_F (-2L_bL_Q+3L_b+4-8\zeta_2)  + \cO(\alpha_s^2) \,, \nbrk
  J_g(b,\mu,\nu)=  \, & 1+\left(\frac{\alpha_s}{4\pi}\right)\left[-2C_A L_b L_Q+\beta_0 L_b+\left(\frac{65}{18}-8\zeta_2\right)C_A-\frac{5}{18} n_f\right]  + \cO(\alpha_s^2) \,.
\end{align}
The complete two-loop results can be found in \cite{Luo:2019hmp,Luo:2019bmw}, and the three-loop results in \cite{Luo:2020epw,Ebert:2020qef}.

The linearly polarized gluon jet function at one loop is finite, which is given by
\begin{align}
    J_g'(b,\mu,\nu)=\left(\frac{\alpha_s}{4\pi}\right) \left(\frac{C_A}{3}-\frac{N_f}{3}\right) + \cO(\alpha_s^2) \,,
\end{align}
and results to two loops can be found in \cite{Luo:2019bmw}.

%%%%%%%%%%%%%%%%%%%%%%
%\subsection{Soft Function}
%%%%%%%%%%%%%%%%%%%%%%
%

\bibliography{TEEC_theory}{}

\providecommand{\href}[2]{#2}\begingroup\raggedright\begin{thebibliography}{100}

\bibitem{Khachatryan:2011dx}
{\scshape CMS} collaboration, V.~Khachatryan et~al., \emph{{First Measurement
  of Hadronic Event Shapes in $pp$ Collisions at $\sqrt {s}=7$ TeV}},
  \href{https://doi.org/10.1016/j.physletb.2011.03.060}{\emph{Phys. Lett. B}
  {\bfseries 699} (2011) 48--67},
  [\href{https://arxiv.org/abs/1102.0068}{{\ttfamily 1102.0068}}].

\bibitem{Aad:2012np}
{\scshape ATLAS} collaboration, G.~Aad et~al., \emph{{Measurement of event
  shapes at large momentum transfer with the ATLAS detector in $pp$ collisions
  at $\sqrt{s}=7$ TeV}},
  \href{https://doi.org/10.1140/epjc/s10052-012-2211-y}{\emph{Eur. Phys. J. C}
  {\bfseries 72} (2012) 2211},
  [\href{https://arxiv.org/abs/1206.2135}{{\ttfamily 1206.2135}}].

\bibitem{Aad:2012fza}
{\scshape ATLAS} collaboration, G.~Aad et~al., \emph{{Measurement of
  charged-particle event shape variables in $\sqrt{s}=7$ TeV proton-proton
  interactions with the ATLAS detector}},
  \href{https://doi.org/10.1103/PhysRevD.88.032004}{\emph{Phys. Rev. D}
  {\bfseries 88} (2013) 032004},
  [\href{https://arxiv.org/abs/1207.6915}{{\ttfamily 1207.6915}}].

\bibitem{Chatrchyan:2013tna}
{\scshape CMS} collaboration, S.~Chatrchyan et~al., \emph{{Event Shapes and
  Azimuthal Correlations in $Z$ + Jets Events in $pp$ Collisions at
  $\sqrt{s}=7$ TeV}},
  \href{https://doi.org/10.1016/j.physletb.2013.04.025}{\emph{Phys. Lett. B}
  {\bfseries 722} (2013) 238--261},
  [\href{https://arxiv.org/abs/1301.1646}{{\ttfamily 1301.1646}}].

\bibitem{Khachatryan:2014ika}
{\scshape CMS} collaboration, V.~Khachatryan et~al., \emph{{Study of Hadronic
  Event-Shape Variables in Multijet Final States in pp Collisions at $\sqrt{s}$
  = 7 TeV}}, \href{https://doi.org/10.1007/JHEP10(2014)087}{\emph{JHEP}
  {\bfseries 10} (2014) 087},
  [\href{https://arxiv.org/abs/1407.2856}{{\ttfamily 1407.2856}}].

\bibitem{ATLAS:2015yaa}
{\scshape ATLAS} collaboration, G.~Aad et~al., \emph{{Measurement of transverse
  energy-energy correlations in multi-jet events in $pp$ collisions at
  $\sqrt{s} = 7$ TeV using the ATLAS detector and determination of the strong
  coupling constant $\alpha_{\mathrm{s}}(m_Z)$}},
  \href{https://doi.org/10.1016/j.physletb.2015.09.050}{\emph{Phys. Lett. B}
  {\bfseries 750} (2015) 427--447},
  [\href{https://arxiv.org/abs/1508.01579}{{\ttfamily 1508.01579}}].

\bibitem{Aaboud:2017fml}
{\scshape ATLAS} collaboration, M.~Aaboud et~al., \emph{{Determination of the
  strong coupling constant $\alpha _\mathrm {s}$ from transverse
  energy\textendash{}energy correlations in multijet events at $\sqrt{s} =
  8~\hbox {TeV}$ using the ATLAS detector}},
  \href{https://doi.org/10.1140/epjc/s10052-017-5442-0}{\emph{Eur. Phys. J. C}
  {\bfseries 77} (2017) 872},
  [\href{https://arxiv.org/abs/1707.02562}{{\ttfamily 1707.02562}}].

\bibitem{CMS:2018svp}
{\scshape CMS} collaboration, A.~M. Sirunyan et~al., \emph{{Event shape
  variables measured using multijet final states in proton-proton collisions at
  $ \sqrt{s}=13 $ TeV}},
  \href{https://doi.org/10.1007/JHEP12(2018)117}{\emph{JHEP} {\bfseries 12}
  (2018) 117}, [\href{https://arxiv.org/abs/1811.00588}{{\ttfamily
  1811.00588}}].

\bibitem{ATLAS:2023tgo}
{\scshape ATLAS} collaboration, G.~Aad et~al., \emph{{Determination of the
  strong coupling constant from transverse energy$-$energy correlations in
  multijet events at $\sqrt{s} = 13$ TeV with the ATLAS detector}},
  \href{https://doi.org/10.1007/JHEP07(2023)085}{\emph{JHEP} {\bfseries 07}
  (2023) 085}, [\href{https://arxiv.org/abs/2301.09351}{{\ttfamily
  2301.09351}}].

\bibitem{Kaplan:2008pt}
D.~E. Kaplan and M.~D. Schwartz, \emph{{Constraining Light Colored Particles
  with Event Shapes}},
  \href{https://doi.org/10.1103/PhysRevLett.101.022002}{\emph{Phys. Rev. Lett.}
  {\bfseries 101} (2008) 022002},
  [\href{https://arxiv.org/abs/0804.2477}{{\ttfamily 0804.2477}}].

\bibitem{Llorente:2018wup}
J.~Llorente and B.~P. Nachman, \emph{{Limits on new coloured fermions using
  precision jet data from the Large Hadron Collider}},
  \href{https://doi.org/10.1016/j.nuclphysb.2018.09.008}{\emph{Nucl. Phys. B}
  {\bfseries 936} (2018) 106--117},
  [\href{https://arxiv.org/abs/1807.00894}{{\ttfamily 1807.00894}}].

\bibitem{Becher:2008cf}
T.~Becher and M.~D. Schwartz, \emph{{A precise determination of $\alpha_s$ from
  LEP thrust data using effective field theory}},
  \href{https://doi.org/10.1088/1126-6708/2008/07/034}{\emph{JHEP} {\bfseries
  07} (2008) 034}, [\href{https://arxiv.org/abs/0803.0342}{{\ttfamily
  0803.0342}}].

\bibitem{Abbate:2010xh}
R.~Abbate, M.~Fickinger, A.~H. Hoang, V.~Mateu and I.~W. Stewart, \emph{{Thrust
  at $N^{3}LL$ with Power Corrections and a Precision Global Fit for
  $\alpha_{s}(mZ)$}},
  \href{https://doi.org/10.1103/PhysRevD.83.074021}{\emph{Phys. Rev. D}
  {\bfseries 83} (2011) 074021},
  [\href{https://arxiv.org/abs/1006.3080}{{\ttfamily 1006.3080}}].

\bibitem{Chien:2010kc}
Y.-T. Chien and M.~D. Schwartz, \emph{{Resummation of heavy jet mass and
  comparison to LEP data}},
  \href{https://doi.org/10.1007/JHEP08(2010)058}{\emph{JHEP} {\bfseries 08}
  (2010) 058}, [\href{https://arxiv.org/abs/1005.1644}{{\ttfamily 1005.1644}}].

\bibitem{Hoang:2014wka}
A.~H. Hoang, D.~W. Kolodrubetz, V.~Mateu and I.~W. Stewart,
  \emph{{$C$-parameter distribution at N$^3$LL' including power corrections}},
  \href{https://doi.org/10.1103/PhysRevD.91.094017}{\emph{Phys. Rev. D}
  {\bfseries 91} (2015) 094017},
  [\href{https://arxiv.org/abs/1411.6633}{{\ttfamily 1411.6633}}].

\bibitem{Duhr:2022yyp}
C.~Duhr, B.~Mistlberger and G.~Vita, \emph{{Four-Loop Rapidity Anomalous
  Dimension and Event Shapes to Fourth Logarithmic Order}},
  \href{https://doi.org/10.1103/PhysRevLett.129.162001}{\emph{Phys. Rev. Lett.}
  {\bfseries 129} (2022) 162001},
  [\href{https://arxiv.org/abs/2205.02242}{{\ttfamily 2205.02242}}].

\bibitem{GehrmannDeRidder:2007hr}
A.~Gehrmann-De~Ridder, T.~Gehrmann, E.~W.~N. Glover and G.~Heinrich,
  \emph{{NNLO corrections to event shapes in e+ e- annihilation}},
  \href{https://doi.org/10.1088/1126-6708/2007/12/094}{\emph{JHEP} {\bfseries
  12} (2007) 094}, [\href{https://arxiv.org/abs/0711.4711}{{\ttfamily
  0711.4711}}].

\bibitem{Gehrmann-DeRidder:2007nzq}
A.~Gehrmann-De~Ridder, T.~Gehrmann, E.~W.~N. Glover and G.~Heinrich,
  \emph{{Second-order QCD corrections to the thrust distribution}},
  \href{https://doi.org/10.1103/PhysRevLett.99.132002}{\emph{Phys. Rev. Lett.}
  {\bfseries 99} (2007) 132002},
  [\href{https://arxiv.org/abs/0707.1285}{{\ttfamily 0707.1285}}].

\bibitem{Weinzierl:2008iv}
S.~Weinzierl, \emph{{NNLO corrections to 3-jet observables in electron-positron
  annihilation}},
  \href{https://doi.org/10.1103/PhysRevLett.101.162001}{\emph{Phys. Rev. Lett.}
  {\bfseries 101} (2008) 162001},
  [\href{https://arxiv.org/abs/0807.3241}{{\ttfamily 0807.3241}}].

\bibitem{Weinzierl:2009ms}
S.~Weinzierl, \emph{{Event shapes and jet rates in electron-positron
  annihilation at NNLO}},
  \href{https://doi.org/10.1088/1126-6708/2009/06/041}{\emph{JHEP} {\bfseries
  06} (2009) 041}, [\href{https://arxiv.org/abs/0904.1077}{{\ttfamily
  0904.1077}}].

\bibitem{Collins:2007nk}
J.~Collins and J.-W. Qiu, \emph{{$k_{T}$ factorization is violated in
  production of high-transverse-momentum particles in hadron-hadron
  collisions}}, \href{https://doi.org/10.1103/PhysRevD.75.114014}{\emph{Phys.
  Rev. D} {\bfseries 75} (2007) 114014},
  [\href{https://arxiv.org/abs/0705.2141}{{\ttfamily 0705.2141}}].

\bibitem{Collins:2007jp}
J.~Collins, \emph{{2-soft-gluon exchange and factorization breaking}},
  \href{https://arxiv.org/abs/0708.4410}{{\ttfamily 0708.4410}}.

\bibitem{Bomhof:2007su}
C.~J. Bomhof, P.~J. Mulders, W.~Vogelsang and F.~Yuan, \emph{{Single-Transverse
  Spin Asymmetry in Dijet Correlations at Hadron Colliders}},
  \href{https://doi.org/10.1103/PhysRevD.75.074019}{\emph{Phys. Rev. D}
  {\bfseries 75} (2007) 074019},
  [\href{https://arxiv.org/abs/hep-ph/0701277}{{\ttfamily hep-ph/0701277}}].

\bibitem{Rogers:2010dm}
T.~C. Rogers and P.~J. Mulders, \emph{{No Generalized TMD-Factorization in
  Hadro-Production of High Transverse Momentum Hadrons}},
  \href{https://doi.org/10.1103/PhysRevD.81.094006}{\emph{Phys. Rev. D}
  {\bfseries 81} (2010) 094006},
  [\href{https://arxiv.org/abs/1001.2977}{{\ttfamily 1001.2977}}].

\bibitem{Buffing:2013dxa}
M.~G.~A. Buffing and P.~J. Mulders, \emph{{Color entanglement for azimuthal
  asymmetries in the Drell-Yan process}},
  \href{https://doi.org/10.1103/PhysRevLett.112.092002}{\emph{Phys. Rev. Lett.}
  {\bfseries 112} (2014) 092002},
  [\href{https://arxiv.org/abs/1309.4681}{{\ttfamily 1309.4681}}].

\bibitem{Gaunt:2014ska}
J.~R. Gaunt, \emph{{Glauber Gluons and Multiple Parton Interactions}},
  \href{https://doi.org/10.1007/JHEP07(2014)110}{\emph{JHEP} {\bfseries 07}
  (2014) 110}, [\href{https://arxiv.org/abs/1405.2080}{{\ttfamily 1405.2080}}].

\bibitem{Zeng:2015iba}
M.~Zeng, \emph{{Drell-Yan process with jet vetoes: breaking of generalized
  factorization}}, \href{https://doi.org/10.1007/JHEP10(2015)189}{\emph{JHEP}
  {\bfseries 10} (2015) 189},
  [\href{https://arxiv.org/abs/1507.01652}{{\ttfamily 1507.01652}}].

\bibitem{Catani:2011st}
S.~Catani, D.~de~Florian and G.~Rodrigo, \emph{{Space-like (versus time-like)
  collinear limits in QCD: Is factorization violated?}},
  \href{https://doi.org/10.1007/JHEP07(2012)026}{\emph{JHEP} {\bfseries 07}
  (2012) 026}, [\href{https://arxiv.org/abs/1112.4405}{{\ttfamily 1112.4405}}].

\bibitem{Schwartz:2017nmr}
M.~D. Schwartz, K.~Yan and H.~X. Zhu, \emph{{Collinear factorization violation
  and effective field theory}},
  \href{https://doi.org/10.1103/PhysRevD.96.056005}{\emph{Phys. Rev. D}
  {\bfseries 96} (2017) 056005},
  [\href{https://arxiv.org/abs/1703.08572}{{\ttfamily 1703.08572}}].

\bibitem{Forshaw:2008cq}
J.~R. Forshaw, A.~Kyrieleis and M.~H. Seymour, \emph{{Super-leading logarithms
  in non-global observables in QCD: Colour basis independent calculation}},
  \href{https://doi.org/10.1088/1126-6708/2008/09/128}{\emph{JHEP} {\bfseries
  09} (2008) 128}, [\href{https://arxiv.org/abs/0808.1269}{{\ttfamily
  0808.1269}}].

\bibitem{Forshaw:2006fk}
J.~R. Forshaw, A.~Kyrieleis and M.~H. Seymour, \emph{{Super-leading logarithms
  in non-global observables in QCD}},
  \href{https://doi.org/10.1088/1126-6708/2006/08/059}{\emph{JHEP} {\bfseries
  08} (2006) 059}, [\href{https://arxiv.org/abs/hep-ph/0604094}{{\ttfamily
  hep-ph/0604094}}].

\bibitem{Martinez:2018ffw}
R.~\'Angeles~Mart\'\i{}nez, M.~De~Angelis, J.~R. Forshaw, S.~Pl\"atzer and
  M.~H. Seymour, \emph{{Soft gluon evolution and non-global logarithms}},
  \href{https://doi.org/10.1007/JHEP05(2018)044}{\emph{JHEP} {\bfseries 05}
  (2018) 044}, [\href{https://arxiv.org/abs/1802.08531}{{\ttfamily
  1802.08531}}].

\bibitem{Angeles-Martinez:2016dph}
R.~\'Angeles~Mart\'\i{}nez, J.~R. Forshaw and M.~H. Seymour, \emph{{Ordering
  multiple soft gluon emissions}},
  \href{https://doi.org/10.1103/PhysRevLett.116.212003}{\emph{Phys. Rev. Lett.}
  {\bfseries 116} (2016) 212003},
  [\href{https://arxiv.org/abs/1602.00623}{{\ttfamily 1602.00623}}].

\bibitem{Forshaw:2012bi}
J.~R. Forshaw, M.~H. Seymour and A.~Siodmok, \emph{{On the Breaking of
  Collinear Factorization in QCD}},
  \href{https://doi.org/10.1007/JHEP11(2012)066}{\emph{JHEP} {\bfseries 11}
  (2012) 066}, [\href{https://arxiv.org/abs/1206.6363}{{\ttfamily 1206.6363}}].

\bibitem{Angeles-Martinez:2015rna}
R.~\'Angeles-Mart\'\i{}nez, J.~R. Forshaw and M.~H. Seymour, \emph{{Coulomb
  gluons and the ordering variable}},
  \href{https://doi.org/10.1007/JHEP12(2015)091}{\emph{JHEP} {\bfseries 12}
  (2015) 091}, [\href{https://arxiv.org/abs/1510.07998}{{\ttfamily
  1510.07998}}].

\bibitem{Schwartz:2018obd}
M.~D. Schwartz, K.~Yan and H.~X. Zhu, \emph{{Factorization Violation and Scale
  Invariance}}, \href{https://doi.org/10.1103/PhysRevD.97.096017}{\emph{Phys.
  Rev. D} {\bfseries 97} (2018) 096017},
  [\href{https://arxiv.org/abs/1801.01138}{{\ttfamily 1801.01138}}].

\bibitem{Rothstein:2016bsq}
I.~Z. Rothstein and I.~W. Stewart, \emph{{An Effective Field Theory for Forward
  Scattering and Factorization Violation}},
  \href{https://doi.org/10.1007/JHEP08(2016)025}{\emph{JHEP} {\bfseries 08}
  (2016) 025}, [\href{https://arxiv.org/abs/1601.04695}{{\ttfamily
  1601.04695}}].

\bibitem{Forshaw:2021fxs}
J.~R. Forshaw and J.~Holguin, \emph{{Coulomb gluons will generally destroy
  coherence}}, \href{https://doi.org/10.1007/JHEP12(2021)084}{\emph{JHEP}
  {\bfseries 12} (2021) 084},
  [\href{https://arxiv.org/abs/2109.03665}{{\ttfamily 2109.03665}}].

\bibitem{Dixon:2019lnw}
L.~J. Dixon, E.~Herrmann, K.~Yan and H.~X. Zhu, \emph{{Soft gluon emission at
  two loops in full color}},
  \href{https://doi.org/10.1007/JHEP05(2020)135}{\emph{JHEP} {\bfseries 05}
  (2020) 135}, [\href{https://arxiv.org/abs/1912.09370}{{\ttfamily
  1912.09370}}].

\bibitem{Alvarez:2023fhi}
M.~Alvarez, J.~Cantero, M.~Czakon, J.~Llorente, A.~Mitov and R.~Poncelet,
  \emph{{NNLO QCD corrections to event shapes at the LHC}},
  \href{https://doi.org/10.1007/JHEP03(2023)129}{\emph{JHEP} {\bfseries 03}
  (2023) 129}, [\href{https://arxiv.org/abs/2301.01086}{{\ttfamily
  2301.01086}}].

\bibitem{Banfi:2004nk}
A.~Banfi, G.~P. Salam and G.~Zanderighi, \emph{{Resummed event shapes at hadron
  - hadron colliders}},
  \href{https://doi.org/10.1088/1126-6708/2004/08/062}{\emph{JHEP} {\bfseries
  08} (2004) 062}, [\href{https://arxiv.org/abs/hep-ph/0407287}{{\ttfamily
  hep-ph/0407287}}].

\bibitem{Banfi:2010xy}
A.~Banfi, G.~P. Salam and G.~Zanderighi, \emph{{Phenomenology of event shapes
  at hadron colliders}},
  \href{https://doi.org/10.1007/JHEP06(2010)038}{\emph{JHEP} {\bfseries 06}
  (2010) 038}, [\href{https://arxiv.org/abs/1001.4082}{{\ttfamily 1001.4082}}].

\bibitem{Stewart:2010pd}
I.~W. Stewart, F.~J. Tackmann and W.~J. Waalewijn, \emph{{The Beam Thrust Cross
  Section for Drell-Yan at NNLL Order}},
  \href{https://doi.org/10.1103/PhysRevLett.106.032001}{\emph{Phys. Rev. Lett.}
  {\bfseries 106} (2011) 032001},
  [\href{https://arxiv.org/abs/1005.4060}{{\ttfamily 1005.4060}}].

\bibitem{Becher:2015lmy}
T.~Becher, X.~Garcia~i Tormo and J.~Piclum, \emph{{Next-to-next-to-leading
  logarithmic resummation for transverse thrust}},
  \href{https://doi.org/10.1103/PhysRevD.93.054038}{\emph{Phys. Rev. D}
  {\bfseries 93} (2016) 054038},
  [\href{https://arxiv.org/abs/1512.00022}{{\ttfamily 1512.00022}}].

\bibitem{Becher:2015gsa}
T.~Becher and X.~Garcia~i Tormo, \emph{{Factorization and resummation for
  transverse thrust}},
  \href{https://doi.org/10.1007/JHEP06(2015)071}{\emph{JHEP} {\bfseries 06}
  (2015) 071}, [\href{https://arxiv.org/abs/1502.04136}{{\ttfamily
  1502.04136}}].

\bibitem{Jouttenus:2013hs}
T.~T. Jouttenus, I.~W. Stewart, F.~J. Tackmann and W.~J. Waalewijn, \emph{{Jet
  mass spectra in Higgs boson plus one jet at next-to-next-to-leading
  logarithmic order}},
  \href{https://doi.org/10.1103/PhysRevD.88.054031}{\emph{Phys. Rev. D}
  {\bfseries 88} (2013) 054031},
  [\href{https://arxiv.org/abs/1302.0846}{{\ttfamily 1302.0846}}].

\bibitem{Alioli:2023rxx}
S.~Alioli, G.~Bell, G.~Billis, A.~Broggio, B.~Dehnadi, M.~A. Lim et~al.,
  \emph{{N$^3$LL resummation of one-jettiness for $Z$-boson plus jet production
  at hadron colliders}},  \href{https://arxiv.org/abs/2312.06496}{{\ttfamily
  2312.06496}}.

\bibitem{Gao:2019ojf}
A.~Gao, H.~T. Li, I.~Moult and H.~X. Zhu, \emph{{Precision QCD Event Shapes at
  Hadron Colliders: The Transverse Energy-Energy Correlator in the Back-to-Back
  Limit}}, \href{https://doi.org/10.1103/PhysRevLett.123.062001}{\emph{Phys.
  Rev. Lett.} {\bfseries 123} (2019) 062001},
  [\href{https://arxiv.org/abs/1901.04497}{{\ttfamily 1901.04497}}].

\bibitem{Ali:2012rn}
A.~Ali, F.~Barreiro, J.~Llorente and W.~Wang, \emph{{Transverse Energy-Energy
  Correlations in Next-to-Leading Order in $\alpha_s$ at the LHC}},
  \href{https://doi.org/10.1103/PhysRevD.86.114017}{\emph{Phys. Rev. D}
  {\bfseries 86} (2012) 114017},
  [\href{https://arxiv.org/abs/1205.1689}{{\ttfamily 1205.1689}}].

\bibitem{Li:2021txc}
H.~T. Li, Y.~Makris and I.~Vitev, \emph{{Energy-energy correlators in Deep
  Inelastic Scattering}},
  \href{https://doi.org/10.1103/PhysRevD.103.094005}{\emph{Phys. Rev. D}
  {\bfseries 103} (2021) 094005},
  [\href{https://arxiv.org/abs/2102.05669}{{\ttfamily 2102.05669}}].

\bibitem{Li:2020bub}
H.~T. Li, I.~Vitev and Y.~J. Zhu, \emph{{Transverse-Energy-Energy Correlations
  in Deep Inelastic Scattering}},
  \href{https://doi.org/10.1007/JHEP11(2020)051}{\emph{JHEP} {\bfseries 11}
  (2020) 051}, [\href{https://arxiv.org/abs/2006.02437}{{\ttfamily
  2006.02437}}].

\bibitem{Kang:2023oqj}
Z.-B. Kang, J.~Penttala, F.~Zhao and Y.~Zhou, \emph{{Transverse Energy-Energy
  Correlators in the Color-Glass Condensate at the Electron-Ion Collider}},
  \href{https://arxiv.org/abs/2311.17142}{{\ttfamily 2311.17142}}.

\bibitem{Cao:2023qat}
H.~Cao, H.~T. Li and Z.~Mi, \emph{{Bjorken $x$ weighted Energy-Energy
  Correlators from the Target Fragmentation Region to the Current Fragmentation
  Region}},  \href{https://arxiv.org/abs/2312.07655}{{\ttfamily 2312.07655}}.

\bibitem{Dixon:2018qgp}
L.~J. Dixon, M.-X. Luo, V.~Shtabovenko, T.-Z. Yang and H.~X. Zhu,
  \emph{{Analytical Computation of Energy-Energy Correlation at Next-to-Leading
  Order in QCD}},
  \href{https://doi.org/10.1103/PhysRevLett.120.102001}{\emph{Phys. Rev. Lett.}
  {\bfseries 120} (2018) 102001},
  [\href{https://arxiv.org/abs/1801.03219}{{\ttfamily 1801.03219}}].

\bibitem{Luo:2019nig}
M.-X. Luo, V.~Shtabovenko, T.-Z. Yang and H.~X. Zhu, \emph{{Analytic
  Next-To-Leading Order Calculation of Energy-Energy Correlation in
  Gluon-Initiated Higgs Decays}},
  \href{https://doi.org/10.1007/JHEP06(2019)037}{\emph{JHEP} {\bfseries 06}
  (2019) 037}, [\href{https://arxiv.org/abs/1903.07277}{{\ttfamily
  1903.07277}}].

\bibitem{Gao:2020vyx}
J.~Gao, V.~Shtabovenko and T.-Z. Yang, \emph{{Energy-energy correlation in
  hadronic Higgs decays: analytic results and phenomenology at NLO}},
  \href{https://doi.org/10.1007/JHEP02(2021)210}{\emph{JHEP} {\bfseries 02}
  (2021) 210}, [\href{https://arxiv.org/abs/2012.14188}{{\ttfamily
  2012.14188}}].

\bibitem{Belitsky:2013xxa}
A.~V. Belitsky, S.~Hohenegger, G.~P. Korchemsky, E.~Sokatchev and A.~Zhiboedov,
  \emph{{From correlation functions to event shapes}},
  \href{https://doi.org/10.1016/j.nuclphysb.2014.04.020}{\emph{Nucl. Phys. B}
  {\bfseries 884} (2014) 305--343},
  [\href{https://arxiv.org/abs/1309.0769}{{\ttfamily 1309.0769}}].

\bibitem{Belitsky:2013bja}
A.~V. Belitsky, S.~Hohenegger, G.~P. Korchemsky, E.~Sokatchev and A.~Zhiboedov,
  \emph{{Event shapes in $\mathcal{N} = 4$ super-Yang-Mills theory}},
  \href{https://doi.org/10.1016/j.nuclphysb.2014.04.019}{\emph{Nucl. Phys. B}
  {\bfseries 884} (2014) 206--256},
  [\href{https://arxiv.org/abs/1309.1424}{{\ttfamily 1309.1424}}].

\bibitem{Belitsky:2013ofa}
A.~V. Belitsky, S.~Hohenegger, G.~P. Korchemsky, E.~Sokatchev and A.~Zhiboedov,
  \emph{{Energy-Energy Correlations in N=4 Supersymmetric Yang-Mills Theory}},
  \href{https://doi.org/10.1103/PhysRevLett.112.071601}{\emph{Phys. Rev. Lett.}
  {\bfseries 112} (2014) 071601},
  [\href{https://arxiv.org/abs/1311.6800}{{\ttfamily 1311.6800}}].

\bibitem{Henn:2019gkr}
J.~M. Henn, E.~Sokatchev, K.~Yan and A.~Zhiboedov, \emph{{Energy-energy
  correlation in $N$=4 super Yang-Mills theory at next-to-next-to-leading
  order}}, \href{https://doi.org/10.1103/PhysRevD.100.036010}{\emph{Phys. Rev.
  D} {\bfseries 100} (2019) 036010},
  [\href{https://arxiv.org/abs/1903.05314}{{\ttfamily 1903.05314}}].

\bibitem{Moult:2018jzp}
I.~Moult and H.~X. Zhu, \emph{{Simplicity from Recoil: The Three-Loop Soft
  Function and Factorization for the Energy-Energy Correlation}},
  \href{https://doi.org/10.1007/JHEP08(2018)160}{\emph{JHEP} {\bfseries 08}
  (2018) 160}, [\href{https://arxiv.org/abs/1801.02627}{{\ttfamily
  1801.02627}}].

\bibitem{Ebert:2020sfi}
M.~A. Ebert, B.~Mistlberger and G.~Vita, \emph{{The Energy-Energy Correlation
  in the back-to-back limit at N$^{3}$LO and N$^{3}$LL'}},
  \href{https://doi.org/10.1007/JHEP08(2021)022}{\emph{JHEP} {\bfseries 08}
  (2021) 022}, [\href{https://arxiv.org/abs/2012.07859}{{\ttfamily
  2012.07859}}].

\bibitem{Dixon:2019uzg}
L.~J. Dixon, I.~Moult and H.~X. Zhu, \emph{{Collinear limit of the
  energy-energy correlator}},
  \href{https://doi.org/10.1103/PhysRevD.100.014009}{\emph{Phys. Rev. D}
  {\bfseries 100} (2019) 014009},
  [\href{https://arxiv.org/abs/1905.01310}{{\ttfamily 1905.01310}}].

\bibitem{Kologlu:2019mfz}
M.~Kologlu, P.~Kravchuk, D.~Simmons-Duffin and A.~Zhiboedov, \emph{{The
  light-ray OPE and conformal colliders}},
  \href{https://doi.org/10.1007/JHEP01(2021)128}{\emph{JHEP} {\bfseries 01}
  (2021) 128}, [\href{https://arxiv.org/abs/1905.01311}{{\ttfamily
  1905.01311}}].

\bibitem{Korchemsky:2019nzm}
G.~P. Korchemsky, \emph{{Energy correlations in the end-point region}},
  \href{https://doi.org/10.1007/JHEP01(2020)008}{\emph{JHEP} {\bfseries 01}
  (2020) 008}, [\href{https://arxiv.org/abs/1905.01444}{{\ttfamily
  1905.01444}}].

\bibitem{Chen:2020vvp}
H.~Chen, I.~Moult, X.~Zhang and H.~X. Zhu, \emph{{Rethinking jets with energy
  correlators: Tracks, resummation, and analytic continuation}},
  \href{https://doi.org/10.1103/PhysRevD.102.054012}{\emph{Phys. Rev. D}
  {\bfseries 102} (2020) 054012},
  [\href{https://arxiv.org/abs/2004.11381}{{\ttfamily 2004.11381}}].

\bibitem{Chen:2023zzh}
H.~Chen, \emph{{QCD Factorization from Light-ray OPE}},
  \href{https://arxiv.org/abs/2311.00350}{{\ttfamily 2311.00350}}.

\bibitem{DelDuca:2016ily}
V.~Del~Duca, C.~Duhr, A.~Kardos, G.~Somogyi, Z.~Sz\H{o}r, Z.~Tr\'ocs\'anyi
  et~al., \emph{{Jet production in the CoLoRFulNNLO method: event shapes in
  electron-positron collisions}},
  \href{https://doi.org/10.1103/PhysRevD.94.074019}{\emph{Phys. Rev. D}
  {\bfseries 94} (2016) 074019},
  [\href{https://arxiv.org/abs/1606.03453}{{\ttfamily 1606.03453}}].

\bibitem{Tulipant:2017ybb}
Z.~Tulip\'ant, A.~Kardos and G.~Somogyi, \emph{{Energy\textendash{}energy
  correlation in electron\textendash{}positron annihilation at NNLL + NNLO
  accuracy}}, \href{https://doi.org/10.1140/epjc/s10052-017-5320-9}{\emph{Eur.
  Phys. J. C} {\bfseries 77} (2017) 749},
  [\href{https://arxiv.org/abs/1708.04093}{{\ttfamily 1708.04093}}].

\bibitem{Chen:2019bpb}
H.~Chen, M.-X. Luo, I.~Moult, T.-Z. Yang, X.~Zhang and H.~X. Zhu, \emph{{Three
  point energy correlators in the collinear limit: symmetries, dualities and
  analytic results}},
  \href{https://doi.org/10.1007/JHEP08(2020)028}{\emph{JHEP} {\bfseries 08}
  (2020) 028}, [\href{https://arxiv.org/abs/1912.11050}{{\ttfamily
  1912.11050}}].

\bibitem{Yan:2022cye}
K.~Yan and X.~Zhang, \emph{{Three-Point Energy Correlator in N=4 Supersymmetric
  Yang-Mills Theory}},
  \href{https://doi.org/10.1103/PhysRevLett.129.021602}{\emph{Phys. Rev. Lett.}
  {\bfseries 129} (2022) 021602},
  [\href{https://arxiv.org/abs/2203.04349}{{\ttfamily 2203.04349}}].

\bibitem{Yang:2022tgm}
T.-Z. Yang and X.~Zhang, \emph{{Analytic Computation of three-point energy
  correlator in QCD}},
  \href{https://doi.org/10.1007/JHEP09(2022)006}{\emph{JHEP} {\bfseries 09}
  (2022) 006}, [\href{https://arxiv.org/abs/2208.01051}{{\ttfamily
  2208.01051}}].

\bibitem{Komiske:2022enw}
P.~T. Komiske, I.~Moult, J.~Thaler and H.~X. Zhu, \emph{{Analyzing N-Point
  Energy Correlators inside Jets with CMS Open Data}},
  \href{https://doi.org/10.1103/PhysRevLett.130.051901}{\emph{Phys. Rev. Lett.}
  {\bfseries 130} (2023) 051901},
  [\href{https://arxiv.org/abs/2201.07800}{{\ttfamily 2201.07800}}].

\bibitem{Holguin:2022epo}
J.~Holguin, I.~Moult, A.~Pathak and M.~Procura, \emph{{New paradigm for
  precision top physics: Weighing the top with energy correlators}},
  \href{https://doi.org/10.1103/PhysRevD.107.114002}{\emph{Phys. Rev. D}
  {\bfseries 107} (2023) 114002},
  [\href{https://arxiv.org/abs/2201.08393}{{\ttfamily 2201.08393}}].

\bibitem{Liu:2022wop}
X.~Liu and H.~X. Zhu, \emph{{Nucleon Energy Correlators}},
  \href{https://doi.org/10.1103/PhysRevLett.130.091901}{\emph{Phys. Rev. Lett.}
  {\bfseries 130} (2023) 091901},
  [\href{https://arxiv.org/abs/2209.02080}{{\ttfamily 2209.02080}}].

\bibitem{Liu:2023aqb}
H.-Y. Liu, X.~Liu, J.-C. Pan, F.~Yuan and H.~X. Zhu, \emph{{Nucleon Energy
  Correlators for the Color Glass Condensate}},
  \href{https://doi.org/10.1103/PhysRevLett.130.181901}{\emph{Phys. Rev. Lett.}
  {\bfseries 130} (2023) 181901},
  [\href{https://arxiv.org/abs/2301.01788}{{\ttfamily 2301.01788}}].

\bibitem{Cao:2023oef}
H.~Cao, X.~Liu and H.~X. Zhu, \emph{{Toward precision measurements of nucleon
  energy correlators in lepton-nucleon collisions}},
  \href{https://doi.org/10.1103/PhysRevD.107.114008}{\emph{Phys. Rev. D}
  {\bfseries 107} (2023) 114008},
  [\href{https://arxiv.org/abs/2303.01530}{{\ttfamily 2303.01530}}].

\bibitem{Devereaux:2023vjz}
K.~Devereaux, W.~Fan, W.~Ke, K.~Lee and I.~Moult, \emph{{Imaging Cold Nuclear
  Matter with Energy Correlators}},
  \href{https://arxiv.org/abs/2303.08143}{{\ttfamily 2303.08143}}.

\bibitem{Andres:2022ovj}
C.~Andres, F.~Dominguez, R.~Kunnawalkam~Elayavalli, J.~Holguin, C.~Marquet and
  I.~Moult, \emph{{Resolving the Scales of the Quark-Gluon Plasma with Energy
  Correlators}},
  \href{https://doi.org/10.1103/PhysRevLett.130.262301}{\emph{Phys. Rev. Lett.}
  {\bfseries 130} (2023) 262301},
  [\href{https://arxiv.org/abs/2209.11236}{{\ttfamily 2209.11236}}].

\bibitem{Andres:2023xwr}
C.~Andres, F.~Dominguez, J.~Holguin, C.~Marquet and I.~Moult, \emph{{A coherent
  view of the quark-gluon plasma from energy correlators}},
  \href{https://doi.org/10.1007/JHEP09(2023)088}{\emph{JHEP} {\bfseries 09}
  (2023) 088}, [\href{https://arxiv.org/abs/2303.03413}{{\ttfamily
  2303.03413}}].

\bibitem{Craft:2022kdo}
E.~Craft, K.~Lee, B.~Me\c{c}aj and I.~Moult, \emph{{Beautiful and Charming
  Energy Correlators}},  \href{https://arxiv.org/abs/2210.09311}{{\ttfamily
  2210.09311}}.

\bibitem{Lee:2022ige}
K.~Lee, B.~Me\c{c}aj and I.~Moult, \emph{{Conformal Colliders Meet the LHC}},
  \href{https://arxiv.org/abs/2205.03414}{{\ttfamily 2205.03414}}.

\bibitem{Bauer:2000ew}
C.~W. Bauer, S.~Fleming and M.~E. Luke, \emph{{Summing Sudakov logarithms in $B
  \to  X_s \gamma $in effective field theory.}},
  \href{https://doi.org/10.1103/PhysRevD.63.014006}{\emph{Phys. Rev. D}
  {\bfseries 63} (2000) 014006},
  [\href{https://arxiv.org/abs/hep-ph/0005275}{{\ttfamily hep-ph/0005275}}].

\bibitem{Bauer:2000yr}
C.~W. Bauer, S.~Fleming, D.~Pirjol and I.~W. Stewart, \emph{{An Effective field
  theory for collinear and soft gluons: Heavy to light decays}},
  \href{https://doi.org/10.1103/PhysRevD.63.114020}{\emph{Phys. Rev. D}
  {\bfseries 63} (2001) 114020},
  [\href{https://arxiv.org/abs/hep-ph/0011336}{{\ttfamily hep-ph/0011336}}].

\bibitem{Bauer:2001ct}
C.~W. Bauer and I.~W. Stewart, \emph{{Invariant operators in collinear
  effective theory}},
  \href{https://doi.org/10.1016/S0370-2693(01)00902-9}{\emph{Phys. Lett. B}
  {\bfseries 516} (2001) 134--142},
  [\href{https://arxiv.org/abs/hep-ph/0107001}{{\ttfamily hep-ph/0107001}}].

\bibitem{Bauer:2001yt}
C.~W. Bauer, D.~Pirjol and I.~W. Stewart, \emph{{Soft collinear factorization
  in effective field theory}},
  \href{https://doi.org/10.1103/PhysRevD.65.054022}{\emph{Phys. Rev. D}
  {\bfseries 65} (2002) 054022},
  [\href{https://arxiv.org/abs/hep-ph/0109045}{{\ttfamily hep-ph/0109045}}].

\bibitem{Almelid:2015jia}
O.~Almelid, C.~Duhr and E.~Gardi, \emph{{Three-loop corrections to the soft
  anomalous dimension in multileg scattering}},
  \href{https://doi.org/10.1103/PhysRevLett.117.172002}{\emph{Phys. Rev. Lett.}
  {\bfseries 117} (2016) 172002},
  [\href{https://arxiv.org/abs/1507.00047}{{\ttfamily 1507.00047}}].

\bibitem{Almelid:2017qju}
O.~Almelid, C.~Duhr, E.~Gardi, A.~McLeod and C.~D. White, \emph{{Bootstrapping
  the QCD soft anomalous dimension}},
  \href{https://doi.org/10.1007/JHEP09(2017)073}{\emph{JHEP} {\bfseries 09}
  (2017) 073}, [\href{https://arxiv.org/abs/1706.10162}{{\ttfamily
  1706.10162}}].

\bibitem{Li:2016ctv}
Y.~Li and H.~X. Zhu, \emph{{Bootstrapping Rapidity Anomalous Dimensions for
  Transverse-Momentum Resummation}},
  \href{https://doi.org/10.1103/PhysRevLett.118.022004}{\emph{Phys. Rev. Lett.}
  {\bfseries 118} (2017) 022004},
  [\href{https://arxiv.org/abs/1604.01404}{{\ttfamily 1604.01404}}].

\bibitem{Moch:2018wjh}
S.~Moch, B.~Ruijl, T.~Ueda, J.~A.~M. Vermaseren and A.~Vogt, \emph{{On quartic
  colour factors in splitting functions and the gluon cusp anomalous
  dimension}},
  \href{https://doi.org/10.1016/j.physletb.2018.06.017}{\emph{Phys. Lett. B}
  {\bfseries 782} (2018) 627--632},
  [\href{https://arxiv.org/abs/1805.09638}{{\ttfamily 1805.09638}}].

\bibitem{Moch:2017uml}
S.~Moch, B.~Ruijl, T.~Ueda, J.~A.~M. Vermaseren and A.~Vogt, \emph{{Four-Loop
  Non-Singlet Splitting Functions in the Planar Limit and Beyond}},
  \href{https://doi.org/10.1007/JHEP10(2017)041}{\emph{JHEP} {\bfseries 10}
  (2017) 041}, [\href{https://arxiv.org/abs/1707.08315}{{\ttfamily
  1707.08315}}].

\bibitem{Davies:2016jie}
J.~Davies, A.~Vogt, B.~Ruijl, T.~Ueda and J.~A.~M. Vermaseren, \emph{{Large-nf
  contributions to the four-loop splitting functions in QCD}},
  \href{https://doi.org/10.1016/j.nuclphysb.2016.12.012}{\emph{Nucl. Phys. B}
  {\bfseries 915} (2017) 335--362},
  [\href{https://arxiv.org/abs/1610.07477}{{\ttfamily 1610.07477}}].

\bibitem{Henn:2019swt}
J.~M. Henn, G.~P. Korchemsky and B.~Mistlberger, \emph{{The full four-loop cusp
  anomalous dimension in $\mathcal{N}=4$ super Yang-Mills and QCD}},
  \href{https://doi.org/10.1007/JHEP04(2020)018}{\emph{JHEP} {\bfseries 04}
  (2020) 018}, [\href{https://arxiv.org/abs/1911.10174}{{\ttfamily
  1911.10174}}].

\bibitem{Gehrmann:2012ze}
T.~Gehrmann, T.~Lubbert and L.~L. Yang, \emph{{Transverse parton distribution
  functions at next-to-next-to-leading order: the quark-to-quark case}},
  \href{https://doi.org/10.1103/PhysRevLett.109.242003}{\emph{Phys. Rev. Lett.}
  {\bfseries 109} (2012) 242003},
  [\href{https://arxiv.org/abs/1209.0682}{{\ttfamily 1209.0682}}].

\bibitem{Gehrmann:2014yya}
T.~Gehrmann, T.~Luebbert and L.~L. Yang, \emph{{Calculation of the transverse
  parton distribution functions at next-to-next-to-leading order}},
  \href{https://doi.org/10.1007/JHEP06(2014)155}{\emph{JHEP} {\bfseries 06}
  (2014) 155}, [\href{https://arxiv.org/abs/1403.6451}{{\ttfamily 1403.6451}}].

\bibitem{Echevarria:2016scs}
M.~G. Echevarria, I.~Scimemi and A.~Vladimirov, \emph{{Unpolarized Transverse
  Momentum Dependent Parton Distribution and Fragmentation Functions at
  next-to-next-to-leading order}},
  \href{https://doi.org/10.1007/JHEP09(2016)004}{\emph{JHEP} {\bfseries 09}
  (2016) 004}, [\href{https://arxiv.org/abs/1604.07869}{{\ttfamily
  1604.07869}}].

\bibitem{Luebbert:2016itl}
T.~L\"ubbert, J.~Oredsson and M.~Stahlhofen, \emph{{Rapidity renormalized TMD
  soft and beam functions at two loops}},
  \href{https://doi.org/10.1007/JHEP03(2016)168}{\emph{JHEP} {\bfseries 03}
  (2016) 168}, [\href{https://arxiv.org/abs/1602.01829}{{\ttfamily
  1602.01829}}].

\bibitem{Luo:2019hmp}
M.-X. Luo, X.~Wang, X.~Xu, L.~L. Yang, T.-Z. Yang and H.~X. Zhu,
  \emph{{Transverse Parton Distribution and Fragmentation Functions at NNLO:
  the Quark Case}}, \href{https://doi.org/10.1007/JHEP10(2019)083}{\emph{JHEP}
  {\bfseries 10} (2019) 083},
  [\href{https://arxiv.org/abs/1908.03831}{{\ttfamily 1908.03831}}].

\bibitem{Luo:2019bmw}
M.-X. Luo, T.-Z. Yang, H.~X. Zhu and Y.~J. Zhu, \emph{{Transverse Parton
  Distribution and Fragmentation Functions at NNLO: the Gluon Case}},
  \href{https://doi.org/10.1007/JHEP01(2020)040}{\emph{JHEP} {\bfseries 01}
  (2020) 040}, [\href{https://arxiv.org/abs/1909.13820}{{\ttfamily
  1909.13820}}].

\bibitem{Anastasiou:2000kg}
C.~Anastasiou, E.~W.~N. Glover, C.~Oleari and M.~E. Tejeda-Yeomans,
  \emph{{Two-loop QCD corrections to the scattering of massless distinct
  quarks}}, \href{https://doi.org/10.1016/S0550-3213(01)00079-7}{\emph{Nucl.
  Phys. B} {\bfseries 601} (2001) 318--340},
  [\href{https://arxiv.org/abs/hep-ph/0010212}{{\ttfamily hep-ph/0010212}}].

\bibitem{Anastasiou:2000ue}
C.~Anastasiou, E.~W.~N. Glover, C.~Oleari and M.~E. Tejeda-Yeomans, \emph{{Two
  loop QCD corrections to massless identical quark scattering}},
  \href{https://doi.org/10.1016/S0550-3213(01)00080-3}{\emph{Nucl. Phys. B}
  {\bfseries 601} (2001) 341--360},
  [\href{https://arxiv.org/abs/hep-ph/0011094}{{\ttfamily hep-ph/0011094}}].

\bibitem{Glover:2001af}
E.~W.~N. Glover, C.~Oleari and M.~E. Tejeda-Yeomans, \emph{{Two loop QCD
  corrections to gluon-gluon scattering}},
  \href{https://doi.org/10.1016/S0550-3213(01)00210-3}{\emph{Nucl. Phys. B}
  {\bfseries 605} (2001) 467--485},
  [\href{https://arxiv.org/abs/hep-ph/0102201}{{\ttfamily hep-ph/0102201}}].

\bibitem{Bern:2002tk}
Z.~Bern, A.~De~Freitas and L.~J. Dixon, \emph{{Two loop helicity amplitudes for
  gluon-gluon scattering in QCD and supersymmetric Yang-Mills theory}},
  \href{https://doi.org/10.1088/1126-6708/2002/03/018}{\emph{JHEP} {\bfseries
  03} (2002) 018}, [\href{https://arxiv.org/abs/hep-ph/0201161}{{\ttfamily
  hep-ph/0201161}}].

\bibitem{Bern:2003ck}
Z.~Bern, A.~De~Freitas and L.~J. Dixon, \emph{{Two loop helicity amplitudes for
  quark gluon scattering in QCD and gluino gluon scattering in supersymmetric
  Yang-Mills theory}},
  \href{https://doi.org/10.1088/1126-6708/2003/06/028}{\emph{JHEP} {\bfseries
  06} (2003) 028}, [\href{https://arxiv.org/abs/hep-ph/0304168}{{\ttfamily
  hep-ph/0304168}}].

\bibitem{Glover:2003cm}
E.~W.~N. Glover and M.~E. Tejeda-Yeomans, \emph{{Two loop QCD helicity
  amplitudes for massless quark massless gauge boson scattering}},
  \href{https://doi.org/10.1088/1126-6708/2003/06/033}{\emph{JHEP} {\bfseries
  06} (2003) 033}, [\href{https://arxiv.org/abs/hep-ph/0304169}{{\ttfamily
  hep-ph/0304169}}].

\bibitem{Glover:2004si}
E.~W.~N. Glover, \emph{{Two loop QCD helicity amplitudes for massless quark
  quark scattering}},
  \href{https://doi.org/10.1088/1126-6708/2004/04/021}{\emph{JHEP} {\bfseries
  04} (2004) 021}, [\href{https://arxiv.org/abs/hep-ph/0401119}{{\ttfamily
  hep-ph/0401119}}].

\bibitem{DeFreitas:2004kmi}
A.~De~Freitas and Z.~Bern, \emph{{Two-loop helicity amplitudes for quark-quark
  scattering in QCD and gluino-gluino scattering in supersymmetric Yang-Mills
  theory}}, \href{https://doi.org/10.1088/1126-6708/2004/09/039}{\emph{JHEP}
  {\bfseries 09} (2004) 039},
  [\href{https://arxiv.org/abs/hep-ph/0409007}{{\ttfamily hep-ph/0409007}}].

\bibitem{Czakon:2021mjy}
M.~Czakon, A.~Mitov and R.~Poncelet, \emph{{Next-to-Next-to-Leading Order Study
  of Three-Jet Production at the LHC}},
  \href{https://doi.org/10.1103/PhysRevLett.127.152001}{\emph{Phys. Rev. Lett.}
  {\bfseries 127} (2021) 152001},
  [\href{https://arxiv.org/abs/2106.05331}{{\ttfamily 2106.05331}}].

\bibitem{Abreu:2023rco}
S.~Abreu, D.~Chicherin, H.~Ita, B.~Page, V.~Sotnikov, W.~Tschernow et~al.,
  \emph{{All Two-Loop Feynman Integrals for Five-Point One-Mass Scattering}},
  \href{https://arxiv.org/abs/2306.15431}{{\ttfamily 2306.15431}}.

\bibitem{Chicherin:2021dyp}
D.~Chicherin, V.~Sotnikov and S.~Zoia, \emph{{Pentagon functions for one-mass
  planar scattering amplitudes}},
  \href{https://doi.org/10.1007/JHEP01(2022)096}{\emph{JHEP} {\bfseries 01}
  (2022) 096}, [\href{https://arxiv.org/abs/2110.10111}{{\ttfamily
  2110.10111}}].

\bibitem{Chicherin:2020oor}
D.~Chicherin and V.~Sotnikov, \emph{{Pentagon Functions for Scattering of Five
  Massless Particles}},
  \href{https://doi.org/10.1007/JHEP12(2020)167}{\emph{JHEP} {\bfseries 20}
  (2020) 167}, [\href{https://arxiv.org/abs/2009.07803}{{\ttfamily
  2009.07803}}].

\bibitem{Badger:2019djh}
S.~Badger, D.~Chicherin, T.~Gehrmann, G.~Heinrich, J.~M. Henn, T.~Peraro
  et~al., \emph{{Analytic form of the full two-loop five-gluon all-plus
  helicity amplitude}},
  \href{https://doi.org/10.1103/PhysRevLett.123.071601}{\emph{Phys. Rev. Lett.}
  {\bfseries 123} (2019) 071601},
  [\href{https://arxiv.org/abs/1905.03733}{{\ttfamily 1905.03733}}].

\bibitem{Chicherin:2018old}
D.~Chicherin, T.~Gehrmann, J.~M. Henn, P.~Wasser, Y.~Zhang and S.~Zoia,
  \emph{{All Master Integrals for Three-Jet Production at
  Next-to-Next-to-Leading Order}},
  \href{https://doi.org/10.1103/PhysRevLett.123.041603}{\emph{Phys. Rev. Lett.}
  {\bfseries 123} (2019) 041603},
  [\href{https://arxiv.org/abs/1812.11160}{{\ttfamily 1812.11160}}].

\bibitem{Chicherin:2018mue}
D.~Chicherin, T.~Gehrmann, J.~M. Henn, N.~A. Lo~Presti, V.~Mitev and P.~Wasser,
  \emph{{Analytic result for the nonplanar hexa-box integrals}},
  \href{https://doi.org/10.1007/JHEP03(2019)042}{\emph{JHEP} {\bfseries 03}
  (2019) 042}, [\href{https://arxiv.org/abs/1809.06240}{{\ttfamily
  1809.06240}}].

\bibitem{Chicherin:2018yne}
D.~Chicherin, T.~Gehrmann, J.~M. Henn, P.~Wasser, Y.~Zhang and S.~Zoia,
  \emph{{Analytic result for a two-loop five-particle amplitude}},
  \href{https://doi.org/10.1103/PhysRevLett.122.121602}{\emph{Phys. Rev. Lett.}
  {\bfseries 122} (2019) 121602},
  [\href{https://arxiv.org/abs/1812.11057}{{\ttfamily 1812.11057}}].

\bibitem{Abreu:2018aqd}
S.~Abreu, L.~J. Dixon, E.~Herrmann, B.~Page and M.~Zeng, \emph{{The two-loop
  five-point amplitude in $\mathcal{N} =4$ super-Yang-Mills theory}},
  \href{https://doi.org/10.1103/PhysRevLett.122.121603}{\emph{Phys. Rev. Lett.}
  {\bfseries 122} (2019) 121603},
  [\href{https://arxiv.org/abs/1812.08941}{{\ttfamily 1812.08941}}].

\bibitem{Gehrmann:2015bfy}
T.~Gehrmann, J.~M. Henn and N.~A. Lo~Presti, \emph{{Analytic form of the
  two-loop planar five-gluon all-plus-helicity amplitude in QCD}},
  \href{https://doi.org/10.1103/PhysRevLett.116.062001}{\emph{Phys. Rev. Lett.}
  {\bfseries 116} (2016) 062001},
  [\href{https://arxiv.org/abs/1511.05409}{{\ttfamily 1511.05409}}].

\bibitem{Abreu:2018jgq}
S.~Abreu, F.~Febres~Cordero, H.~Ita, B.~Page and V.~Sotnikov, \emph{{Planar
  Two-Loop Five-Parton Amplitudes from Numerical Unitarity}},
  \href{https://doi.org/10.1007/JHEP11(2018)116}{\emph{JHEP} {\bfseries 11}
  (2018) 116}, [\href{https://arxiv.org/abs/1809.09067}{{\ttfamily
  1809.09067}}].

\bibitem{Abreu:2017hqn}
S.~Abreu, F.~Febres~Cordero, H.~Ita, B.~Page and M.~Zeng, \emph{{Planar
  Two-Loop Five-Gluon Amplitudes from Numerical Unitarity}},
  \href{https://doi.org/10.1103/PhysRevD.97.116014}{\emph{Phys. Rev. D}
  {\bfseries 97} (2018) 116014},
  [\href{https://arxiv.org/abs/1712.03946}{{\ttfamily 1712.03946}}].

\bibitem{Abreu:2018zmy}
S.~Abreu, J.~Dormans, F.~Febres~Cordero, H.~Ita and B.~Page, \emph{{Analytic
  Form of Planar Two-Loop Five-Gluon Scattering Amplitudes in QCD}},
  \href{https://doi.org/10.1103/PhysRevLett.122.082002}{\emph{Phys. Rev. Lett.}
  {\bfseries 122} (2019) 082002},
  [\href{https://arxiv.org/abs/1812.04586}{{\ttfamily 1812.04586}}].

\bibitem{Luo:2019szz}
M.-x. Luo, T.-Z. Yang, H.~X. Zhu and Y.~J. Zhu, \emph{{Quark Transverse Parton
  Distribution at the Next-to-Next-to-Next-to-Leading Order}},
  \href{https://doi.org/10.1103/PhysRevLett.124.092001}{\emph{Phys. Rev. Lett.}
  {\bfseries 124} (2020) 092001},
  [\href{https://arxiv.org/abs/1912.05778}{{\ttfamily 1912.05778}}].

\bibitem{Luo:2020epw}
M.-x. Luo, T.-Z. Yang, H.~X. Zhu and Y.~J. Zhu, \emph{{Unpolarized quark and
  gluon TMD PDFs and FFs at N$^{3}$LO}},
  \href{https://doi.org/10.1007/JHEP06(2021)115}{\emph{JHEP} {\bfseries 06}
  (2021) 115}, [\href{https://arxiv.org/abs/2012.03256}{{\ttfamily
  2012.03256}}].

\bibitem{Ebert:2020qef}
M.~A. Ebert, B.~Mistlberger and G.~Vita, \emph{{TMD fragmentation functions at
  N$^{3}$LO}}, \href{https://doi.org/10.1007/JHEP07(2021)121}{\emph{JHEP}
  {\bfseries 07} (2021) 121},
  [\href{https://arxiv.org/abs/2012.07853}{{\ttfamily 2012.07853}}].

\bibitem{Ebert:2020yqt}
M.~A. Ebert, B.~Mistlberger and G.~Vita, \emph{{Transverse momentum dependent
  PDFs at N$^3$LO}}, \href{https://doi.org/10.1007/JHEP09(2020)146}{\emph{JHEP}
  {\bfseries 09} (2020) 146},
  [\href{https://arxiv.org/abs/2006.05329}{{\ttfamily 2006.05329}}].

\bibitem{Moult:2022xzt}
I.~Moult, H.~X. Zhu and Y.~J. Zhu, \emph{{The four loop QCD rapidity anomalous
  dimension}}, \href{https://doi.org/10.1007/JHEP08(2022)280}{\emph{JHEP}
  {\bfseries 08} (2022) 280},
  [\href{https://arxiv.org/abs/2205.02249}{{\ttfamily 2205.02249}}].

\bibitem{Bargiela:2021wuy}
P.~Bargiela, F.~Caola, A.~von Manteuffel and L.~Tancredi, \emph{{Three-loop
  helicity amplitudes for diphoton production in gluon fusion}},
  \href{https://doi.org/10.1007/JHEP02(2022)153}{\emph{JHEP} {\bfseries 02}
  (2022) 153}, [\href{https://arxiv.org/abs/2111.13595}{{\ttfamily
  2111.13595}}].

\bibitem{Caola:2021izf}
F.~Caola, A.~Chakraborty, G.~Gambuti, A.~von Manteuffel and L.~Tancredi,
  \emph{{Three-Loop Gluon Scattering in QCD and the Gluon Regge Trajectory}},
  \href{https://doi.org/10.1103/PhysRevLett.128.212001}{\emph{Phys. Rev. Lett.}
  {\bfseries 128} (2022) 212001},
  [\href{https://arxiv.org/abs/2112.11097}{{\ttfamily 2112.11097}}].

\bibitem{Caola:2021rqz}
F.~Caola, A.~Chakraborty, G.~Gambuti, A.~von Manteuffel and L.~Tancredi,
  \emph{{Three-loop helicity amplitudes for four-quark scattering in massless
  QCD}}, \href{https://doi.org/10.1007/JHEP10(2021)206}{\emph{JHEP} {\bfseries
  10} (2021) 206}, [\href{https://arxiv.org/abs/2108.00055}{{\ttfamily
  2108.00055}}].

\bibitem{Caola:2020dfu}
F.~Caola, A.~Von~Manteuffel and L.~Tancredi, \emph{{Diphoton Amplitudes in
  Three-Loop Quantum Chromodynamics}},
  \href{https://doi.org/10.1103/PhysRevLett.126.112004}{\emph{Phys. Rev. Lett.}
  {\bfseries 126} (2021) 112004},
  [\href{https://arxiv.org/abs/2011.13946}{{\ttfamily 2011.13946}}].

\bibitem{Collins:1981uk}
J.~C. Collins and D.~E. Soper, \emph{{Back-To-Back Jets in QCD}},
  \href{https://doi.org/10.1016/0550-3213(81)90339-4}{\emph{Nucl. Phys. B}
  {\bfseries 193} (1981) 381}.

\bibitem{Collins:1981va}
J.~C. Collins and D.~E. Soper, \emph{{Back-To-Back Jets: Fourier Transform from
  B to K-Transverse}},
  \href{https://doi.org/10.1016/0550-3213(82)90453-9}{\emph{Nucl. Phys. B}
  {\bfseries 197} (1982) 446--476}.

\bibitem{Collins:1981ta}
J.~C. Collins and G.~F. Sterman, \emph{{Soft Partons in {QCD}}},
  \href{https://doi.org/10.1016/0550-3213(81)90370-9}{\emph{Nucl. Phys. B}
  {\bfseries 185} (1981) 172--188}.

\bibitem{Collins:1984kg}
J.~C. Collins, D.~E. Soper and G.~F. Sterman, \emph{{Transverse Momentum
  Distribution in Drell-Yan Pair and W and Z Boson Production}},
  \href{https://doi.org/10.1016/0550-3213(85)90479-1}{\emph{Nucl. Phys. B}
  {\bfseries 250} (1985) 199--224}.

\bibitem{Collins:1985ue}
J.~C. Collins, D.~E. Soper and G.~F. Sterman, \emph{{Factorization for Short
  Distance Hadron - Hadron Scattering}},
  \href{https://doi.org/10.1016/0550-3213(85)90565-6}{\emph{Nucl. Phys. B}
  {\bfseries 261} (1985) 104--142}.

\bibitem{Collins:1988ig}
J.~C. Collins, D.~E. Soper and G.~F. Sterman, \emph{{Soft Gluons and
  Factorization}},
  \href{https://doi.org/10.1016/0550-3213(88)90130-7}{\emph{Nucl. Phys. B}
  {\bfseries 308} (1988) 833--856}.

\bibitem{Collins:1989gx}
J.~C. Collins, D.~E. Soper and G.~F. Sterman, \emph{{Factorization of Hard
  Processes in QCD}},
  \href{https://doi.org/10.1142/9789814503266_0001}{\emph{Adv. Ser. Direct.
  High Energy Phys.} {\bfseries 5} (1989) 1--91},
  [\href{https://arxiv.org/abs/hep-ph/0409313}{{\ttfamily hep-ph/0409313}}].

\bibitem{Basham:1979gh}
C.~L. Basham, L.~S. Brown, S.~D. Ellis and S.~T. Love, \emph{{Energy
  Correlations in Perturbative Quantum Chromodynamics: A Conjecture for All
  Orders}}, \href{https://doi.org/10.1016/0370-2693(79)90601-4}{\emph{Phys.
  Lett. B} {\bfseries 85} (1979) 297--299}.

\bibitem{Basham:1978zq}
C.~L. Basham, L.~S. Brown, S.~D. Ellis and S.~T. Love, \emph{{Energy
  Correlations in electron-Positron Annihilation in Quantum Chromodynamics:
  Asymptotically Free Perturbation Theory}},
  \href{https://doi.org/10.1103/PhysRevD.19.2018}{\emph{Phys. Rev. D}
  {\bfseries 19} (1979) 2018}.

\bibitem{Basham:1978bw}
C.~L. Basham, L.~S. Brown, S.~D. Ellis and S.~T. Love, \emph{{Energy
  Correlations in electron - Positron Annihilation: Testing QCD}},
  \href{https://doi.org/10.1103/PhysRevLett.41.1585}{\emph{Phys. Rev. Lett.}
  {\bfseries 41} (1978) 1585}.

\bibitem{Basham:1977iq}
C.~L. Basham, L.~S. Brown, S.~D. Ellis and S.~T. Love, \emph{{Electron -
  Positron Annihilation Energy Pattern in Quantum Chromodynamics:
  Asymptotically Free Perturbation Theory}},
  \href{https://doi.org/10.1103/PhysRevD.17.2298}{\emph{Phys. Rev. D}
  {\bfseries 17} (1978) 2298}.

\bibitem{Ali:1984yp}
A.~Ali, E.~Pietarinen and W.~J. Stirling, \emph{{Transverse Energy-energy
  Correlations: A Test of Perturbative {QCD} for the Proton - Anti-proton
  Collider}}, \href{https://doi.org/10.1016/0370-2693(84)90283-1}{\emph{Phys.
  Lett. B} {\bfseries 141} (1984) 447--454}.

\bibitem{Kalinowski:1980wea}
J.~Kalinowski, K.~Konishi, P.~N. Scharbach and T.~R. Taylor, \emph{{RESOLVING
  QCD JETS BEYOND LEADING ORDER: QUARK DECAY PROBABILITIES}},
  \href{https://doi.org/10.1016/0550-3213(81)90352-7}{\emph{Nucl. Phys. B}
  {\bfseries 181} (1981) 253--276}.

\bibitem{Konishi:1979cb}
K.~Konishi, A.~Ukawa and G.~Veneziano, \emph{{Jet Calculus: A Simple Algorithm
  for Resolving QCD Jets}},
  \href{https://doi.org/10.1016/0550-3213(79)90053-1}{\emph{Nucl. Phys. B}
  {\bfseries 157} (1979) 45--107}.

\bibitem{Richards:1982te}
D.~G. Richards, W.~J. Stirling and S.~D. Ellis, \emph{{Second Order Corrections
  to the Energy-energy Correlation Function in Quantum Chromodynamics}},
  \href{https://doi.org/10.1016/0370-2693(82)90275-1}{\emph{Phys. Lett. B}
  {\bfseries 119} (1982) 193--197}.

\bibitem{Hartanto:2019uvl}
H.~B. Hartanto, S.~Badger, C.~Br\o{}nnum-Hansen and T.~Peraro, \emph{{A
  numerical evaluation of planar two-loop helicity amplitudes for a W-boson
  plus four partons}},
  \href{https://doi.org/10.1007/JHEP09(2019)119}{\emph{JHEP} {\bfseries 09}
  (2019) 119}, [\href{https://arxiv.org/abs/1906.11862}{{\ttfamily
  1906.11862}}].

\bibitem{Banfi:2002vw}
A.~Banfi, G.~Marchesini and G.~Smye, \emph{{Azimuthal correlation in DIS}},
  \href{https://doi.org/10.1088/1126-6708/2002/04/024}{\emph{JHEP} {\bfseries
  04} (2002) 024}, [\href{https://arxiv.org/abs/hep-ph/0203150}{{\ttfamily
  hep-ph/0203150}}].

\bibitem{Buffing:2018ggv}
M.~G.~A. Buffing, Z.-B. Kang, K.~Lee and X.~Liu, \emph{{A transverse momentum
  dependent framework for back-to-back photon+jet production}},
  \href{https://arxiv.org/abs/1812.07549}{{\ttfamily 1812.07549}}.

\bibitem{Chien:2019gyf}
Y.-T. Chien, D.~Y. Shao and B.~Wu, \emph{{Resummation of Boson-Jet Correlation
  at Hadron Colliders}},
  \href{https://doi.org/10.1007/JHEP11(2019)025}{\emph{JHEP} {\bfseries 11}
  (2019) 025}, [\href{https://arxiv.org/abs/1905.01335}{{\ttfamily
  1905.01335}}].

\bibitem{Chien:2022wiq}
Y.-T. Chien, R.~Rahn, D.~Y. Shao, W.~J. Waalewijn and B.~Wu, \emph{{Precision
  boson-jet azimuthal decorrelation at hadron colliders}},
  \href{https://doi.org/10.1007/JHEP02(2023)256}{\emph{JHEP} {\bfseries 02}
  (2023) 256}, [\href{https://arxiv.org/abs/2205.05104}{{\ttfamily
  2205.05104}}].

\bibitem{Chien:2020hzh}
Y.-T. Chien, R.~Rahn, S.~Schrijnder~van Velzen, D.~Y. Shao, W.~J. Waalewijn and
  B.~Wu, \emph{{Recoil-free azimuthal angle for precision boson-jet
  correlation}},
  \href{https://doi.org/10.1016/j.physletb.2021.136124}{\emph{Phys. Lett. B}
  {\bfseries 815} (2021) 136124},
  [\href{https://arxiv.org/abs/2005.12279}{{\ttfamily 2005.12279}}].

\bibitem{Stewart:2009yx}
I.~W. Stewart, F.~J. Tackmann and W.~J. Waalewijn, \emph{{Factorization at the
  LHC: From PDFs to Initial State Jets}},
  \href{https://doi.org/10.1103/PhysRevD.81.094035}{\emph{Phys. Rev. D}
  {\bfseries 81} (2010) 094035},
  [\href{https://arxiv.org/abs/0910.0467}{{\ttfamily 0910.0467}}].

\bibitem{Georgi:1977mg}
H.~Georgi and H.~D. Politzer, \emph{{Quark Decay Functions and Heavy Hadron
  Production in QCD}},
  \href{https://doi.org/10.1016/0550-3213(78)90269-9}{\emph{Nucl. Phys. B}
  {\bfseries 136} (1978) 445--460}.

\bibitem{Ellis:1978ty}
R.~K. Ellis, H.~Georgi, M.~Machacek, H.~D. Politzer and G.~G. Ross,
  \emph{{Perturbation Theory and the Parton Model in QCD}},
  \href{https://doi.org/10.1016/0550-3213(79)90105-6}{\emph{Nucl. Phys. B}
  {\bfseries 152} (1979) 285--329}.

\bibitem{Collins:1981uw}
J.~C. Collins and D.~E. Soper, \emph{{Parton Distribution and Decay
  Functions}}, \href{https://doi.org/10.1016/0550-3213(82)90021-9}{\emph{Nucl.
  Phys. B} {\bfseries 194} (1982) 445--492}.

\bibitem{Collins:2011zzd}
J.~Collins, \emph{{Foundations of Perturbative QCD}}, vol.~32 of
  \emph{Cambridge Monographs on Particle Physics, Nuclear Physics and
  Cosmology}.
\newblock Cambridge University Press, 7, 2023,
  \href{https://doi.org/10.1017/9781009401845}{10.1017/9781009401845}.

\bibitem{Moult:2015aoa}
I.~Moult, I.~W. Stewart, F.~J. Tackmann and W.~J. Waalewijn, \emph{{Employing
  Helicity Amplitudes for Resummation}},
  \href{https://doi.org/10.1103/PhysRevD.93.094003}{\emph{Phys. Rev. D}
  {\bfseries 93} (2016) 094003},
  [\href{https://arxiv.org/abs/1508.02397}{{\ttfamily 1508.02397}}].

\bibitem{Kunszt:1993sd}
Z.~Kunszt, A.~Signer and Z.~Trocsanyi, \emph{{One loop helicity amplitudes for
  all 2 ---\ensuremath{>} 2 processes in QCD and N=1 supersymmetric Yang-Mills
  theory}}, \href{https://doi.org/10.1016/0550-3213(94)90456-1}{\emph{Nucl.
  Phys. B} {\bfseries 411} (1994) 397--442},
  [\href{https://arxiv.org/abs/hep-ph/9305239}{{\ttfamily hep-ph/9305239}}].

\bibitem{Kelley:2010fn}
R.~Kelley and M.~D. Schwartz, \emph{{1-loop matching and NNLL resummation for
  all partonic 2 to 2 processes in QCD}},
  \href{https://doi.org/10.1103/PhysRevD.83.045022}{\emph{Phys. Rev. D}
  {\bfseries 83} (2011) 045022},
  [\href{https://arxiv.org/abs/1008.2759}{{\ttfamily 1008.2759}}].

\bibitem{Broggio:2014hoa}
A.~Broggio, A.~Ferroglia, B.~D. Pecjak and Z.~Zhang, \emph{{NNLO hard functions
  in massless QCD}}, \href{https://doi.org/10.1007/JHEP12(2014)005}{\emph{JHEP}
  {\bfseries 12} (2014) 005},
  [\href{https://arxiv.org/abs/1409.5294}{{\ttfamily 1409.5294}}].

\bibitem{Stewart:2010qs}
I.~W. Stewart, F.~J. Tackmann and W.~J. Waalewijn, \emph{{The Quark Beam
  Function at NNLL}},
  \href{https://doi.org/10.1007/JHEP09(2010)005}{\emph{JHEP} {\bfseries 09}
  (2010) 005}, [\href{https://arxiv.org/abs/1002.2213}{{\ttfamily 1002.2213}}].

\bibitem{Alioli:2016wqt}
S.~Alioli, C.~W. Bauer, S.~Guns and F.~J. Tackmann, \emph{{Underlying event
  sensitive observables in Drell-Yan production using GENEVA}},
  \href{https://doi.org/10.1140/epjc/s10052-016-4458-1}{\emph{Eur. Phys. J. C}
  {\bfseries 76} (2016) 614},
  [\href{https://arxiv.org/abs/1605.07192}{{\ttfamily 1605.07192}}].

\bibitem{Cacciari:2009dp}
M.~Cacciari, G.~P. Salam and S.~Sapeta, \emph{{On the characterisation of the
  underlying event}},
  \href{https://doi.org/10.1007/JHEP04(2010)065}{\emph{JHEP} {\bfseries 04}
  (2010) 065}, [\href{https://arxiv.org/abs/0912.4926}{{\ttfamily 0912.4926}}].

\bibitem{Sjostrand:2006za}
T.~Sjostrand, S.~Mrenna and P.~Z. Skands, \emph{{PYTHIA 6.4 Physics and
  Manual}}, \href{https://doi.org/10.1088/1126-6708/2006/05/026}{\emph{JHEP}
  {\bfseries 05} (2006) 026},
  [\href{https://arxiv.org/abs/hep-ph/0603175}{{\ttfamily hep-ph/0603175}}].

\bibitem{Sjostrand:2007gs}
T.~Sjostrand, S.~Mrenna and P.~Z. Skands, \emph{{A Brief Introduction to PYTHIA
  8.1}}, \href{https://doi.org/10.1016/j.cpc.2008.01.036}{\emph{Comput. Phys.
  Commun.} {\bfseries 178} (2008) 852--867},
  [\href{https://arxiv.org/abs/0710.3820}{{\ttfamily 0710.3820}}].

\bibitem{Stewart:2014nna}
I.~W. Stewart, F.~J. Tackmann and W.~J. Waalewijn, \emph{{Dissecting Soft
  Radiation with Factorization}},
  \href{https://doi.org/10.1103/PhysRevLett.114.092001}{\emph{Phys. Rev. Lett.}
  {\bfseries 114} (2015) 092001},
  [\href{https://arxiv.org/abs/1405.6722}{{\ttfamily 1405.6722}}].

\bibitem{Gehrmann:2023jyv}
T.~Gehrmann, P.~Jakub\v{c}\'\i{}k, C.~C. Mella, N.~Syrrakos and L.~Tancredi,
  \emph{{Planar three-loop QCD helicity amplitudes for V+jet production at
  hadron colliders}},
  \href{https://doi.org/10.1016/j.physletb.2023.138369}{\emph{Phys. Lett. B}
  {\bfseries 848} (2024) 138369},
  [\href{https://arxiv.org/abs/2307.15405}{{\ttfamily 2307.15405}}].

\bibitem{Henn:2023vbd}
J.~M. Henn, J.~Lim and W.~J. Torres~Bobadilla, \emph{{First look at the
  evaluation of three-loop non-planar Feynman diagrams for Higgs plus jet
  production}}, \href{https://doi.org/10.1007/JHEP05(2023)026}{\emph{JHEP}
  {\bfseries 05} (2023) 026},
  [\href{https://arxiv.org/abs/2302.12776}{{\ttfamily 2302.12776}}].

\bibitem{Bell:2018vaa}
G.~Bell, R.~Rahn and J.~Talbert, \emph{{Two-loop anomalous dimensions of
  generic dijet soft functions}},
  \href{https://doi.org/10.1016/j.nuclphysb.2018.09.026}{\emph{Nucl. Phys. B}
  {\bfseries 936} (2018) 520--541},
  [\href{https://arxiv.org/abs/1805.12414}{{\ttfamily 1805.12414}}].

\bibitem{Bell:2018oqa}
G.~Bell, R.~Rahn and J.~Talbert, \emph{{Generic dijet soft functions at
  two-loop order: correlated emissions}},
  \href{https://doi.org/10.1007/JHEP07(2019)101}{\emph{JHEP} {\bfseries 07}
  (2019) 101}, [\href{https://arxiv.org/abs/1812.08690}{{\ttfamily
  1812.08690}}].

\bibitem{Bell:2018mkk}
G.~Bell, B.~Dehnadi, T.~Mohrmann and R.~Rahn, \emph{{Automated Calculation of
  ${\pmb N}$-jet Soft Functions}},
  \href{https://doi.org/10.22323/1.303.0044}{\emph{PoS} {\bfseries LL2018}
  (2018) 044}, [\href{https://arxiv.org/abs/1808.07427}{{\ttfamily
  1808.07427}}].

\bibitem{Bell:2023yso}
G.~Bell, B.~Dehnadi, T.~Mohrmann and R.~Rahn, \emph{{The NNLO soft function for
  N-jettiness in hadronic collisions}},
  \href{https://arxiv.org/abs/2312.11626}{{\ttfamily 2312.11626}}.

\bibitem{Jin:2019dho}
S.~Jin and X.~Liu, \emph{{Two-loop $N$-jettiness soft function for $pp \to 2j$
  production}}, \href{https://doi.org/10.1103/PhysRevD.99.114017}{\emph{Phys.
  Rev. D} {\bfseries 99} (2019) 114017},
  [\href{https://arxiv.org/abs/1901.10935}{{\ttfamily 1901.10935}}].

\bibitem{Dixon:2017nat}
L.~J. Dixon, \emph{{The Principle of Maximal Transcendentality and the
  Four-Loop Collinear Anomalous Dimension}},
  \href{https://doi.org/10.1007/JHEP01(2018)075}{\emph{JHEP} {\bfseries 01}
  (2018) 075}, [\href{https://arxiv.org/abs/1712.07274}{{\ttfamily
  1712.07274}}].

\bibitem{Li:2014bfa}
Y.~Li, A.~von Manteuffel, R.~M. Schabinger and H.~X. Zhu, \emph{{N$^3$LO Higgs
  boson and Drell-Yan production at threshold: The one-loop two-emission
  contribution}}, \href{https://doi.org/10.1103/PhysRevD.90.053006}{\emph{Phys.
  Rev. D} {\bfseries 90} (2014) 053006},
  [\href{https://arxiv.org/abs/1404.5839}{{\ttfamily 1404.5839}}].

\bibitem{Vladimirov:2017ksc}
A.~Vladimirov, \emph{{Structure of rapidity divergences in multi-parton
  scattering soft factors}},
  \href{https://doi.org/10.1007/JHEP04(2018)045}{\emph{JHEP} {\bfseries 04}
  (2018) 045}, [\href{https://arxiv.org/abs/1707.07606}{{\ttfamily
  1707.07606}}].

\bibitem{Vladimirov:2016dll}
A.~A. Vladimirov, \emph{{Correspondence between Soft and Rapidity Anomalous
  Dimensions}},
  \href{https://doi.org/10.1103/PhysRevLett.118.062001}{\emph{Phys. Rev. Lett.}
  {\bfseries 118} (2017) 062001},
  [\href{https://arxiv.org/abs/1610.05791}{{\ttfamily 1610.05791}}].

\bibitem{Li:2016axz}
Y.~Li, D.~Neill and H.~X. Zhu, \emph{{An exponential regulator for rapidity
  divergences}},
  \href{https://doi.org/10.1016/j.nuclphysb.2020.115193}{\emph{Nucl. Phys. B}
  {\bfseries 960} (2020) 115193},
  [\href{https://arxiv.org/abs/1604.00392}{{\ttfamily 1604.00392}}].

\bibitem{Kidonakis:1998bk}
N.~Kidonakis, G.~Oderda and G.~F. Sterman, \emph{{Threshold resummation for
  dijet cross-sections}},
  \href{https://doi.org/10.1016/S0550-3213(98)00243-0}{\emph{Nucl. Phys. B}
  {\bfseries 525} (1998) 299--332},
  [\href{https://arxiv.org/abs/hep-ph/9801268}{{\ttfamily hep-ph/9801268}}].

\bibitem{Kidonakis:1998nf}
N.~Kidonakis, G.~Oderda and G.~F. Sterman, \emph{{Evolution of color exchange
  in QCD hard scattering}},
  \href{https://doi.org/10.1016/S0550-3213(98)00441-6}{\emph{Nucl. Phys. B}
  {\bfseries 531} (1998) 365--402},
  [\href{https://arxiv.org/abs/hep-ph/9803241}{{\ttfamily hep-ph/9803241}}].

\bibitem{Aybat:2006mz}
S.~M. Aybat, L.~J. Dixon and G.~F. Sterman, \emph{{The Two-loop soft anomalous
  dimension matrix and resummation at next-to-next-to leading pole}},
  \href{https://doi.org/10.1103/PhysRevD.74.074004}{\emph{Phys. Rev. D}
  {\bfseries 74} (2006) 074004},
  [\href{https://arxiv.org/abs/hep-ph/0607309}{{\ttfamily hep-ph/0607309}}].

\bibitem{Aybat:2006wq}
S.~M. Aybat, L.~J. Dixon and G.~F. Sterman, \emph{{The Two-loop anomalous
  dimension matrix for soft gluon exchange}},
  \href{https://doi.org/10.1103/PhysRevLett.97.072001}{\emph{Phys. Rev. Lett.}
  {\bfseries 97} (2006) 072001},
  [\href{https://arxiv.org/abs/hep-ph/0606254}{{\ttfamily hep-ph/0606254}}].

\bibitem{Korchemsky:1987wg}
G.~P. Korchemsky and A.~V. Radyushkin, \emph{{Renormalization of the Wilson
  Loops Beyond the Leading Order}},
  \href{https://doi.org/10.1016/0550-3213(87)90277-X}{\emph{Nucl. Phys. B}
  {\bfseries 283} (1987) 342--364}.

\bibitem{Li:2014afw}
Y.~Li, A.~von Manteuffel, R.~M. Schabinger and H.~X. Zhu, \emph{{Soft-virtual
  corrections to Higgs production at N$^3$LO}},
  \href{https://doi.org/10.1103/PhysRevD.91.036008}{\emph{Phys. Rev. D}
  {\bfseries 91} (2015) 036008},
  [\href{https://arxiv.org/abs/1412.2771}{{\ttfamily 1412.2771}}].

\bibitem{Chiu:2011qc}
J.-y. Chiu, A.~Jain, D.~Neill and I.~Z. Rothstein, \emph{{The Rapidity
  Renormalization Group}},
  \href{https://doi.org/10.1103/PhysRevLett.108.151601}{\emph{Phys. Rev. Lett.}
  {\bfseries 108} (2012) 151601},
  [\href{https://arxiv.org/abs/1104.0881}{{\ttfamily 1104.0881}}].

\bibitem{Chiu:2012ir}
J.-Y. Chiu, A.~Jain, D.~Neill and I.~Z. Rothstein, \emph{{A Formalism for the
  Systematic Treatment of Rapidity Logarithms in Quantum Field Theory}},
  \href{https://doi.org/10.1007/JHEP05(2012)084}{\emph{JHEP} {\bfseries 05}
  (2012) 084}, [\href{https://arxiv.org/abs/1202.0814}{{\ttfamily 1202.0814}}].

\bibitem{Gatheral:1983cz}
J.~G.~M. Gatheral, \emph{{Exponentiation of Eikonal Cross-sections in
  Nonabelian Gauge Theories}},
  \href{https://doi.org/10.1016/0370-2693(83)90112-0}{\emph{Phys. Lett. B}
  {\bfseries 133} (1983) 90--94}.

\bibitem{Frenkel:1984pz}
J.~Frenkel and J.~C. Taylor, \emph{{NONABELIAN EIKONAL EXPONENTIATION}},
  \href{https://doi.org/10.1016/0550-3213(84)90294-3}{\emph{Nucl. Phys. B}
  {\bfseries 246} (1984) 231--245}.

\bibitem{Collins:2004nx}
J.~C. Collins and A.~Metz, \emph{{Universality of soft and collinear factors in
  hard-scattering factorization}},
  \href{https://doi.org/10.1103/PhysRevLett.93.252001}{\emph{Phys. Rev. Lett.}
  {\bfseries 93} (2004) 252001},
  [\href{https://arxiv.org/abs/hep-ph/0408249}{{\ttfamily hep-ph/0408249}}].

\bibitem{Catani:1999ss}
S.~Catani and M.~Grazzini, \emph{{Infrared factorization of tree level QCD
  amplitudes at the next-to-next-to-leading order and beyond}},
  \href{https://doi.org/10.1016/S0550-3213(99)00778-6}{\emph{Nucl. Phys. B}
  {\bfseries 570} (2000) 287--325},
  [\href{https://arxiv.org/abs/hep-ph/9908523}{{\ttfamily hep-ph/9908523}}].

\bibitem{Catani:2000pi}
S.~Catani and M.~Grazzini, \emph{{The soft gluon current at one loop order}},
  \href{https://doi.org/10.1016/S0550-3213(00)00572-1}{\emph{Nucl. Phys. B}
  {\bfseries 591} (2000) 435--454},
  [\href{https://arxiv.org/abs/hep-ph/0007142}{{\ttfamily hep-ph/0007142}}].

\bibitem{Falcioni:2019nxk}
G.~Falcioni, E.~Gardi and C.~Milloy, \emph{{Relating amplitude and PDF
  factorisation through Wilson-line geometries}},
  \href{https://doi.org/10.1007/JHEP11(2019)100}{\emph{JHEP} {\bfseries 11}
  (2019) 100}, [\href{https://arxiv.org/abs/1909.00697}{{\ttfamily
  1909.00697}}].

\bibitem{Dixon:2016epj}
L.~J. Dixon and I.~Esterlis, \emph{{All orders results for self-crossing Wilson
  loops mimicking double parton scattering}},
  \href{https://doi.org/10.1007/JHEP07(2016)116}{\emph{JHEP} {\bfseries 07}
  (2016) 116}, [\href{https://arxiv.org/abs/1602.02107}{{\ttfamily
  1602.02107}}].

\bibitem{Catani:1996vz}
S.~Catani and M.~H. Seymour, \emph{{A General algorithm for calculating jet
  cross-sections in NLO QCD}},
  \href{https://doi.org/10.1016/S0550-3213(96)00589-5}{\emph{Nucl. Phys. B}
  {\bfseries 485} (1997) 291--419},
  [\href{https://arxiv.org/abs/hep-ph/9605323}{{\ttfamily hep-ph/9605323}}].

\bibitem{Dixon:2009gx}
L.~J. Dixon, \emph{{Matter Dependence of the Three-Loop Soft Anomalous
  Dimension Matrix}},
  \href{https://doi.org/10.1103/PhysRevD.79.091501}{\emph{Phys. Rev. D}
  {\bfseries 79} (2009) 091501},
  [\href{https://arxiv.org/abs/0901.3414}{{\ttfamily 0901.3414}}].

\bibitem{Dixon:2008gr}
L.~J. Dixon, L.~Magnea and G.~F. Sterman, \emph{{Universal structure of
  subleading infrared poles in gauge theory amplitudes}},
  \href{https://doi.org/10.1088/1126-6708/2008/08/022}{\emph{JHEP} {\bfseries
  08} (2008) 022}, [\href{https://arxiv.org/abs/0805.3515}{{\ttfamily
  0805.3515}}].

\bibitem{Gardi:2009zv}
E.~Gardi and L.~Magnea, \emph{{Infrared singularities in QCD amplitudes}},
  \href{https://doi.org/10.1393/ncc/i2010-10528-x}{\emph{Nuovo Cim. C}
  {\bfseries 32N5-6} (2009) 137--157},
  [\href{https://arxiv.org/abs/0908.3273}{{\ttfamily 0908.3273}}].

\bibitem{Gardi:2009qi}
E.~Gardi and L.~Magnea, \emph{{Factorization constraints for soft anomalous
  dimensions in QCD scattering amplitudes}},
  \href{https://doi.org/10.1088/1126-6708/2009/03/079}{\emph{JHEP} {\bfseries
  03} (2009) 079}, [\href{https://arxiv.org/abs/0901.1091}{{\ttfamily
  0901.1091}}].

\bibitem{Dixon:2009ur}
L.~J. Dixon, E.~Gardi and L.~Magnea, \emph{{On soft singularities at three
  loops and beyond}},
  \href{https://doi.org/10.1007/JHEP02(2010)081}{\emph{JHEP} {\bfseries 02}
  (2010) 081}, [\href{https://arxiv.org/abs/0910.3653}{{\ttfamily 0910.3653}}].

\bibitem{Becher:2009cu}
T.~Becher and M.~Neubert, \emph{{Infrared singularities of scattering
  amplitudes in perturbative QCD}},
  \href{https://doi.org/10.1103/PhysRevLett.102.162001}{\emph{Phys. Rev. Lett.}
  {\bfseries 102} (2009) 162001},
  [\href{https://arxiv.org/abs/0901.0722}{{\ttfamily 0901.0722}}].

\bibitem{Becher:2009qa}
T.~Becher and M.~Neubert, \emph{{On the Structure of Infrared Singularities of
  Gauge-Theory Amplitudes}},
  \href{https://doi.org/10.1088/1126-6708/2009/06/081}{\emph{JHEP} {\bfseries
  06} (2009) 081}, [\href{https://arxiv.org/abs/0903.1126}{{\ttfamily
  0903.1126}}].

\bibitem{Gardi:2010rn}
E.~Gardi, E.~Laenen, G.~Stavenga and C.~D. White, \emph{{Webs in multiparton
  scattering using the replica trick}},
  \href{https://doi.org/10.1007/JHEP11(2010)155}{\emph{JHEP} {\bfseries 11}
  (2010) 155}, [\href{https://arxiv.org/abs/1008.0098}{{\ttfamily 1008.0098}}].

\bibitem{DelDuca:2011ae}
V.~Del~Duca, C.~Duhr, E.~Gardi, L.~Magnea and C.~D. White, \emph{{The Infrared
  structure of gauge theory amplitudes in the high-energy limit}},
  \href{https://doi.org/10.1007/JHEP12(2011)021}{\emph{JHEP} {\bfseries 12}
  (2011) 021}, [\href{https://arxiv.org/abs/1109.3581}{{\ttfamily 1109.3581}}].

\bibitem{Gardi:2011wa}
E.~Gardi and C.~D. White, \emph{{General properties of multiparton webs: Proofs
  from combinatorics}},
  \href{https://doi.org/10.1007/JHEP03(2011)079}{\emph{JHEP} {\bfseries 03}
  (2011) 079}, [\href{https://arxiv.org/abs/1102.0756}{{\ttfamily 1102.0756}}].

\bibitem{Gardi:2011yz}
E.~Gardi, J.~M. Smillie and C.~D. White, \emph{{On the renormalization of
  multiparton webs}},
  \href{https://doi.org/10.1007/JHEP09(2011)114}{\emph{JHEP} {\bfseries 09}
  (2011) 114}, [\href{https://arxiv.org/abs/1108.1357}{{\ttfamily 1108.1357}}].

\bibitem{Bret:2011xm}
V.~Del~Duca, C.~Duhr, E.~Gardi, L.~Magnea and C.~D. White, \emph{{An infrared
  approach to Reggeization}},
  \href{https://doi.org/10.1103/PhysRevD.85.071104}{\emph{Phys. Rev. D}
  {\bfseries 85} (2012) 071104},
  [\href{https://arxiv.org/abs/1108.5947}{{\ttfamily 1108.5947}}].

\bibitem{Ahrens:2012qz}
V.~Ahrens, M.~Neubert and L.~Vernazza, \emph{{Structure of Infrared
  Singularities of Gauge-Theory Amplitudes at Three and Four Loops}},
  \href{https://doi.org/10.1007/JHEP09(2012)138}{\emph{JHEP} {\bfseries 09}
  (2012) 138}, [\href{https://arxiv.org/abs/1208.4847}{{\ttfamily 1208.4847}}].

\bibitem{DelDuca:2013ara}
V.~Del~Duca, G.~Falcioni, L.~Magnea and L.~Vernazza, \emph{{High-energy QCD
  amplitudes at two loops and beyond}},
  \href{https://doi.org/10.1016/j.physletb.2014.03.033}{\emph{Phys. Lett. B}
  {\bfseries 732} (2014) 233--240},
  [\href{https://arxiv.org/abs/1311.0304}{{\ttfamily 1311.0304}}].

\bibitem{Caron-Huot:2013fea}
S.~Caron-Huot, \emph{{When does the gluon reggeize?}},
  \href{https://doi.org/10.1007/JHEP05(2015)093}{\emph{JHEP} {\bfseries 05}
  (2015) 093}, [\href{https://arxiv.org/abs/1309.6521}{{\ttfamily 1309.6521}}].

\bibitem{Gardi:2013saa}
E.~Gardi, \emph{{From Webs to Polylogarithms}},
  \href{https://doi.org/10.1007/JHEP04(2014)044}{\emph{JHEP} {\bfseries 04}
  (2014) 044}, [\href{https://arxiv.org/abs/1310.5268}{{\ttfamily 1310.5268}}].

\bibitem{Gardi:2013ita}
E.~Gardi, J.~M. Smillie and C.~D. White, \emph{{The Non-Abelian Exponentiation
  theorem for multiple Wilson lines}},
  \href{https://doi.org/10.1007/JHEP06(2013)088}{\emph{JHEP} {\bfseries 06}
  (2013) 088}, [\href{https://arxiv.org/abs/1304.7040}{{\ttfamily 1304.7040}}].

\bibitem{Falcioni:2014pka}
G.~Falcioni, E.~Gardi, M.~Harley, L.~Magnea and C.~D. White, \emph{{Multiple
  Gluon Exchange Webs}},
  \href{https://doi.org/10.1007/JHEP10(2014)010}{\emph{JHEP} {\bfseries 10}
  (2014) 010}, [\href{https://arxiv.org/abs/1407.3477}{{\ttfamily 1407.3477}}].

\bibitem{Becher:2019avh}
T.~Becher and M.~Neubert, \emph{{Infrared singularities of scattering
  amplitudes and N$^{3}$LL resummation for $n$-jet processes}},
  \href{https://doi.org/10.1007/JHEP01(2020)025}{\emph{JHEP} {\bfseries 01}
  (2020) 025}, [\href{https://arxiv.org/abs/1908.11379}{{\ttfamily
  1908.11379}}].

\bibitem{Henn:2016jdu}
J.~M. Henn and B.~Mistlberger, \emph{{Four-Gluon Scattering at Three Loops,
  Infrared Structure, and the Regge Limit}},
  \href{https://doi.org/10.1103/PhysRevLett.117.171601}{\emph{Phys. Rev. Lett.}
  {\bfseries 117} (2016) 171601},
  [\href{https://arxiv.org/abs/1608.00850}{{\ttfamily 1608.00850}}].

\bibitem{Buras:1991jm}
A.~J. Buras, M.~Jamin, M.~E. Lautenbacher and P.~H. Weisz, \emph{{Effective
  Hamiltonians for $\Delta S = 1$ and $\Delta B = 1$ nonleptonic decays beyond
  the leading logarithmic approximation}},
  \href{https://doi.org/10.1016/0550-3213(92)90345-C}{\emph{Nucl. Phys. B}
  {\bfseries 370} (1992) 69--104}.

\bibitem{Chen:2020adz}
H.~Chen, I.~Moult and H.~X. Zhu, \emph{{Quantum Interference in Jet
  Substructure from Spinning Gluons}},
  \href{https://doi.org/10.1103/PhysRevLett.126.112003}{\emph{Phys. Rev. Lett.}
  {\bfseries 126} (2021) 112003},
  [\href{https://arxiv.org/abs/2011.02492}{{\ttfamily 2011.02492}}].

\bibitem{Chen:2021gdk}
H.~Chen, I.~Moult and H.~X. Zhu, \emph{{Spinning gluons from the QCD light-ray
  OPE}}, \href{https://doi.org/10.1007/JHEP08(2022)233}{\emph{JHEP} {\bfseries
  08} (2022) 233}, [\href{https://arxiv.org/abs/2104.00009}{{\ttfamily
  2104.00009}}].

\bibitem{Li:2023gkh}
X.~L. Li, X.~Liu, F.~Yuan and H.~X. Zhu, \emph{{Illuminating nucleon-gluon
  interference via calorimetric asymmetry}},
  \href{https://doi.org/10.1103/PhysRevD.108.L091502}{\emph{Phys. Rev. D}
  {\bfseries 108} (2023) L091502},
  [\href{https://arxiv.org/abs/2308.10942}{{\ttfamily 2308.10942}}].

\bibitem{Nagy:2001fj}
Z.~Nagy, \emph{{Three jet cross-sections in hadron hadron collisions at
  next-to-leading order}},
  \href{https://doi.org/10.1103/PhysRevLett.88.122003}{\emph{Phys. Rev. Lett.}
  {\bfseries 88} (2002) 122003},
  [\href{https://arxiv.org/abs/hep-ph/0110315}{{\ttfamily hep-ph/0110315}}].

\bibitem{Nagy:2003tz}
Z.~Nagy, \emph{{Next-to-leading order calculation of three jet observables in
  hadron hadron collision}},
  \href{https://doi.org/10.1103/PhysRevD.68.094002}{\emph{Phys. Rev. D}
  {\bfseries 68} (2003) 094002},
  [\href{https://arxiv.org/abs/hep-ph/0307268}{{\ttfamily hep-ph/0307268}}].

\bibitem{Cacciari:2008gp}
M.~Cacciari, G.~P. Salam and G.~Soyez, \emph{{The anti-$k_t$ jet clustering
  algorithm}}, \href{https://doi.org/10.1088/1126-6708/2008/04/063}{\emph{JHEP}
  {\bfseries 04} (2008) 063},
  [\href{https://arxiv.org/abs/0802.1189}{{\ttfamily 0802.1189}}].

\bibitem{Butterworth:2015oua}
J.~Butterworth et~al., \emph{{PDF4LHC recommendations for LHC Run II}},
  \href{https://doi.org/10.1088/0954-3899/43/2/023001}{\emph{J. Phys. G}
  {\bfseries 43} (2016) 023001},
  [\href{https://arxiv.org/abs/1510.03865}{{\ttfamily 1510.03865}}].

\bibitem{Ebert:2018gsn}
M.~A. Ebert, I.~Moult, I.~W. Stewart, F.~J. Tackmann, G.~Vita and H.~X. Zhu,
  \emph{{Subleading power rapidity divergences and power corrections for
  q$_{T}$}}, \href{https://doi.org/10.1007/JHEP04(2019)123}{\emph{JHEP}
  {\bfseries 04} (2019) 123},
  [\href{https://arxiv.org/abs/1812.08189}{{\ttfamily 1812.08189}}].

\bibitem{Moult:2019vou}
I.~Moult, G.~Vita and K.~Yan, \emph{{Subleading power resummation of rapidity
  logarithms: the energy-energy correlator in $ \mathcal{N} $ = 4 SYM}},
  \href{https://doi.org/10.1007/JHEP07(2020)005}{\emph{JHEP} {\bfseries 07}
  (2020) 005}, [\href{https://arxiv.org/abs/1912.02188}{{\ttfamily
  1912.02188}}].

\bibitem{Chen:2023wah}
H.~Chen, X.~Zhou and H.~X. Zhu, \emph{{Power corrections to energy flow
  correlations from large spin perturbation}},
  \href{https://doi.org/10.1007/JHEP10(2023)132}{\emph{JHEP} {\bfseries 10}
  (2023) 132}, [\href{https://arxiv.org/abs/2301.03616}{{\ttfamily
  2301.03616}}].

\bibitem{Agarwal:2023suw}
B.~Agarwal, F.~Buccioni, F.~Devoto, G.~Gambuti, A.~von Manteuffel and
  L.~Tancredi, \emph{{Five-Parton Scattering in QCD at Two Loops}},
  \href{https://arxiv.org/abs/2311.09870}{{\ttfamily 2311.09870}}.

\bibitem{DeLaurentis:2023nss}
G.~De~Laurentis, H.~Ita, M.~Klinkert and V.~Sotnikov, \emph{{Double-Virtual
  NNLO QCD Corrections for Five-Parton Scattering: The Gluon Channel}},
  \href{https://arxiv.org/abs/2311.10086}{{\ttfamily 2311.10086}}.

\bibitem{DeLaurentis:2023izi}
G.~De~Laurentis, H.~Ita and V.~Sotnikov, \emph{{Double-Virtual NNLO QCD
  Corrections for Five-Parton Scattering: The Quark Channels}},
  \href{https://arxiv.org/abs/2311.18752}{{\ttfamily 2311.18752}}.

\bibitem{Dokshitzer:1999sh}
Y.~L. Dokshitzer, G.~Marchesini and B.~R. Webber, \emph{{Nonperturbative
  effects in the energy energy correlation}},
  \href{https://doi.org/10.1088/1126-6708/1999/07/012}{\emph{JHEP} {\bfseries
  07} (1999) 012}, [\href{https://arxiv.org/abs/hep-ph/9905339}{{\ttfamily
  hep-ph/9905339}}].

\bibitem{Fiore:1992sa}
R.~Fiore, A.~Quartarolo and L.~Trentadue, \emph{{Energy-energy correlation for
  Theta ---\ensuremath{>} 180-degrees at LEP}},
  \href{https://doi.org/10.1016/0370-2693(92)91545-K}{\emph{Phys. Lett. B}
  {\bfseries 294} (1992) 431--435}.

\bibitem{deFlorian:2004mp}
D.~de~Florian and M.~Grazzini, \emph{{The Back-to-back region in e+ e-
  energy-energy correlation}},
  \href{https://doi.org/10.1016/j.nuclphysb.2004.10.051}{\emph{Nucl. Phys. B}
  {\bfseries 704} (2005) 387--403},
  [\href{https://arxiv.org/abs/hep-ph/0407241}{{\ttfamily hep-ph/0407241}}].

\bibitem{Schindler:2023cww}
S.~T. Schindler, I.~W. Stewart and Z.~Sun, \emph{{Renormalons in the
  energy-energy correlator}},
  \href{https://doi.org/10.1007/JHEP10(2023)187}{\emph{JHEP} {\bfseries 10}
  (2023) 187}, [\href{https://arxiv.org/abs/2305.19311}{{\ttfamily
  2305.19311}}].

\bibitem{Chang:2013rca}
H.-M. Chang, M.~Procura, J.~Thaler and W.~J. Waalewijn, \emph{{Calculating
  Track-Based Observables for the LHC}},
  \href{https://doi.org/10.1103/PhysRevLett.111.102002}{\emph{Phys. Rev. Lett.}
  {\bfseries 111} (2013) 102002},
  [\href{https://arxiv.org/abs/1303.6637}{{\ttfamily 1303.6637}}].

\bibitem{Chang:2013iba}
H.-M. Chang, M.~Procura, J.~Thaler and W.~J. Waalewijn, \emph{{Calculating
  Track Thrust with Track Functions}},
  \href{https://doi.org/10.1103/PhysRevD.88.034030}{\emph{Phys. Rev. D}
  {\bfseries 88} (2013) 034030},
  [\href{https://arxiv.org/abs/1306.6630}{{\ttfamily 1306.6630}}].

\bibitem{Li:2021zcf}
Y.~Li, I.~Moult, S.~S. van Velzen, W.~J. Waalewijn and H.~X. Zhu,
  \emph{{Extending Precision Perturbative QCD with Track Functions}},
  \href{https://doi.org/10.1103/PhysRevLett.128.182001}{\emph{Phys. Rev. Lett.}
  {\bfseries 128} (2022) 182001},
  [\href{https://arxiv.org/abs/2108.01674}{{\ttfamily 2108.01674}}].

\bibitem{Jaarsma:2022kdd}
M.~Jaarsma, Y.~Li, I.~Moult, W.~Waalewijn and H.~X. Zhu, \emph{{Renormalization
  group flows for track function moments}},
  \href{https://doi.org/10.1007/JHEP06(2022)139}{\emph{JHEP} {\bfseries 06}
  (2022) 139}, [\href{https://arxiv.org/abs/2201.05166}{{\ttfamily
  2201.05166}}].

\bibitem{Chen:2022pdu}
H.~Chen, M.~Jaarsma, Y.~Li, I.~Moult, W.~J. Waalewijn and H.~X. Zhu,
  \emph{{Multi-collinear splitting kernels for track function evolution}},
  \href{https://doi.org/10.1007/JHEP07(2023)185}{\emph{JHEP} {\bfseries 07}
  (2023) 185}, [\href{https://arxiv.org/abs/2210.10058}{{\ttfamily
  2210.10058}}].

\bibitem{Chen:2022muj}
H.~Chen, M.~Jaarsma, Y.~Li, I.~Moult, W.~J. Waalewijn and H.~X. Zhu,
  \emph{{Collinear Parton Dynamics Beyond DGLAP}},
  \href{https://arxiv.org/abs/2210.10061}{{\ttfamily 2210.10061}}.

\bibitem{Tarasov:1980au}
O.~V. Tarasov, A.~A. Vladimirov and A.~Y. Zharkov, \emph{{The Gell-Mann-Low
  Function of QCD in the Three Loop Approximation}},
  \href{https://doi.org/10.1016/0370-2693(80)90358-5}{\emph{Phys. Lett. B}
  {\bfseries 93} (1980) 429--432}.

\bibitem{Larin:1993tp}
S.~A. Larin and J.~A.~M. Vermaseren, \emph{{The Three loop QCD Beta function
  and anomalous dimensions}},
  \href{https://doi.org/10.1016/0370-2693(93)91441-O}{\emph{Phys. Lett. B}
  {\bfseries 303} (1993) 334--336},
  [\href{https://arxiv.org/abs/hep-ph/9302208}{{\ttfamily hep-ph/9302208}}].

\bibitem{Moch:2004pa}
S.~Moch, J.~A.~M. Vermaseren and A.~Vogt, \emph{{The Three loop splitting
  functions in QCD: The Nonsinglet case}},
  \href{https://doi.org/10.1016/j.nuclphysb.2004.03.030}{\emph{Nucl. Phys. B}
  {\bfseries 688} (2004) 101--134},
  [\href{https://arxiv.org/abs/hep-ph/0403192}{{\ttfamily hep-ph/0403192}}].

\bibitem{Moch:2005id}
S.~Moch, J.~A.~M. Vermaseren and A.~Vogt, \emph{{The Quark form-factor at
  higher orders}},
  \href{https://doi.org/10.1088/1126-6708/2005/08/049}{\emph{JHEP} {\bfseries
  08} (2005) 049}, [\href{https://arxiv.org/abs/hep-ph/0507039}{{\ttfamily
  hep-ph/0507039}}].

\bibitem{Moch:2005tm}
S.~Moch, J.~A.~M. Vermaseren and A.~Vogt, \emph{{Three-loop results for quark
  and gluon form-factors}},
  \href{https://doi.org/10.1016/j.physletb.2005.08.067}{\emph{Phys. Lett. B}
  {\bfseries 625} (2005) 245--252},
  [\href{https://arxiv.org/abs/hep-ph/0508055}{{\ttfamily hep-ph/0508055}}].

\bibitem{Idilbi:2005ni}
A.~Idilbi, X.-d. Ji, J.-P. Ma and F.~Yuan, \emph{{Threshold resummation for
  Higgs production in effective field theory}},
  \href{https://doi.org/10.1103/PhysRevD.73.077501}{\emph{Phys. Rev. D}
  {\bfseries 73} (2006) 077501},
  [\href{https://arxiv.org/abs/hep-ph/0509294}{{\ttfamily hep-ph/0509294}}].

\bibitem{Idilbi:2006dg}
A.~Idilbi, X.-d. Ji and F.~Yuan, \emph{{Resummation of threshold logarithms in
  effective field theory for DIS, Drell-Yan and Higgs production}},
  \href{https://doi.org/10.1016/j.nuclphysb.2006.07.002}{\emph{Nucl. Phys. B}
  {\bfseries 753} (2006) 42--68},
  [\href{https://arxiv.org/abs/hep-ph/0605068}{{\ttfamily hep-ph/0605068}}].

\bibitem{Becher:2006mr}
T.~Becher, M.~Neubert and B.~D. Pecjak, \emph{{Factorization and Momentum-Space
  Resummation in Deep-Inelastic Scattering}},
  \href{https://doi.org/10.1088/1126-6708/2007/01/076}{\emph{JHEP} {\bfseries
  01} (2007) 076}, [\href{https://arxiv.org/abs/hep-ph/0607228}{{\ttfamily
  hep-ph/0607228}}].

\end{thebibliography}\endgroup
%\Bibliography{new_note}{}
\bibliographystyle{JHEP}

\end{document}